\pdfoutput=1
\documentclass[12pt,a4paper]{article}
\usepackage{ifthen} 
\usepackage{booktabs,dcolumn}
\usepackage{tabularx}
\usepackage{verbatim}
\usepackage{float}
\usepackage{fixmath}
\usepackage{upgreek}

\usepackage{graphicx}
\usepackage{caption}
\usepackage{subfig}
\newboolean{pdflatex}
\setboolean{pdflatex}{true} 

\newboolean{articletitles}
\setboolean{articletitles}{true} 

\newboolean{uprightparticles}
\setboolean{uprightparticles}{false} 

\newboolean{inbibliography}
\setboolean{inbibliography}{false} 

\def\paperauthors{LHCb collaboration} 
\def\paperasciititle{Measurements of DACP in charmless four-body Lb and Xib decays} 
\def\papertitle{Measurements of \CP asymmetries in charmless four-body \Lb and \Xibz decays} 
\def\paperkeywords{{High Energy Physics}, {LHCb}} 
\def\papercopyright{2019 CERN for the benefit of the LHCb collaboration.}
\def\paperlicence{CC-BY-4.0 licence}
\def\paperlicenceurl{https://creativecommons.org/licenses/by/4.0/}

\usepackage[top=1in, bottom=1.25in, left=1in, right=1in]{geometry}

\usepackage{microtype}
\usepackage{lineno}  
\usepackage{xspace} 
\usepackage{caption} 

\usepackage{graphicx}  
\usepackage{color}
\usepackage{colortbl}
\graphicspath{{./figs/}} 
\DeclareGraphicsExtensions{.pdf,.PDF,png,.PNG}

\usepackage{amsmath} 
\usepackage{amssymb}
\usepackage{amsfonts}
\usepackage{upgreek} 

\newcommand*\patchAmsMathEnvironmentForLineno[1]{%
\expandafter\let\csname old#1\expandafter\endcsname\csname #1\endcsname
\expandafter\let\csname oldend#1\expandafter\endcsname\csname
end#1\endcsname
 \renewenvironment{#1}%
   {\linenomath\csname old#1\endcsname}%
   {\csname oldend#1\endcsname\endlinenomath}%
}
\newcommand*\patchBothAmsMathEnvironmentsForLineno[1]{%
  \patchAmsMathEnvironmentForLineno{#1}%
  \patchAmsMathEnvironmentForLineno{#1*}%
}
\AtBeginDocument{%
\patchBothAmsMathEnvironmentsForLineno{equation}%
\patchBothAmsMathEnvironmentsForLineno{align}%
\patchBothAmsMathEnvironmentsForLineno{flalign}%
\patchBothAmsMathEnvironmentsForLineno{alignat}%
\patchBothAmsMathEnvironmentsForLineno{gather}%
\patchBothAmsMathEnvironmentsForLineno{multline}%
\patchBothAmsMathEnvironmentsForLineno{eqnarray}%
}

\usepackage{hyperxmp}

\usepackage[pdftex,
            pdfauthor={\paperauthors},
            pdftitle={\paperasciititle},
            pdfkeywords={\paperkeywords},
            pdfcopyright={Copyright (C) \papercopyright},
            pdflicenseurl={\paperlicenceurl}]{hyperref}

\usepackage[colorinlistoftodos,textsize=scriptsize]{todonotes}

\usepackage[all]{hypcap} 

\usepackage{xspace} 
\usepackage{upgreek}

\newcommand{\offsetoverline}[2][0.1em]{\kern #1\overline{\kern -#1 #2}}%

\def\lhcb   {\mbox{LHCb}\xspace}

\def\MagUp {\mbox{\em Mag\kern -0.05em Up}\xspace}

\ifthenelse{\boolean{uprightparticles}}%
{
 
 \def\Pgamma      {\ensuremath{\upgamma}\xspace}

 \def\Peta        {\ensuremath{\upeta}\xspace}

 \def\Ppi         {\ensuremath{\uppi}\xspace}                 
                  
 \def\Prho        {\ensuremath{\uprho}\xspace}

 \def\Pphi        {\ensuremath{\upphi}\xspace}

 \def\Ppsi        {\ensuremath{\uppsi}\xspace}

 \def\PDelta      {\ensuremath{\Delta}\xspace}                 
 \def\PXi         {\ensuremath{\Xi}\xspace}                 
 \def\PLambda     {\ensuremath{\Lambda}\xspace}                 
 \def\PSigma      {\ensuremath{\Sigma}\xspace}                 
 \def\POmega      {\ensuremath{\Omega}\xspace}                 
 \def\PUpsilon    {\ensuremath{\Upsilon}\xspace}

 \def\PB      {\ensuremath{\mathrm{B}}\xspace}                 
                  
 \def\PD      {\ensuremath{\mathrm{D}}\xspace}

 \def\PJ      {\ensuremath{\mathrm{J}}\xspace}                 
 \def\PK      {\ensuremath{\mathrm{K}}\xspace}

 \def\PX      {\ensuremath{\mathrm{X}}\xspace}

 \def\Pb      {\ensuremath{\mathrm{b}}\xspace}                 
 \def\Pc      {\ensuremath{\mathrm{c}}\xspace}                 
 \def\Pd      {\ensuremath{\mathrm{d}}\xspace}

 \def\Ph      {\ensuremath{\mathrm{h}}\xspace}                 
 \def\Pi      {\ensuremath{\mathrm{i}}\xspace}

 \def\Pp      {\ensuremath{\mathrm{p}}\xspace}

 \def\Ps      {\ensuremath{\mathrm{s}}\xspace}                 
                  
 \def\Pu      {\ensuremath{\mathrm{u}}\xspace}

}
{
 
 \def\Pgamma      {\ensuremath{\gamma}\xspace}

 \def\Peta        {\ensuremath{\eta}\xspace}

 \def\Ppi         {\ensuremath{\pi}\xspace}                 
                  
 \def\Prho        {\ensuremath{\rho}\xspace}

 \def\Pphi        {\ensuremath{\phi}\xspace}

 \def\Ppsi        {\ensuremath{\psi}\xspace}                 
                  
 \mathchardef\PDelta="7101
 \mathchardef\PXi="7104
 \mathchardef\PLambda="7103
 \mathchardef\PSigma="7106
 \mathchardef\POmega="710A
 \mathchardef\PUpsilon="7107
                  
 \def\PB      {\ensuremath{B}\xspace}                 
                  
 \def\PD      {\ensuremath{D}\xspace}

 \def\PJ      {\ensuremath{J}\xspace}                 
 \def\PK      {\ensuremath{K}\xspace}

 \def\PX      {\ensuremath{X}\xspace}

 \def\Pb      {\ensuremath{b}\xspace}                 
 \def\Pc      {\ensuremath{c}\xspace}                 
 \def\Pd      {\ensuremath{d}\xspace}

 \def\Ph      {\ensuremath{h}\xspace}                 
 \def\Pi      {\ensuremath{i}\xspace}

 \def\Pp      {\ensuremath{p}\xspace}

 \def\Ps      {\ensuremath{s}\xspace}                 
                  
 \def\Pu      {\ensuremath{u}\xspace}

}

\makeatletter
\ifcase \@ptsize \relax
  \newcommand{\miniscule}{\@setfontsize\miniscule{4}{5}}
\or
  \newcommand{\miniscule}{\@setfontsize\miniscule{5}{6}}
\or
  \newcommand{\miniscule}{\@setfontsize\miniscule{5}{6}}
\fi
\makeatother

\DeclareRobustCommand{\optbar}[1]{\shortstack{{\miniscule (\rule[.5ex]{1.25em}{.18mm})}
  \\ [-.7ex] $#1$}}

\def\g      {{\ensuremath{\Pgamma}}\xspace}

\def\uquark    {{\ensuremath{\Pu}}\xspace}

\def\dquark    {{\ensuremath{\Pd}}\xspace}

\def\squark    {{\ensuremath{\Ps}}\xspace}

\def\cquark    {{\ensuremath{\Pc}}\xspace}
\def\cquarkbar {{\ensuremath{\overline \cquark}}\xspace}

\def\bquark    {{\ensuremath{\Pb}}\xspace}

\def\pion   {{\ensuremath{\Ppi}}\xspace}
\def\piz    {{\ensuremath{\pion^0}}\xspace}
\def\pip    {{\ensuremath{\pion^+}}\xspace}
\def\pim    {{\ensuremath{\pion^-}}\xspace}

\def\pimp   {{\ensuremath{\pion^\mp}}\xspace}

\def\rhoz     {{\ensuremath{\rhomeson^0}}\xspace}

\def\kaon    {{\ensuremath{\PK}}\xspace}
  \def\Kbar    {{\kern 0.2em\overline{\kern -0.2em \PK}{}}\xspace}

\def\KorKbar {\kern 0.18em\optbar{\kern -0.18em K}{}\xspace}

\def\Kp      {{\ensuremath{\kaon^+}}\xspace}
\def\Km      {{\ensuremath{\kaon^-}}\xspace}
\def\Kpm     {{\ensuremath{\kaon^\pm}}\xspace}

\def\KS      {{\ensuremath{\kaon^0_{\mathrm{S}}}}\xspace}

\def\Kstarz  {{\ensuremath{\kaon^*(892)^{0}}}\xspace}

\newcommand{\etapr}{\ensuremath{\Peta^{\prime}}\xspace}
\newcommand{\phiz}{\ensuremath{\Pphi}\xspace}

  \def\Dbar    {{\kern 0.2em\overline{\kern -0.2em \PD}{}}\xspace}
\def\D       {{\ensuremath{\PD}}\xspace}

\def\DorDbar {\kern 0.18em\optbar{\kern -0.18em D}{}\xspace}
\def\Dz      {{\ensuremath{\D^0}}\xspace}

\def\Dp      {{\ensuremath{\D^+}}\xspace}

\def\Dsp     {{\ensuremath{\D^+_\squark}}\xspace}

\def\B       {{\ensuremath{\PB}}\xspace}
\def\Bbar    {{\ensuremath{\kern 0.18em\overline{\kern -0.18em \PB}{}}}\xspace}

\def\BorBbar    {\kern 0.18em\optbar{\kern -0.18em B}{}\xspace}

\def\Bd      {{\ensuremath{\B^0}}\xspace}
\def\Bs      {{\ensuremath{\B^0_\squark}}\xspace}

\def\jpsi     {{\ensuremath{{\PJ\mskip -3mu/\mskip -2mu\Ppsi\mskip 2mu}}}\xspace}

\def\Y#1S{\ensuremath{\PUpsilon{(#1S)}}\xspace}

\def\Lz          {{\ensuremath{\PLambda}}\xspace}

\def\LorLbar     {\kern 0.18em\optbar{\kern -0.18em \PLambda}{}\xspace}

\def\Xires       {{\ensuremath{\PXi}}\xspace}

\def\Lc          {{\ensuremath{\Lz^+_\cquark}}\xspace}

\def\Xicp        {{\ensuremath{\Xires^+_\cquark}}\xspace}

\def\Lb           {{\ensuremath{\Lz^0_\bquark}}\xspace}

\def\Xibz         {{\ensuremath{\Xires^0_\bquark}}\xspace}

\newcommand{\decay}[2]{\mbox{\ensuremath{#1\!\to #2}}\xspace}         

\def\to                 {\ensuremath{\rightarrow}\xspace}

\def\CP                {{\ensuremath{C\!P}}\xspace}

\newcommand{\ACP}{{\ensuremath{{\mathcal{A}}^{\CP}}}\xspace}

\def\AT#1     {\ensuremath{A_{\mathrm{T}}^{#1}}\xspace}           

\def\C#1      {\ensuremath{\mathcal{C}_{#1}}\xspace}                       
\def\Cp#1     {\ensuremath{\mathcal{C}_{#1}^{'}}\xspace}                    
\def\Ceff#1   {\ensuremath{\mathcal{C}_{#1}^{\mathrm{(eff)}}}\xspace}        
\def\Cpeff#1  {\ensuremath{\mathcal{C}_{#1}^{'\mathrm{(eff)}}}\xspace}       
\def\Ope#1    {\ensuremath{\mathcal{O}_{#1}}\xspace}                       
\def\Opep#1   {\ensuremath{\mathcal{O}_{#1}^{'}}\xspace}                    

\newcommand{\nospaceunit}[1]{\ensuremath{\text{#1}}}       
\newcommand{\aunit}[1]{\ensuremath{\text{\,#1}}}       

\newcommand{\tev}{\aunit{Te\kern -0.1em V}\xspace}
\newcommand{\gev}{\aunit{Ge\kern -0.1em V}\xspace}
\newcommand{\mev}{\aunit{Me\kern -0.1em V}\xspace}
\newcommand{\kev}{\aunit{ke\kern -0.1em V}\xspace}
\newcommand{\ev}{\aunit{e\kern -0.1em V}\xspace}
\newcommand{\mevc}{\ensuremath{\aunit{Me\kern -0.1em V\!/}c}\xspace}
\newcommand{\gevc}{\ensuremath{\aunit{Ge\kern -0.1em V\!/}c}\xspace}
\newcommand{\mevcc}{\ensuremath{\aunit{Me\kern -0.1em V\!/}c^2}\xspace}
\newcommand{\gevcc}{\ensuremath{\aunit{Ge\kern -0.1em V\!/}c^2}\xspace}

\def\mum  {\ensuremath{\,\upmu\nospaceunit{m}}\xspace}

\def\fb   {\ensuremath{\aunit{fb}}\xspace}
\def\invfb   {\ensuremath{\fb^{-1}}\xspace}

\newcommand{\chisq}{\ensuremath{\chi^2}\xspace}

\newcommand{\chisqip}{\ensuremath{\chi^2_{\text{IP}}}\xspace}

\newcommand{\chisqvtx}{\ensuremath{\chi^2_{\text{vtx}}}\xspace}

\def\gsim{{~\raise.15em\hbox{$>$}\kern-.85em
          \lower.35em\hbox{$\sim$}~}\xspace}
\def\lsim{{~\raise.15em\hbox{$<$}\kern-.85em
          \lower.35em\hbox{$\sim$}~}\xspace}

\def\pt         {\ensuremath{p_{\mathrm{T}}}\xspace}

\def\ptot       {\ensuremath{p}\xspace}

\def\evtgen     {\mbox{\textsc{EvtGen}}\xspace}

\def\geant      {\mbox{\textsc{Geant4}}\xspace}

\def\photos     {\mbox{\textsc{Photos}}\xspace}

\def\pythia     {\mbox{\textsc{Pythia}}\xspace}

\xspace

\def\tell1  {TELL1\xspace}
\def\ukl1   {UKL1\xspace}

\newcommand{\ie}{\mbox{\itshape i.e.}\xspace}

\usepackage{cite} 
\usepackage{mciteplus}

\def\LbToppipipi              {\mbox{\ensuremath{\Lb\to \Pp\Ppi^-\Ppi^+\Ppi^-}}\xspace}
\def\LbTopKpipi               {\mbox{\ensuremath{\Lb\to \Pp\PK^-\Ppi^+\Ppi^-}}\xspace}

\def\LbTopKKpi                {\mbox{\ensuremath{\Lb\to \Pp\PK^-\PK^+\Ppi^-}}\xspace}

\def\LbTopKKK                 {\mbox{\ensuremath{\Lb\to \Pp\PK^-\PK^+\PK^-}}\xspace}
\def\XibzTopKpipi             {\mbox{\ensuremath{\Xibz\to \Pp\PK^-\Ppi^+\Ppi^-}}\xspace}

\def\XibzTopKpiK              {\mbox{\ensuremath{\Xibz\to \Pp\PK^-\Ppi^+\PK^-}}\xspace}

\def\XbTophhhz                {\mbox{\ensuremath{\Xb\to \Pp\Ph\Ph'\Ph''}}\xspace}

\def\Xb                       {\mbox{\ensuremath{\PX^0_\bquark}}\xspace}
\def\Xbbar                    {\mbox{\ensuremath{\overline{\PX}{}^0_\bquark}}\xspace}

\def\pipi                     {\mbox{\ensuremath{\Ppi\Ppi}}\xspace}

\def\kk                       {\mbox{\ensuremath{\PK\PK}}\xspace}

\def\ppipipi                  {\mbox{\ensuremath{\Pp\Ppi^-\Ppi^+\Ppi^-}}\xspace}
\def\pKpipi                   {\mbox{\ensuremath{\Pp\PK^-\Ppi^+\Ppi^-}}\xspace}

\def\pKKpi                    {\mbox{\ensuremath{\Pp\PK^-\PK^+\Ppi^-}}\xspace}
\def\pKpiK                    {\mbox{\ensuremath{\Pp\PK^-\Ppi^+\PK^-}}\xspace}

\def\pKKK                     {\mbox{\ensuremath{\Pp\PK^-\PK^+\PK^-}}\xspace}
\def\ppipipib                 {$\mathbold{\Pp\Ppi^-\Ppi^+\Ppi^-}$\xspace}
\def\pKpipib                  {$\mathbold{\Pp\PK^-\Ppi^+\Ppi^-}$\xspace}

\def\pKKpib                   {$\mathbold{\Pp\PK^-\PK^+\Ppi^-}$\xspace}
\def\pKpiKb                   {$\mathbold{\Pp\PK^-\Ppi^+\PK^-}$\xspace}

\def\pKKKb                    {$\mathbold{\Pp\PK^-\PK^+\PK^-}$\xspace}

\def\chisqfd                  {\mbox{\ensuremath{\chi^2_{\rm FD}}}\xspace}

\def\ACP                      {\mbox{\ensuremath{{\cal A}^{\CP}}}\xspace}

\def\dACP                     {\mbox{\ensuremath{\PDelta\ACP}}\xspace}

\def\ACPnoC                   {\mbox{\ensuremath{{\cal A}^{\CP}_{\rm no\text{-}c}}}\xspace}
\def\ACPC                   {\mbox{\ensuremath{{\cal A}^{\CP}_{\rm c}}}\xspace}

\def\LbToLcpiLcToppipi        {\mbox{\ensuremath{\Lb\to(\Lc\to\Pp\Ppi^-\Ppi^+)\Ppi^-}}\xspace}
\def\LbToLcpiLcTopKpi         {\mbox{\ensuremath{\Lb\to(\Lc\to\Pp\PK^-\Ppi^+)\Ppi^-}}\xspace}

\def\XibzToXicpiXicTopKpi     {\mbox{\ensuremath{\Xibz\to(\Xicp\to\Pp\PK^-\Ppi^+)\Ppi^-}}\xspace}

\def\LbTopKetap               {\mbox{\ensuremath{\Lb\to\Pp\PK^-\etapr}}\xspace}
\def\LbToppietap              {\mbox{\ensuremath{\Lb\to\Pp\Ppi^-\etapr}}\xspace}
\def\EtapTopipig              {\mbox{\ensuremath{\etapr\to\Ppi^+\Ppi^-\gamma}}\xspace}
\def\LbTopKpipipiz            {\mbox{\ensuremath{\Lb\to\Pp\PK^-\Ppi^+\Ppi^-\Ppi^0}}\xspace}
\def\LbToppipipipiz           {\mbox{\ensuremath{\Lb\to\Pp\Ppi^-\Ppi^+\Ppi^-\Ppi^0}}\xspace}
\def\LbTopKKpipiz             {\mbox{\ensuremath{\Lb\to\Pp\PK^-\PK^+\Ppi^-\Ppi^0}}\xspace}
\def\LbTopKKKpiz              {\mbox{\ensuremath{\Lb\to\Pp\PK^-\PK^+\PK^-\Ppi^0}}\xspace}
\def\LbTopKKKg                {\mbox{\ensuremath{\Lb\to\Pp\PK^-\PK^+\PK^-\g}}\xspace}

\def\DzorDzbar                {\kern 0.18em\optbar{\kern -0.18em \Dz}{}\xspace}

\def\pid                      {\mbox{\ensuremath{\mathrm{PID}}}\xspace}

\def\Y                        {\mbox{\ensuremath{\mathcal{N}}}\xspace}

\def\LbTopaone                {\mbox{\ensuremath{\Lb \to \Pp\aone(1260)^-}}\xspace}
\def\LbToNstarRhoOrFz         {\mbox{\ensuremath{\Lb \to N(1520)^0\rhoz}}\xspace}
\def\LbToDeltapipi            {\mbox{\ensuremath{\Lb \to \PDelta(1232)^{++}\Ppi^{-}\Ppi^{-}}}\xspace}

\def\LbToNstarKstar           {\mbox{\ensuremath{\Lb \to N(1520)^0\Kstarz}}\xspace}
\def\LbToLstarRhoOrFz         {\mbox{\ensuremath{\Lb \to \PLambda(1520)\rhoz}}\xspace}

\def\LbToDeltaKpi             {\mbox{\ensuremath{\Lb \to \PDelta(1232)^{++}\PK^{-}\Ppi^{-}}}\xspace}
\def\LbTopKone                {\mbox{\ensuremath{\Lb \to \Pp\Kone(1410)^-}}\xspace}

\def\LbToLstarPhi             {\mbox{\ensuremath{\Lb \to \PLambda(1520)\phiz}}\xspace}
\def\LbTopKPhi                {\mbox{\ensuremath{\Lb \to (\Pp \PK^{-})_\text{\tiny high-mass} \phiz}}\xspace}

\def\aone     {{\mbox{\ensuremath{a_{1}}}}\xspace}
\def\Kone     {{\mbox{\ensuremath{K_{1}}}}\xspace}

\ifthenelse{\boolean{uprightparticles}}%
{\def\Prho      {\ensuremath{\uprho}\xspace}
 \def\Pphi      {\ensuremath{\upphi}\xspace}
}
{\def\Prho      {\ensuremath{\rho}\xspace}
 \def\Pphi      {\ensuremath{\phi}\xspace}
}
\def\rhoz   {\ensuremath{\Prho(770)^0}\xspace}

\def\phiz   {\ensuremath{\Pphi(1020)}\xspace}

\def\had  {\ensuremath{\Ph}\xspace}

\def\BdorBdbar    {\ensuremath{\kern 0.18em\optbar{\kern -0.18em B}{}^0}\xspace}
\def\BsorBsbar    {\ensuremath{\kern 0.18em\optbar{\kern  0.06em B_s}{}^0}\xspace}

\def\pipi  {\ensuremath{\pion^+\pion^-}\xspace}

\def\kk      {\ensuremath{\Kp\Km}\xspace}

\def\BdtoKsKK   {\decay{\Bd}{\KS \Kp \Km}}
\def\BdtoKsPiPi   {\decay{\Bd}{\KS \pip \pim}}

\def\BdtoKsKpPim   {\decay{\Bd}{\KS \Kp \pim}}
\def\BdtoKsPipKm   {\decay{\Bd}{\KS \Km \pip}}

\def\BstoKsKK   {\decay{\Bs}{\KS \Kp \Km}}
\def\BstoKsPiPi   {\decay{\Bs}{\KS \pip \pim}}

\def\BstoKsKpPim   {\decay{\Bs}{\KS \Kp \pim}}
\def\BstoKsPipKm   {\decay{\Bs}{\KS \Km \pip}}

\def\KsPiPi{\ensuremath{\KS \pip \pim}\xspace}

\def\KsKK{\ensuremath{\KS \Kp \Km}\xspace}

\def\KsKpPim{\ensuremath{\KS \Kp \pim}\xspace}
\def\KsPipKm{\ensuremath{\KS \pip \Km}\xspace}

\def\LL   {Long-Long\xspace}
\def\DD   {Down-Down\xspace}

\def \fitCombShape{Polynomial}
\def \fitCombModel{from5150}

\newcommand{\plotIfExists}[2]{
  \IfFileExists{#1}{\includegraphics[width=#2\textwidth]{#1}}{\includegraphics[width=#2\textwidth]{#1}}
}    
\newcommand{\plotOne}[2]{
  \ifthenelse{\boolean{pdflatex}}{
    \plotIfExists{#1.pdf}{#2}
  }{
    \plotIfExists{#1.eps}{#2}
  }
}

\newcommand{\plotDataFitResults}[3]{ 
  \begin{figure}[p]
    \begin{center}
      \ifthenelse{\equal{\fitCombShape}{Exponential}}
                 {\def \tempFitCombShape{Exponential}}
                 {\def \tempFitCombShape{Polynomial}}
                 
                 \plotOne{figs/FitResults/#1/\fitCombModel-Louis-\tempFitCombShape-StrongPcut-#1-KSKK#2_#3-Standard-DoubleCB}{0.35}
                 \plotOne{figs/FitResults/#1/\fitCombModel-Louis-\tempFitCombShape-StrongPcut-#1-KSKK#2_#3-Standard-DoubleCB_log}{0.35}\\
                 \plotOne{figs/FitResults/#1/\fitCombModel-Louis-\tempFitCombShape-StrongPcut-#1-KSKpi#2_#3-Standard-DoubleCB}{0.35}
                 \plotOne{figs/FitResults/#1/\fitCombModel-Louis-\tempFitCombShape-StrongPcut-#1-KSKpi#2_#3-Standard-DoubleCB_log}{0.35}\\
                 \plotOne{figs/FitResults/#1/\fitCombModel-Louis-\tempFitCombShape-StrongPcut-#1-KSpiK#2_#3-Standard-DoubleCB}{0.35}
                 \plotOne{figs/FitResults/#1/\fitCombModel-Louis-\tempFitCombShape-StrongPcut-#1-KSpiK#2_#3-Standard-DoubleCB_log}{0.35}\\
                 \plotOne{figs/FitResults/#1/\fitCombModel-Louis-\tempFitCombShape-StrongPcut-#1-KSpipi#2_#3-Standard-DoubleCB}{0.35}
                 \plotOne{figs/FitResults/#1/\fitCombModel-Louis-\tempFitCombShape-StrongPcut-#1-KSpipi#2_#3-Standard-DoubleCB_log}{0.35}\\
                 \ifthenelse{\equal{#2}{DD}}{
                   \caption{Results of the simultaneous fit to data (\DD, #3) with the \MakeLowercase{#1} BDT optimisation. The modes \KsKK, \KsKpPim, \KsPipKm and \KsPiPi are shown from top to bottom. The left-hand side plots show the results with a linear scale and the right-hand side with a logarithmic scale.}
                 }{
                   \caption{Results of the simultaneous fit to data (\LL, #3) with the \MakeLowercase{#1} BDT optimisation. The modes \KsKK, \KsKpPim, \KsPipKm and \KsPiPi are shown from top to bottom. The left-hand side plots show the results with a linear scale and the right-hand side with a logarithmic scale.}
                 }
                 \label{fig:FitResult:#1:#2:#3}
    \end{center}
  \end{figure}
}
\newcommand{\plotSignalMCFitResults}[3]{
  \begin{figure}[!htbp]
    \begin{center}
      \plotOne{figs/FitModel/Signal/#1/Bd2KSKK#2_#3-MCFit-Louis-DoubleCB-Standard_log}{0.35}
      \plotOne{figs/FitModel/Signal/#1/Bs2KSKK#2_#3-MCFit-Louis-DoubleCB-Standard_log}{0.35}\\
      \plotOne{figs/FitModel/Signal/#1/Bd2KSKpi#2_#3-MCFit-Louis-DoubleCB-Standard_log}{0.35}
      \plotOne{figs/FitModel/Signal/#1/Bs2KSKpi#2_#3-MCFit-Louis-DoubleCB-Standard_log}{0.35}\\
      \plotOne{figs/FitModel/Signal/#1/Bd2KSpiK#2_#3-MCFit-Louis-DoubleCB-Standard_log}{0.35}
      \plotOne{figs/FitModel/Signal/#1/Bs2KSpiK#2_#3-MCFit-Louis-DoubleCB-Standard_log}{0.35}\\
      \plotOne{figs/FitModel/Signal/#1/Bd2KSpipi#2_#3-MCFit-Louis-DoubleCB-Standard_log}{0.35}
      \plotOne{figs/FitModel/Signal/#1/Bs2KSpipi#2_#3-MCFit-Louis-DoubleCB-Standard_log}{0.35}\\
      \ifthenelse{\equal{#2}{DD}}{
        \caption{Result of the simultaneous fit to simulated samples of the signal decays for \DD \KS reconstruction mode, using the \MakeLowercase{#1} optimisation of the BDT, and shown using a logarithmic scale. \KsKK, \KsKpPim, \KsPipKm, and \KsPiPi are shown from top to bottom, while \Bd decays are shown on the left and \Bs decays on the right. }
      }
                 {
                   \caption{Result of the simultaneous fit to simulated samples of the signal decays for \LL \KS reconstruction mode, using the \MakeLowercase{#1} optimisation of the BDT, and shown using a logarithmic scale. \KsKK, \KsKpPim, \KsPipKm, and \KsPiPi are shown from top to bottom, while \Bd decays are shown on the left and \Bs decays on the right. }
                 }
                 \label{fig:FitModel:Signal:#1:#2:#3}
    \end{center}
\end{figure}
}

\newcommand{\plotSplots}[3]{
  \ifthenelse{\equal{#3}{log}}
             {\def\suffix{FitStandard_log}}
             {\def\suffix{FitStandard}}
             \ifthenelse{\equal{#3}{log}}
                        {\def\scale{logarithmic}\xspace}
                        {\def\scale{linear}\xspace}
                        \begin{figure}[!htbp]
                          \begin{center}
                            \plotOne{figs/FitResults/sWeights/#1/sWeights-from5150-Louis-PolSlopes-StrongPcut-#1-KSKKDD_#2_\suffix}{0.35}
                            \plotOne{figs/FitResults/sWeights/#1/sWeights-from5150-Louis-PolSlopes-StrongPcut-#1-KSKKLL_#2_\suffix}{0.35}\\
                            \plotOne{figs/FitResults/sWeights/#1/sWeights-from5150-Louis-PolSlopes-StrongPcut-#1-KSKpiDD_#2_\suffix}{0.35}
                            \plotOne{figs/FitResults/sWeights/#1/sWeights-from5150-Louis-PolSlopes-StrongPcut-#1-KSKpiLL_#2_\suffix}{0.35}\\
                            \plotOne{figs/FitResults/sWeights/#1/sWeights-from5150-Louis-PolSlopes-StrongPcut-#1-KSpiKDD_#2_\suffix}{0.35}
                            \plotOne{figs/FitResults/sWeights/#1/sWeights-from5150-Louis-PolSlopes-StrongPcut-#1-KSpiKLL_#2_\suffix}{0.35}\\
                            \plotOne{figs/FitResults/sWeights/#1/sWeights-from5150-Louis-PolSlopes-StrongPcut-#1-KSpipiDD_#2_\suffix}{0.35}
                            \plotOne{figs/FitResults/sWeights/#1/sWeights-from5150-Louis-PolSlopes-StrongPcut-#1-KSpipiLL_#2_\suffix}{0.35}\\
                            \caption{Result of the invariant mass fits used for the sWeights extraction with the \MakeLowercase{#1} BDT optimisation on #2 data (\scale~scale). \KsKK, \KsKpPim, \KsPipKm, and \KsPiPi are shown from top to bottom, \DD on the left, \LL on the right.}
                            \label{fig:FitResult:Splots:#1:#2:#3}
                          \end{center}
                        \end{figure}
}
\newcommand{\plotSplotsDalitz}[3]{ 
  \ifthenelse{\equal{\fitCombShape}{Exponential}}
             {\def \tempFitCombShape{Exponential}}
             {\def \tempFitCombShape{PolSlopes}}
             
             \begin{figure}[!htbp]
               \begin{center}
                 \plotOne{figs/FitResults/sWeights/#1/sWeights-\fitCombModel-Louis-\tempFitCombShape-StrongPcut-#1-KSKKDD_#3_#2_Dalitz}{0.35}
                 \plotOne{figs/FitResults/sWeights/#1/sWeights-\fitCombModel-Louis-\tempFitCombShape-StrongPcut-#1-KSKKLL_#3_#2_Dalitz}{0.35}\\
                 \plotOne{figs/FitResults/sWeights/#1/sWeights-\fitCombModel-Louis-\tempFitCombShape-StrongPcut-#1-KSKpiDD_#3_#2_Dalitz}{0.35}
                 \plotOne{figs/FitResults/sWeights/#1/sWeights-\fitCombModel-Louis-\tempFitCombShape-StrongPcut-#1-KSKpiLL_#3_#2_Dalitz}{0.35}\\
                 \plotOne{figs/FitResults/sWeights/#1/sWeights-\fitCombModel-Louis-\tempFitCombShape-StrongPcut-#1-KSpiKDD_#3_#2_Dalitz}{0.35}
                 \plotOne{figs/FitResults/sWeights/#1/sWeights-\fitCombModel-Louis-\tempFitCombShape-StrongPcut-#1-KSpiKLL_#3_#2_Dalitz}{0.35}\\
                 \plotOne{figs/FitResults/sWeights/#1/sWeights-\fitCombModel-Louis-\tempFitCombShape-StrongPcut-#1-KSpipiDD_#3_#2_Dalitz}{0.35}
                 \plotOne{figs/FitResults/sWeights/#1/sWeights-\fitCombModel-Louis-\tempFitCombShape-StrongPcut-#1-KSpipiLL_#3_#2_Dalitz}{0.35}\\
                 \ifthenelse{\equal{#2}{Bd}}{
                   \caption{Distribution of \Bd signal sWeights extracted with the \MakeLowercase{#1} BDT optimisation in #3 data. \KsKpPim, \KsPipKm, and \KsPiPi are shown from top to bottom, \DD on the left, \LL on the right. The corrections due to the presence of species with fixed yields is not applied here.}
                 }{
                   \caption{Distribution of \Bs signal sWeights extracted with the \MakeLowercase{#1} BDT optimisation in #3 data. \KsKpPim, \KsPipKm, and \KsPiPi are shown from top to bottom, \DD on the left, \LL on the right. The corrections due to the presence of species with fixed yields is not applied here.}
                 } 
                 \label{fig:FitResult:Splots:#1:#3:Dalitz#2}
               \end{center}
             \end{figure}
}

\newcommand{\makeSystTabular}[2]{
  \begin{table}[!htbp]
    \begin{center}
      \ifthenelse{\equal{#2}{KK}}
                 {       
                   \caption{Systematic uncertainties originating from the fit model of the \KsKK modes (using \MakeLowercase{#1} BDT optimisation). The numbers are rounded to the upper integer value, except for the total yield. The total systematic error is defined as the sum in quadrature of all the components.}
                 }{}
                 \ifthenelse{\equal{#2}{Kpi}}
                            {
                              \caption{Systematic uncertainties originating from the fit model of the \KsKpPim modes (using \MakeLowercase{#1} BDT optimisation). The numbers are rounded to the upper integer value, except for the total yield. The total systematic error is defined as the sum in quadrature of all the components.}
                            }{}
                            \ifthenelse{\equal{#2}{piK}}        
                                       {
                                         \caption{Systematic uncertainties originating from the fit model of the \KsPipKm modes (using \MakeLowercase{#1} BDT optimisation). The numbers are rounded to the upper integer value, except for the total yield. The total systematic error is defined as the sum in quadrature of all the components.}
                                       }{}
                                       \ifthenelse{\equal{#2}{pipi}}
                                                  {
                                                    \caption{Systematic uncertainties originating from the fit model of the \KsPiPi modes (using \MakeLowercase{#1} BDT optimisation). The numbers are rounded to the upper integer value, except for the total yield. The total systematic error is defined as the sum in quadrature of all the components.}
                                                  }{}
                                                  \label{Table:FitSyst:#1:#2}
                                                  \resizebox{\textwidth}{!}{
                                                    \input{tables/SystI-PolSlopes-from5150-#1-#2.txt}
                                                  }
    \end{center}
  \end{table}
  
}

\newcommand{\makeInvMass}[1]{
  \ifthenelse{\equal{#1}{Bd2KSpipi} \or \equal{#1}{Bs2KSpipi}}              {\def\myInvMass {pipi}}
             {\ifthenelse{\equal{#1}{Bd2KSpiK} \or \equal{#1}{Bs2KSpiK}}    {\def\myInvMass {piK}}
               {\ifthenelse{\equal{#1}{Bd2KSKpi} \or \equal{#1}{Bs2KSKpi}}  {\def\myInvMass {Kpi}}
                 {\ifthenelse{\equal{#1}{Bd2KSKK} \or \equal{#1}{Bs2KSKK}}  {\def\myInvMass {KK}}{\def\myInvMass ERROR}}
               }
             }

}
\newcommand{\makeMode}[1]{
  \ifthenelse{\equal{#1}{Bd2KSpipi}}                     {\def\myMode {\BdtoKsPiPi}}
             {\ifthenelse{\equal{#1}{Bd2KSpiK}}          {\def\myMode {\BdtoKsPipKm}}
               {\ifthenelse{\equal{#1}{Bd2KSKpi}}        {\def\myMode {\BdtoKsKpPim}}
                 {\ifthenelse{\equal{#1}{Bd2KSKK}}       {\def\myMode {\BdtoKsKK}}
                   {\ifthenelse{\equal{#1}{Bs2KSpipi}}           {\def\myMode {\BstoKsPiPi}}
                     {\ifthenelse{\equal{#1}{Bs2KSpiK}}          {\def\myMode {\BstoKsPipKm}}
                       {\ifthenelse{\equal{#1}{Bs2KSKpi}}        {\def\myMode {\BstoKsKpPim}}
                         {\ifthenelse{\equal{#1}{Bs2KSKK}}       {\def\myMode {\BstoKsKK}}{\def\myMode{ERROR}}
                         }
                       }
                     }
                   }
                 }
               }
             }
}

\newcommand{\plotGeomEff}[3]{
  \makeInvMass{#2}
  \makeMode{#2}
  \ifthenelse{\equal{#1}{2011}}{\def\redYear {2011}}{\def\redYear {2012}}
  \centering
  \includegraphics[width=0.45\textwidth]{figs/EfficiencyMaps/#1/Geometry/NoSel/#2/Efficiency_Map_NoSel_#2_#3_#1_\myInvMass.pdf.eps}
  \includegraphics[width=0.45\textwidth]{figs/EfficiencyMaps/#1/Geometry/NoSel/#2/Spline_Eff_NoSel_#2_#3_#1_\myInvMass.pdf.eps}
  \includegraphics[width=0.45\textwidth]{figs/EfficiencyMaps/#1/Geometry/NoSel/#2/Efficiency_errorHi_Map_NoSel_#2_#3_#1_\myInvMass.pdf.eps}
  \includegraphics[width=0.45\textwidth]{figs/EfficiencyMaps/#1/Geometry/NoSel/#2/Efficiency_errorLo_Map_NoSel_#2_#3_#1_\myInvMass.pdf.eps}
  \caption{
    (Top) $\epsilon^{\rm geom}$ as a function of the \myMode
    square Dalitz plot position obtained from \redYear-conditions generator-level signal MC:
    (left) the raw histogram,
    (right) smoothed using a 2D cubic spline.
    (Bottom) the (left) upper and (right) lower uncertainties on the histogram bins.
    Uncertainties are due to MC statistics.
  }
  \label{fig:geoeff:GeoEff:#1:#2:#3}
}

\newcommand{\plotTrackCorr}[4]{
  \makeInvMass{#2}
  \makeMode{#2}
  \centering
  \includegraphics[width=0.45\textwidth]{figs/EfficiencyMaps/#1/Tracking/#4/#2/Trk_AllEff-#4-#2-#3-#1-\myInvMass.pdf.eps}
  \includegraphics[width=0.45\textwidth]{figs/EfficiencyMaps/#1/Tracking/#4/#2/Spline_Eff_#4_#2_#3_#1_\myInvMass.pdf.eps}\\
  \includegraphics[width=0.45\textwidth]{figs/EfficiencyMaps/#1/Tracking/#4/#2/Tracking_errorHi_bootstrap_#4_#2_#3_\myInvMass_#1.pdf.eps}
  \includegraphics[width=0.45\textwidth]{figs/EfficiencyMaps/#1/Tracking/#4/#2/Tracking_errorHi_bootstrap_#4_#2_#3_\myInvMass_#1.pdf.eps}
  \caption{
    (Top) Combined tracking efficiency corrections in the \myMode signal mode for #3 and #1 configuration:
    (left) the raw histogram obtained from MC simulation and (right) smoothed using a 2D cubic spline.  
    (Bottom) the (left) upper and (right) lower uncertainties on the histogram bins.
  }
  \label{fig:DPefficiency-trk-all-#1-#2-#3-#4}
}

\newcommand{\plotTrackCorrMode}[1]{
  \begin{figure}[!htb]
    \plotTrackCorr{2011}{#1}{DD}{Loose}
  \end{figure}
  \begin{figure}[!htb]
    \plotTrackCorr{2011}{#1}{LL}{Loose}
  \end{figure}
}

\newcommand{\plotTrigCorr}[4]{
  \makeInvMass{#1}
  \makeMode{#1}
  \ifthenelse{\equal{#4}{TOS}}{
    \def\trigComment {$\epsilon^{\rm L0TOS|sel\&geom}_{\rm data}/\epsilon^{\rm L0TOS|sel\&geom}_{\rm MC}$}}{
    \def\trigComment {$\epsilon^{\rm !L0TOS|sel\&geom}_{\rm data}/\epsilon^{\rm !L0TOS|sel\&geom}_{\rm MC}$}}
  
  \centering
  \includegraphics[width=0.45\textwidth]{figs/EfficiencyMaps/2011/L0#4/#3/#1/L0#4-Correction-#3-#1-#2-2011-\myInvMass.pdf.eps}
  \includegraphics[width=0.45\textwidth]{figs/EfficiencyMaps/2011/L0#4/#3/#1/Spline_Eff_#3_#1_#2_2011_\myInvMass.pdf.eps}\\
  \includegraphics[width=0.45\textwidth]{figs/EfficiencyMaps/2011/L0#4/#3/#1/L0#4_errorHi_combined_#3_#1_#2_\myInvMass_2011.pdf.eps}
  \includegraphics[width=0.45\textwidth]{figs/EfficiencyMaps/2011/L0#4/#3/#1/L0#4_errorLo_combined_#3_#1_#2_\myInvMass_2011.pdf.eps}
  \caption{
    (Top) \trigComment across the \myMode #2 square Dalitz plot (2011+2012 combined):
    (left) the raw histogram obtained using the procedure described in the text and (right) smoothed using a 2D cubic spline.  
    (Bottom) the (left) upper and (right) lower uncertainties on the histogram bins.
  }
  \label{fig:DPcorrection-#1-#2-#3-#4}
}

\newcommand{\plotSelEff}[5]{
  \makeInvMass{#2}
  \makeMode{#2}
  \ifthenelse{\equal{#5}{TOS}}{\def\selComment {{\tt L0Hadron\_TOS}}}{\def\selComment {{\tt L0Global\_TIS\&\&!L0Hadron\_TOS}}}
  \centering
  \includegraphics[width=0.45\textwidth]{figs/EfficiencyMaps/#1/Sel#5/#4/#2/#2-Eff-#4-#2-#3-#1-\myInvMass.pdf.eps}
  \includegraphics[width=0.45\textwidth]{figs/EfficiencyMaps/#1/Sel#5/#4/#2/Spline_Eff_#4_#2_#3_#1_\myInvMass.pdf.eps}\\
  \includegraphics[width=0.45\textwidth]{figs/EfficiencyMaps/#1/Sel#5/#4/#2/#2-ErrorHi-#4-#2-#3-#1-\myInvMass.pdf.eps}
  \includegraphics[width=0.45\textwidth]{figs/EfficiencyMaps/#1/Sel#5/#4/#2/#2-ErrorLo-#4-#2-#3-#1-\myInvMass.pdf.eps}
  \caption{
    (Top) $\epsilon^{\rm sel|geom}$ across the \myMode #3 #1 square Dalitz plot for \selComment \MakeLowercase{#4} BDT candidates:
    (left) the raw histogram obtained using the procedure described in the text and (right) smoothed using a 2D cubic spline.  
    (Bottom) the (left) upper and (right) lower uncertainties on the histogram bins.
  }
  \label{fig:DPefficiency-sel-#1-#2-#3-#4-#5}
}

\newcommand{\plotSelEffMode}[2]{
  \begin{figure}[!htb]
    \plotSelEff{2011}{#1}{DD}{#1}{TOS}
  \end{figure}
  \begin{figure}[!htb]
    \plotSelEff{2011}{#1}{DD}{#1}{TIS}
  \end{figure}
  \begin{figure}[!htb]
    \plotSelEff{2011}{#1}{LL}{#1}{TOS}
  \end{figure}
  \begin{figure}[!htb]
    \plotSelEff{2011}{#1}{LL}{#1}{TIS}
  \end{figure}
}

\usepackage{longtable} 
\usepackage{multirow}
\usepackage{placeins}

\begin{document}

\renewcommand{\thefootnote}{\fnsymbol{footnote}}
\setcounter{footnote}{1}

\begin{titlepage}
\pagenumbering{roman}

\vspace*{-1.5cm}
\centerline{\large EUROPEAN ORGANIZATION FOR NUCLEAR RESEARCH (CERN)}
\vspace*{1.5cm}
\noindent
\begin{tabular*}{\linewidth}{lc@{\extracolsep{\fill}}r@{\extracolsep{0pt}}}
\ifthenelse{\boolean{pdflatex}}
{\vspace*{-1.5cm}\mbox{\!\!\!\includegraphics[width=.14\textwidth]{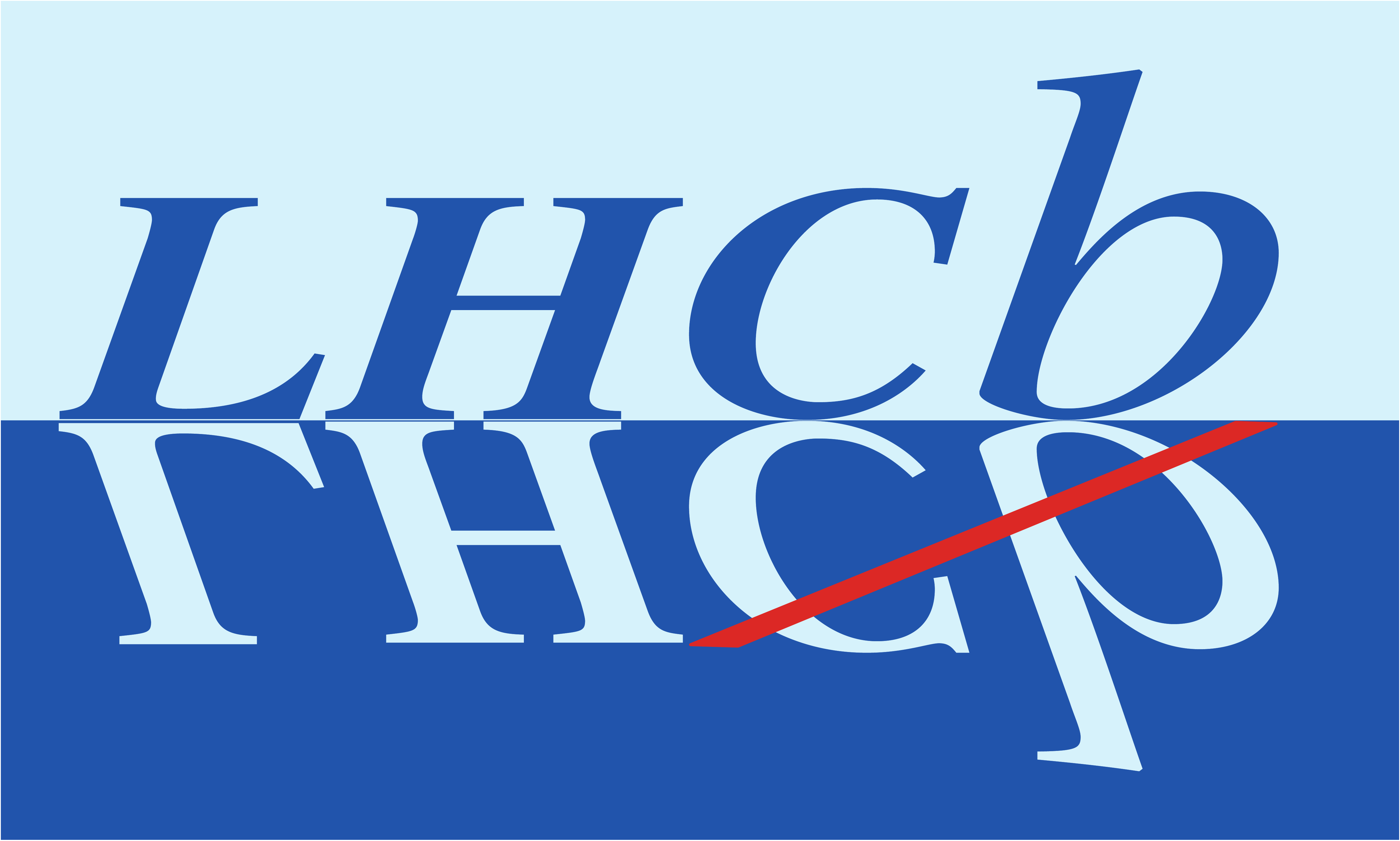}} & &}%
{\vspace*{-1.2cm}\mbox{\!\!\!\includegraphics[width=.12\textwidth]{lhcb-logo.eps}} & &}%
\\ & & CERN-EP-2019-013 \\  
 & & LHCb-PAPER-2018-044 \\  
 & & 23 September  2019 \\
 & & \\
\end{tabular*}

\vspace*{4.0cm}

{\normalfont\bfseries\boldmath\huge
\begin{center}
  \papertitle 
\end{center}
}

\vspace*{2.0cm}

\begin{center}
\paperauthors\footnote{Authors are listed at the end of this paper.}
\end{center}

\vspace{\fill}

\begin{abstract}
  \noindent A search for \CP violation in charmless four-body decays of \Lb and \Xibz baryons with a proton and three charged mesons in the final state is performed. To cancel out production and detection charge-asymmetry effects, the search is carried out by measuring the difference between the \CP asymmetries in a charmless decay and in a decay with an intermediate charmed baryon with the same particles in the final state. The data sample used was recorded in 2011 and 2012 with the \lhcb detector and corresponds to an integrated luminosity of $3 \invfb$. A total of 18 \CP asymmetries are considered, either accounting for the full phase space of the decays or exploring specific regions of the decay kinematics. No significant \CP-violation effect is observed in any of the measurements.   
\end{abstract}

\vspace*{2.0cm}

\begin{center}
  Published in Eur.~Phys.~J.~C (2019) 79: 745
\end{center}

\vspace{\fill}

{\footnotesize 
\centerline{\copyright~\papercopyright ~\href{\paperlicenceurl}{\paperlicence}.}}
\vspace*{2mm}

\end{titlepage}


\newpage
\setcounter{page}{2}
\mbox{~}
%
%
%
%

\cleardoublepage

\renewcommand{\thefootnote}{\arabic{footnote}}
\setcounter{footnote}{0}
\pagestyle{plain} 
\setcounter{page}{1}
\pagenumbering{arabic}
%
\section{Introduction}
\label{sec:introduction}

All measurements of \CP violation performed so far are consistent with the predictions of the Standard Model (SM)~\cite{CKMfitter2015}.  Nonvanishing \CP-violating asymmetries have been observed in the decays of both $K$ and \B mesons~\cite{PDG2018}. In contrast, \CP violation has not been observed in baryon decays, although some indications for nonvanishing \CP asymmetries in \bquark-flavoured baryon decays have been reported  by the \lhcb collaboration~\cite{LHCb-PAPER-2013-061,LHCb-PAPER-2014-020,LHCb-PAPER-2016-030,LHCb-PAPER-2016-004}.

The abundant production of  \Lb and \Xibz baryons\footnote{The inclusion of charge conjugate processes is implied throughout this paper, unless stated otherwise.} in proton-proton collisions at the Large Hadron Collider (LHC) gives the \lhcb experiment the opportunity to study multibody charmless decays of \bquark-flavoured baryons.  In particular, \Lb and \Xibz baryon decays to charmless four-body final states were observed by the \lhcb collaboration and their branching fractions measured~\cite{LHCb-PAPER-2017-034}. Their large yields enable measurements of \CP-violating asymmetries to be performed with a precision at the level of a few percent.  

This search follows the successful path of the observation of large  \CP-violating  asymmetries in multibody charmless decays of charged and neutral \B mesons by \lhcb~\cite{LHCb-PAPER-2013-051,LHCb-PAPER-2017-033,LHCb-PAPER-2013-027,LHCb-PAPER-2014-044}. These decays proceed simultaneously through the charged-current $\bquark \to \uquark$ transition and neutral-current $\bquark \to \squark,\dquark$ transitions, and the resulting interference exhibits a weak-phase difference. Furthermore, and analogously to the aforementioned charmless multibody \B-meson decays,  charmless multibody decays of \bquark-flavoured baryons contain rich resonance structures, both in the two- or three-body baryonic invariant-mass spectra (\ie $\Pp\PK^-$, $\Pp\Ppi^-$, $\Pp\Ppi^+$, $\Pp\Ppi^+\Ppi^-$ and $\Pp\PK^-\Ppi^+$) and in the two- or three-body nonbaryonic ones (\ie the \pipi, \Kpm\pimp, \kk, $\pi^+\pi^{-}\pi^{\pm}$ and $K^{\pm}\pi^+\pi^-$). Consequently, \CP asymmetries might be enhanced due to the strong-phase differences induced by the interference patterns between these transitions in the mass regions around resonances. The charmless \bquark-baryon decays studied in this paper are hence well suited for a potential first observation of \CP violation in the baryon sector. However, the presence of these strong phases, that are difficult to predict, would make a potential observation of \CP violation difficult to interpret in terms of the weak phase of the Cabibbo-Kobayashi-Maskawa (CKM) quark-mixing matrix~\cite{PhysRevLett.10.531, Kobayashi:1973fv}. 

This work focuses on a search for \CP violation in \XbTophhhz charmless decays, where \Xb stands either for \Lb or \Xibz and $h^{(\prime,\, \prime\prime)}$ stand either for a pion or a kaon. Six decays are studied, namely \LbToppipipi, \LbTopKpipi, \LbTopKKpi, \LbTopKKK, \XibzTopKpipi and \XibzTopKpiK. The \CP asymmetry is defined as 
\begin{equation}
\ACP \equiv \frac{\mathrm{\Gamma} (\Xb \to f) - \mathrm{\Gamma}(\Xbbar \to \overline{f})}{\mathrm{\Gamma} (\Xb \to f) + \mathrm{\Gamma}(\Xbbar \to \overline{f})},
\end{equation}
where $\mathrm{\Gamma}(\Xb \to f)$ is the partial width of the given decay. The \CP asymmetry measurement relies on counting the number of reconstructed particle and antiparticle decays and includes therefore experimental charge-asymmetric effects such as the track detection efficiency or $b$-baryon production asymmetries. They are cancelled out to first order by comparing the \CP asymmetries of the signal modes to those of charmed decays that lead to the same or very similar final states and for which no measurable \CP violation is expected in the SM~\cite{LHCb-PAPER-2014-004}.  
The decays \LbToLcpiLcTopKpi, \LbToLcpiLcToppipi  and \XibzToXicpiXicTopKpi are thus reconstructed with the same selection as the corresponding charmless signals. The \CP-violating observable considered in this work is then referred to as $\dACP \equiv \ACPnoC - \ACPC$, where  $\ACPnoC$ ($\ACPC$) is the asymmetry measured in the charmless (charmed) decays. The decays of interest are reported in Table~\ref{tab:modesconsidered}.

\begin{table}[tb]
  \centering
  \renewcommand{\arraystretch}{1.2}
  \setlength{\tabcolsep}{14pt}
  \caption{Four-body charmless and charmed decays considered in this analysis. The difference of \CP-asymmetries measured for the charmless modes and for the control channels results in \dACP measurements. For each observable, the choice of the control channel is aiming at cancelling at first order production and detection asymmetries. Given the data samples at hand, it is not possible to meet both criteria for the signal decay \XibzTopKpiK : the choice of the Cabibbo-favoured decay \XibzToXicpiXicTopKpi as a control channel requires in turn to correct the corresponding \dACP for the kaon-detection asymmetry.}
  \label{tab:modesconsidered}
  \begin{tabular}{cc}
  \hline
  Charmless mode & Control channel \\ 
  \hline
  \LbToppipipi & \LbToLcpiLcToppipi \\
  \LbTopKpipi & \LbToLcpiLcTopKpi \\
  \LbTopKKpi & \LbToLcpiLcToppipi \\
  \LbTopKKK & \LbToLcpiLcTopKpi \\
  \XibzTopKpipi & \XibzToXicpiXicTopKpi \\ 
  \XibzTopKpiK & \XibzToXicpiXicTopKpi \\
  \hline
  \end{tabular}
\end{table}

In addition to \dACP measurements integrated over all of the four-body phase space, specific regions of the space are studied 
in order to search for local \CP asymmetries. 

The same final states have been used by the LHCb experiment to search for \CP violation using triple product asymmetries~\cite{LHCb-PAPER-2016-030,LHCb-PAPER-2018-001}. The latter technique and the \dACP measurements exhibit different sensitivity to \CP violation~\cite{PhysRevD.92.076013}, which makes the two approaches complementary.

\section{Detector and data set}
\label{sec:Detector}

The analysis is performed using $pp$ collision data
recorded with the LHCb detector, corresponding to an integrated luminosity of
1.0\invfb at a centre-of-mass energy of 7\tev in 2011 and 2.0\invfb at a
centre-of-mass energy of 8\tev in 2012.
The \lhcb detector~\cite{Alves:2008zz,LHCb-DP-2014-002} is a single-arm forward
spectrometer covering the \mbox{pseudorapidity} range $2<\eta <5$,
designed for the study of particles containing \bquark or \cquark
quarks. The detector includes a high-precision tracking system
consisting of a silicon-strip vertex detector surrounding the $pp$
interaction region,
a large-area silicon-strip detector located
upstream of a dipole magnet with a bending power of about
$4{\mathrm{\,Tm}}$, and three stations of silicon-strip detectors and straw
drift tubes
placed downstream of the magnet.
The tracking system provides a measurement of the momentum, \ptot, of charged particles with
a relative uncertainty that varies from 0.5\% at low momentum to 1.0\% at 200\gevc.
The minimum distance of a track to a primary vertex (PV), the impact parameter (IP), 
is measured with a resolution of $(15+29/\pt)\mum$,
where \pt is the component of the momentum transverse to the beam, in\,\gevc.
Different types of charged hadrons are distinguished using information
from two ring-imaging Cherenkov detectors.
Photons, electrons and hadrons are identified by a calorimeter system consisting of
scintillating-pad and preshower detectors, an electromagnetic
and a hadronic calorimeter. Muons are identified by a
system composed of alternating layers of iron and multiwire
proportional chambers.

Simulation is used to investigate backgrounds from other
\bquark-hadron decays and also to study the detection and reconstruction
efficiencies of the signals.
In the simulation, $pp$ collisions are generated using
\pythia~\cite{Sjostrand:2006za,*Sjostrand:2007gs} with a specific \lhcb
configuration~\cite{LHCb-PROC-2010-056}.
Decays of hadronic particles are described by \evtgen~\cite{Lange:2001uf}
in which final-state radiation is generated using
\photos~\cite{Golonka:2005pn}.
The interactions of the generated particles with the detector, and its
response, are implemented using the \geant toolkit~\cite{Allison:2006ve,
*Agostinelli:2002hh} as described in Ref.~\cite{LHCb-PROC-2011-006}.

\section{Trigger and selection requirements}
\label{sec:Selection}

The selection follows most of the strategy described in Ref.~\cite{LHCb-PAPER-2017-034}. 
The online event selection is performed by a trigger\cite{LHCb-DP-2012-004} 
that consists of a hardware stage, based on information from the calorimeter and muon systems, followed by a software stage, in which all charged particles with $\pt>500\,(300)\mevc$ are reconstructed for 2011\,(2012) data. 
At the hardware-trigger stage, events are required to include a muon or a dimuon with high transverse momentum  or a hadron, photon or electron with 
high transverse energy. The software trigger reconstructs charged particles with transverse momentum  $\pt>500\,(300)\mevc$ for 2011\,(2012) data and requires a two-, three- or four-track secondary vertex with significant 
displacement from all primary $pp$ interaction vertices. At least one charged particle must have transverse 
momentum $\pt > 1.7\,(1.6)\gevc$ for 2011\,(2012) data and be inconsistent with originating from any PV. A multivariate 
algorithm~\cite{BBDT} is used for the identification of secondary vertices consistent with the decay of a \bquark hadron. 
In the offline selection, trigger signals are associated with reconstructed particles. Selection requirements can therefore be made on the trigger selection itself
and on whether the decision was due to the signal candidate, other particles produced in the $pp$ collision, or a combination of both. 

The events passing the trigger requirements are filtered in two stages. Initial requirements are applied to further reduce the size of the data sample before a 
multivariate selection is applied. Selection requirements based on topological variables, such as the flight distance of the 
\bquark baryon, are used as the main discriminants. In order to preserve the phase space of the decays of interest, only 
loose requirements are placed on the transverse momenta of the decay products, $\pt >250\mevc$. 

%
%

Neutral \bquark-baryon candidates, hereafter denoted as \Xb,  are formed from a proton candidate selected with 
particle identification (PID) requirements and three additional charged tracks. When more than one PV is reconstructed, the \Xb candidate is associated to the PV with the smallest value of $\chisqip$, where $\chisqip$ is the difference 
in $\chisq$ of a given PV reconstructed with and without the considered  candidate.  Each of the four tracks of the final state is required to have $\chisqip > 16$ and $3<\ptot<100\gevc$. Beyond 100 \gevc, there is little pion/kaon/proton discrimination. 
The  \Xb candidates are then required to form a vertex with a fit quality $\chisqvtx < 20$ and to be significantly separated from any PV with $\chisqfd > 50$, where $\chisqfd$ is the square of the flight-distance significance. To remove backgrounds from higher-multiplicity decays, the difference in \chisqvtx when adding any other track must be greater than 4. The  \Xb  candidates must have a transverse momentum $p_{\rm T}(\Xb)$ greater than $1.5\gevc$ and an invariant mass within the range $5340 < m(phh^{\prime}h^{\prime \prime}) < 6400 \mevcc$. 
They are further required to be consistent with originating from a PV, quantified by both  $\chisqip < 16$ and the cosine of the angle $\theta_{\rm DIR}$ between the reconstructed momentum of the  $b$ hadron and the vector defined by the associated PV and the decay vertex be greater than 0.999.  Finally, PID requirements are applied to provide discrimination between kaons and pions in order to assign the candidates to one of the five different final-state hypotheses \ppipipi, \pKpipi, \pKKpi, \pKpiK and \pKKK. 

%
%

There are three main categories of background that contribute significantly in the selected invariant-mass regions: the so-called signal cross-feed background, resulting from a misidentification of one or more final-state particles in a charmless baryon decay, which can therefore be reconstructed as another charmless decay with 
a different mass hypothesis;  the charmless decays of neutral \B mesons 
to final states containing four charged mesons, where a  pion or a kaon is misidentified as a proton; and the combinatorial background, which results from a random 
association of unrelated tracks. The pion and kaon PID requirements, that define mutually exclusive samples, are optimised to reduce the cross-feed background, 
and hence to maximise the significance of the signal. The charmless \B-meson decays are identified by reconstructing the invariant-mass distributions of candidates 
using the pion or kaon mass instead of the proton mass hypothesis, in the high-mass sidebands defined as $m_{\rm sideband} < m(phh^{\prime}h^{\prime \prime}) < 6400 \mevcc$, where 
$m_{\rm sideband} = 5680 \mevcc$ for \ppipipi and \pKKpi final states, and $m_{\rm sideband} = 5840 \mevcc$ for \pKpipi, \pKpiK and \pKKK final states. This 
background contribution is reduced by the optimisation of the proton PID requirement.

To reject combinatorial background, multivariate discriminants based on a boosted decision tree~(BDT)~\cite{Breiman} with the AdaBoost algorithm~\cite{AdaBoost} have been 
designed. Candidates from simulated \LbToppipipi  decays and the high-mass sideband are used as the signal and background training samples, respectively. This high-mass 
sideband region is chosen such that the sample is free of cross-feed background.  The samples are divided into two data-taking periods and  further subdivided into two equally sized 
subsamples.  Each subsample is then used to train an independent discriminant. The BDT trained on one subsample is used to select candidates 
from the other subsample, in order to avoid a possible bias in the selection. 
 
The BDTs have the following quantities as inputs:  \pt, $\eta$, \chisqip, \chisqfd, $\cos \theta_{\rm DIR}$, and \chisqvtx of the \Xb candidate; the smallest change in the  \bquark-baryon \chisqvtx 
when adding any other track from the event; the sum of the \chisqip of the four tracks of the final state; and the  \pt asymmetry
\begin{equation}
p_{\rm T}^{\rm asym} = \frac{p_{\rm T}(\Xb) - p_{\rm T}^{\rm cone}}{p_{\rm T}(\Xb) + p_{\rm T}^{\rm cone}} \,,
\end{equation}
where $p_{\rm T}^{\rm cone}$ is the transverse component of the vector sum of all particle momenta inside a cone around the  \bquark-baryon candidate 
direction, of radius $R \equiv \sqrt{\delta \eta^2 +\delta \phi^2} =1.5$, where  $\delta \eta$ and $\delta \phi$  are the difference in pseudorapidity and azimuthal angle (expressed in radians) around the beam direction, between the momentum vector of the track under consideration and that of the $b$-hadron candidate.
 The distribution of $p_{\rm T}^{\rm asym}$ for the signal candidates is enhanced towards high values. The BDT output is determined to be uncorrelated with the position in phase space of the decays of interest. The selection requirement placed on the output of the BDTs is optimised for the six decays of interest by minimising the uncertainties on the \CP-asymmetry differences.

%
%
A number of background contributions consisting of fully reconstructed \bquark-baryon decays into the two-body $\Lc \had$, $\Xicp \had$, three-body $\D p \had$ or  $(\cquark\cquarkbar) p \had$ 
combinations, where $(\cquark\cquarkbar)$ represents a charmonium resonance, may produce the same final state as the signal. Hence, they have similar invariant-mass distribution of the \bquark-baryon candidate as the signal along with a similar selection efficiency.
The presence of a misidentified hadron in the \D , \Lc and \Xicp decay also produces peaking background under the signal.  Therefore, the following decay channels are explicitly reconstructed
under the relevant particle hypotheses and vetoed by means of a requirement on the resulting invariant mass, in all spectra: $\Lc \to pK^-\pi^+$, $\Lc \to  p\pi^+\pi^-$, $\Lc \to  pK^+K^-$, $\Xicp \to pK^-\pi^+$, $\Dp \to K^-\pi^+\pi^+$, $\Dsp \to K^-K^+\pi^+$, $\Dz \to K^{-}\pi^{+}$, $\Dz \to \pi^+\pi^-$, $\Dz \to K^+K^-$, $\jpsi \to \pi^+\pi^-$ and $\jpsi \to K^+K^-$. The decays of other possible broad charmonium resonances to $\pi^+\pi^-$ and $K^+K^-$ are retained as potential interfering amplitudes with the charmless amplitudes under study.   

%
%

The same set of trigger, \pid and BDT requirements is applied to the control modes \LbToLcpiLcTopKpi, \LbToLcpiLcToppipi and \XibzToXicpiXicTopKpi to cancel out most of 
the systematic effects related to the selection criteria. Candidates whose $pK^-\pi^+$ or $p\pi^-\pi^+$ invariant mass is in the range $[2213, 2313] \mevcc $ for $\Lc$ and 
$[2437, 2497] \mevcc$ for $\Xicp$, are retained as control channels candidates. Events outside these intervals belong to the corresponding signal spectrum, 
again ensuring statistically independent samples for the simultaneous fit. 

%
%

The fraction of events containing more than one candidate is below the percent level.  The candidate to be retained in each event is chosen randomly and reproducibly.

\section{Simultaneous fit}
\label{sec:fit}

A simultaneous unbinned extended maximum-likelihood fit is performed to the invariant-mass distributions of the $\bquark$-hadron candidates under each 
of the mass hypotheses for the signal and control channel final-state tracks. The data samples are split according to the charge 
of the proton and to the year of data taking. Furthermore, data are split according to the hardware trigger conditions, in order to correct raw measurements for charge-asymmetric trigger efficiencies. The components of the model include, in addition to signal decays, partially reconstructed five-body \Xb decays, signal and background cross-feeds, four- and five-body decays of \B mesons and combinatorial background. 
The independent data samples obtained for each final state are fitted simultaneously. For each sample, the likelihood is expressed as
\begin{equation}
 \displaystyle \ln{\cal L}  =  \sum_{i} \ln \left( {\sum_{j} N_j P_{j,i}} \right) -\sum_{j}N_j  
\end{equation}
\noindent where $N_{j}$ is the number of events related to the component $j$ and $P_{j,i}$ is the probability distribution function for component $j$ evaluated at the mass of the candidate $i$. 

%
%

\subsection{Fit model} 
\label{sec:model}


The signal decays are modelled as the sum of two Crystal Ball (CB) functions~\cite{Skwarnicki:1986xj} that share peak positions and widths but have independent power-law tails on opposite sides of the peak.  The \Lb mass parameter is free in the fit and  shared among the \Lb decays. The difference between the fitted \Xibz and \Lb masses is also a shared parameter and is constrained to the value reported in Ref.~\cite{PDG2018} by using a Gaussian function.  

The width parameter for \LbTopKpipi decays measured in the 2012 data-taking sample is found to be $16.47 \pm 0.22 \mevcc$ and is chosen as reference. The ratio of the experimental widths of the signal decay functions is constrained using Gaussian prior probability distributions multiplying the likelihood function, with parameters obtained from a fit to simulated events.  The other parameters of the CB components are obtained from a simultaneous fit to simulated samples, and are fixed to those values in the  fits to the data. 


The cross-feed backgrounds are modelled by the sum of two CB functions, whose parameters are determined from simulated samples weighted to match the performances of the particle identification algorithm as measured in the data. All cases resulting from the misidentification of either one or two of the final-state particles are considered. The yield of each misidentified decay is constrained to the yield of the corresponding correctly identified decay and the known misidentification probabilities. These constraints are implemented using Gaussian prior probability distributions multiplying the likelihood function. Their mean values are obtained from the ratio of selection efficiencies and their widths include uncertainties originating from the finite size of the simulated events samples as well as the systematic uncertainties related to the determination of the PID efficiencies.


The backgrounds resulting from four- or five-body decays of \B mesons are identified in each spectrum by a dedicated fit to the candidates in the high-mass sideband, reconstructed under the hypothesis of the kaon mass for the proton candidates. The relative yield of each decay is then constrained in the simultaneous fit from its observed abundance in the high-mass sidebands. The invariant-mass distributions are modelled by the sum of two CB functions, whose parameters are determined from simulation.  


Partially reconstructed backgrounds where a neutral pion is not reconstructed, such as $\Lb, \Xibz \to phh^{\prime}h^{\prime \prime} \piz$, are modelled by means of generalised ARGUS functions~\cite{Albrecht:1990cs} convolved with a Gaussian resolution function. The Gaussian width is taken as the signal \LbTopKpipi width parameter. The parameters of the ARGUS function are shared among all invariant-mass spectra and are determined directly from the fit, except for the threshold, which is given by $m(X_b) - m(\pi^0)$.  Partially reconstructed decays with a missing photon such as \LbToppietap and  \LbTopKetap decays, with \EtapTopipig, are modelled separately using the same functional form but where the parameters are fixed from simulation. The \LbTopKpipipiz decay modes where a charged pion is misidentified as a kaon can significantly contribute to the \pKKpi and \pKpiK spectra. They are modelled with an 
empirical function determined from the partially reconstructed background candidates in the control channel. 

Finally, the combinatorial background is modelled by a linear function whose slope is shared among the invariant-mass spectra. 

\subsection{The ensemble of measurements}

The following three categories of measurements have been considered {\it a priori} (before any evaluation of the data) to search for global and local effects of \CP violation.  

\begin{itemize}

\item \CP asymmetries are measured, considering the whole selected phase space of the decay candidates. 
\item \CP asymmetries are also measured in the phase-space region of low invariant mass on the baryonic pair (\textit{i.e.} $p\pi^{\pm}$ or $pK^-$) and low invariant mass on the pairing of the two other tracks. The ensemble of measurements that are performed with this phase-space selection is hereafter referred to as LBM (Low $2\!\times\!2$-Body Mass) measurements. The invariant mass of the baryonic pair is required to be lower than 2 \gevcc while the invariant-mass requirements on the two remaining tracks depends on whether it is a $\pi^+\pi^-$ pair, a $K^{\pm}\pi^{\mp}$ or a $KK$ pair. These values are chosen  to include several known resonances, in particular $f_0(1500)$ resonance for $\pi^+\pi^-$, the broad scalar $K_0^*(1430)^0$ resonance for $K^+\pi^-$ and the $f_2^{\prime}(1525)$ resonance for $K^+K^-$. Only the modes with the largest signal yields are considered, namely \LbToppipipi, \LbTopKpipi and \LbTopKKK decays. The two-body low-mass distributions are displayed in Fig.~\ref{fig:phsp}.
Several resonant structures are observed, and correspond to baryon resonances like $\Lambda(1520)$, $\Delta(1232)^{++}$ and $N(1520)$ or meson resonances like $K^*(892)^0$, $\rhoz$ or $\phiz$. This phase-space selection focuses therefore on low-invariant-mass resonances (both mesonic and baryonic) as well as low-invariant-mass nonresonant components of the amplitudes. The latter have been shown to generate large \CP-violating asymmetries in analogous \B-meson decays~\cite{LHCb-PAPER-2013-051}.

\begin{figure}[!tb]
\centering
\begin{tabular}{@{}c@{}c}
\includegraphics[width = .49\textwidth]{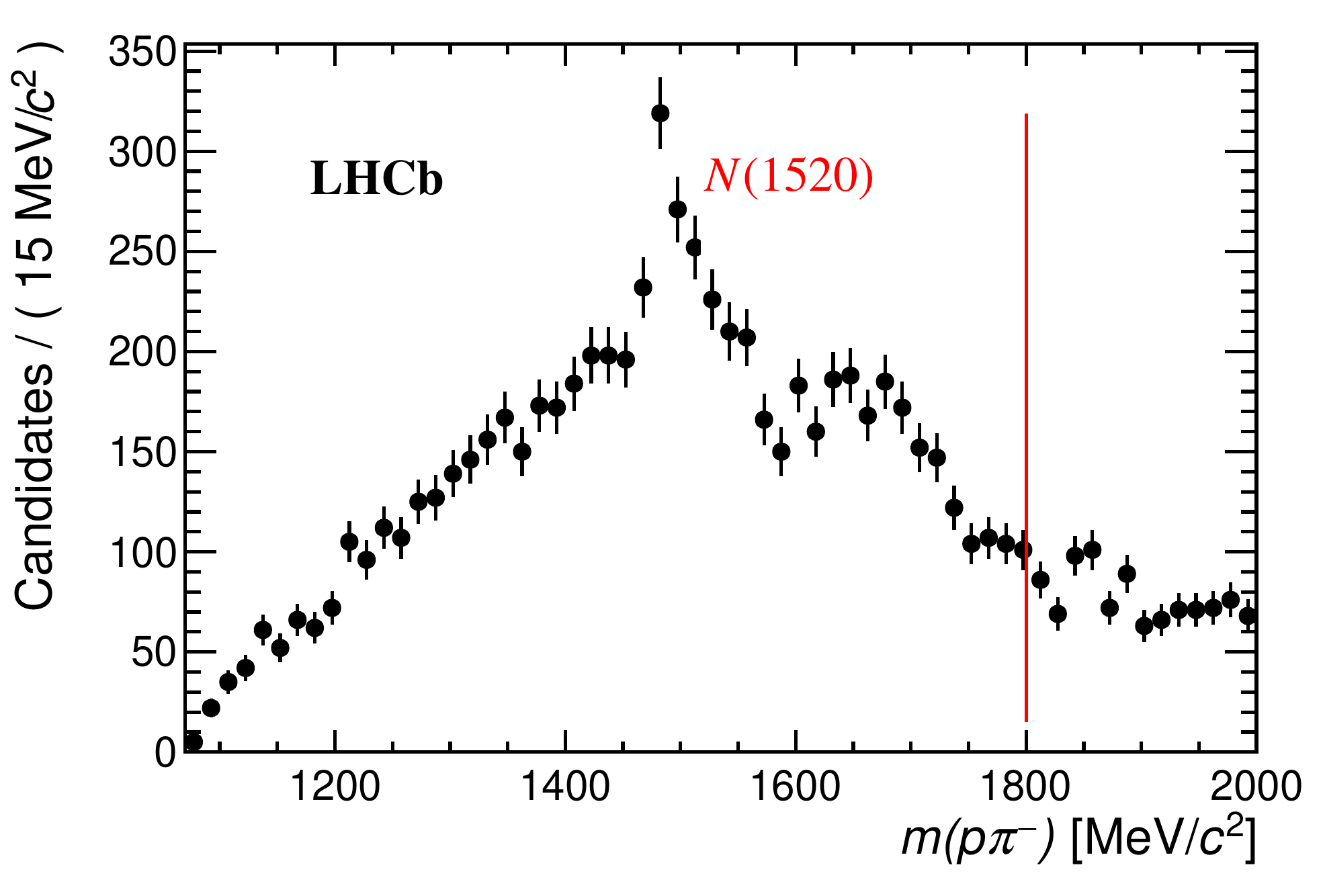} & \includegraphics[width = .49\textwidth]{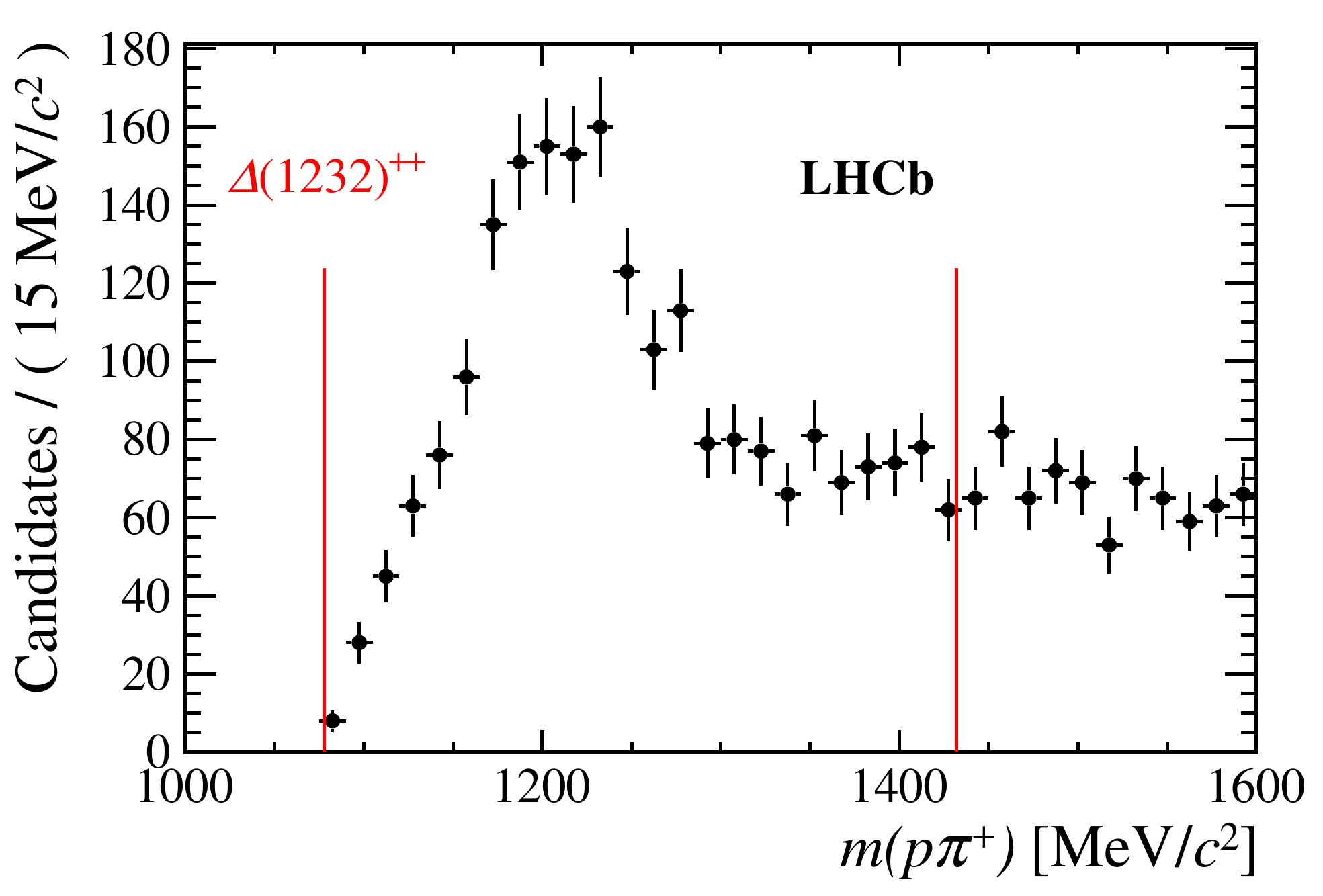} \\
\small (a) & \small (b) \\
\includegraphics[width = .49\textwidth]{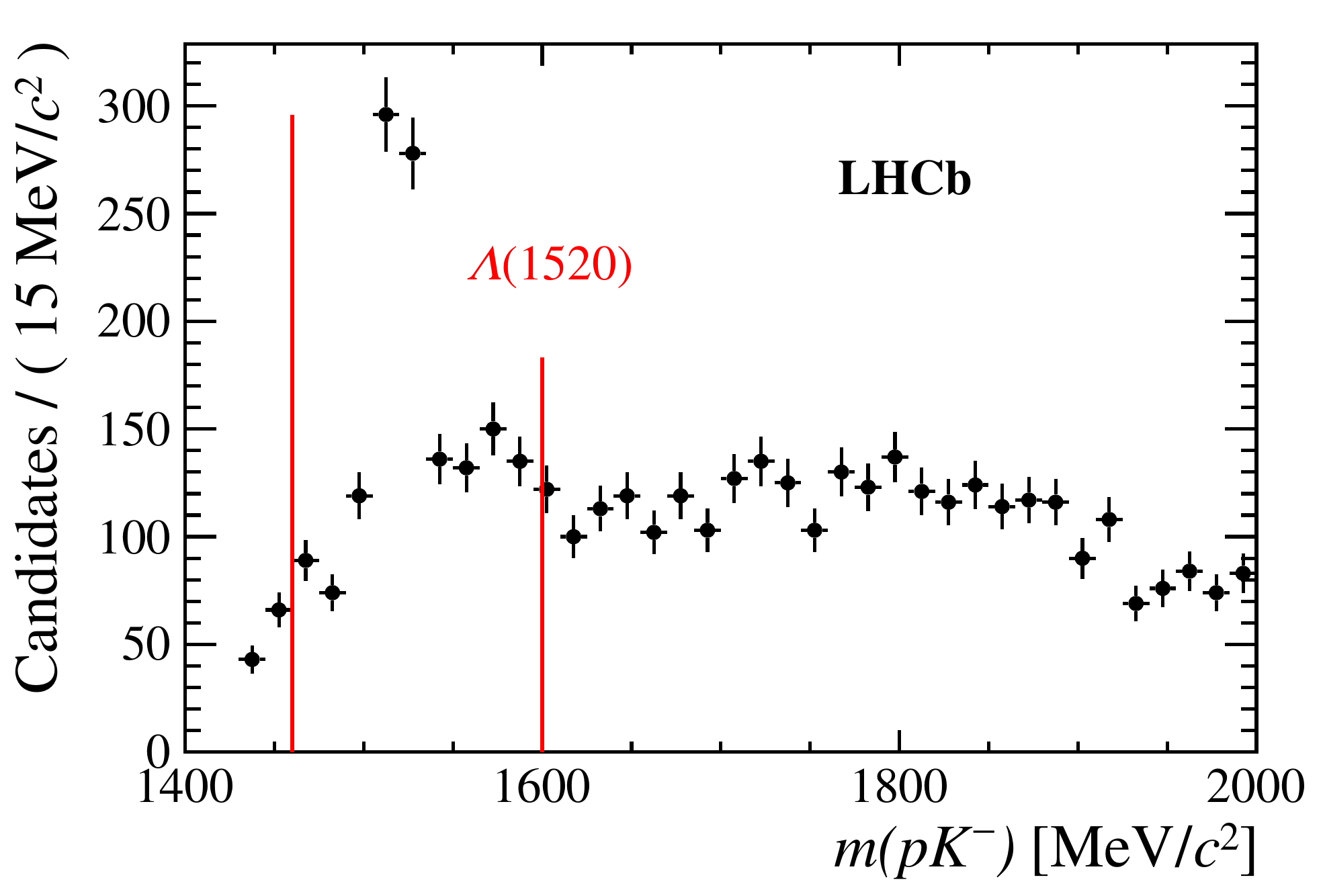} &\includegraphics[width = .49\textwidth]{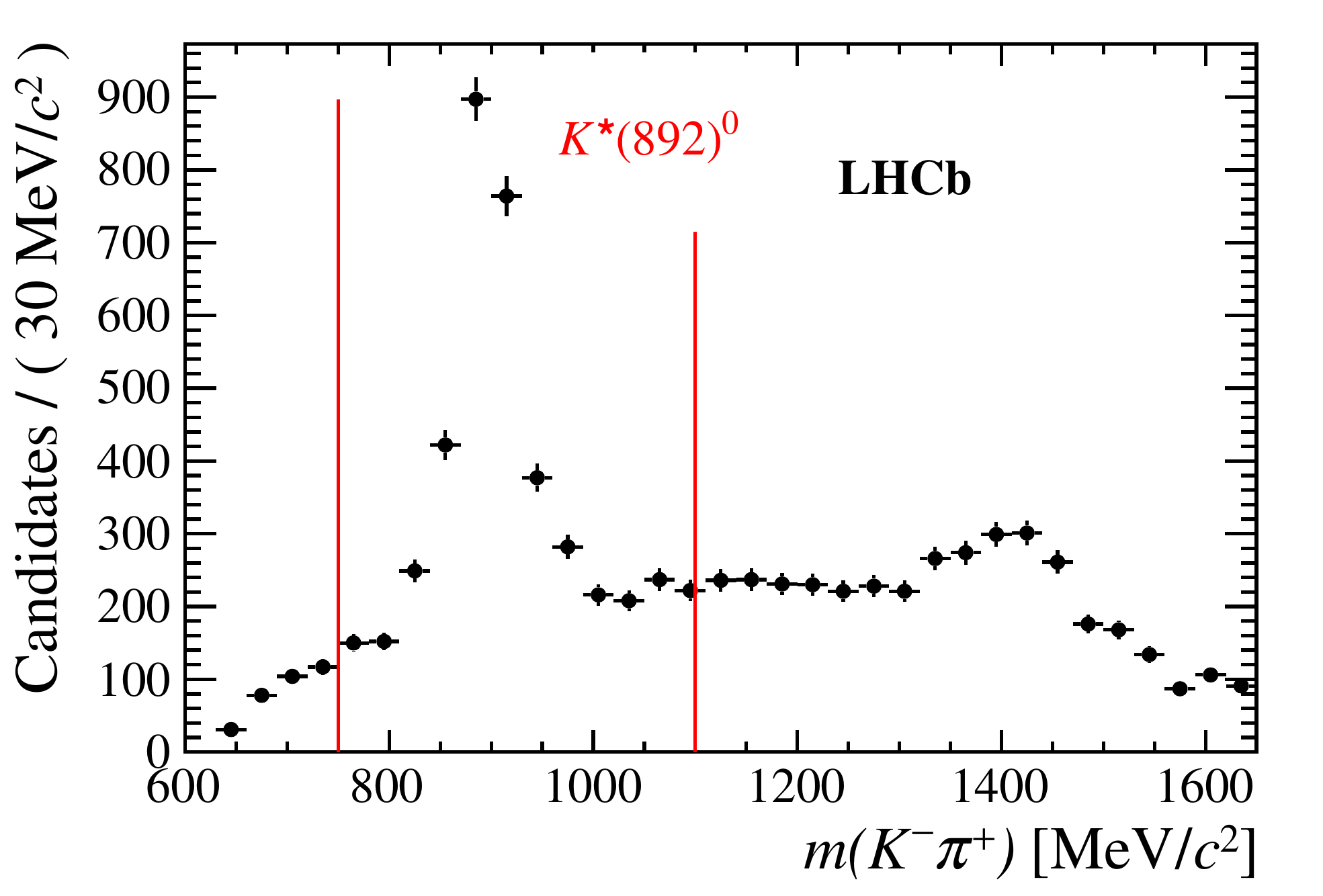} \\
\small (c) & \small (d) \\
\includegraphics[width = .49\textwidth]{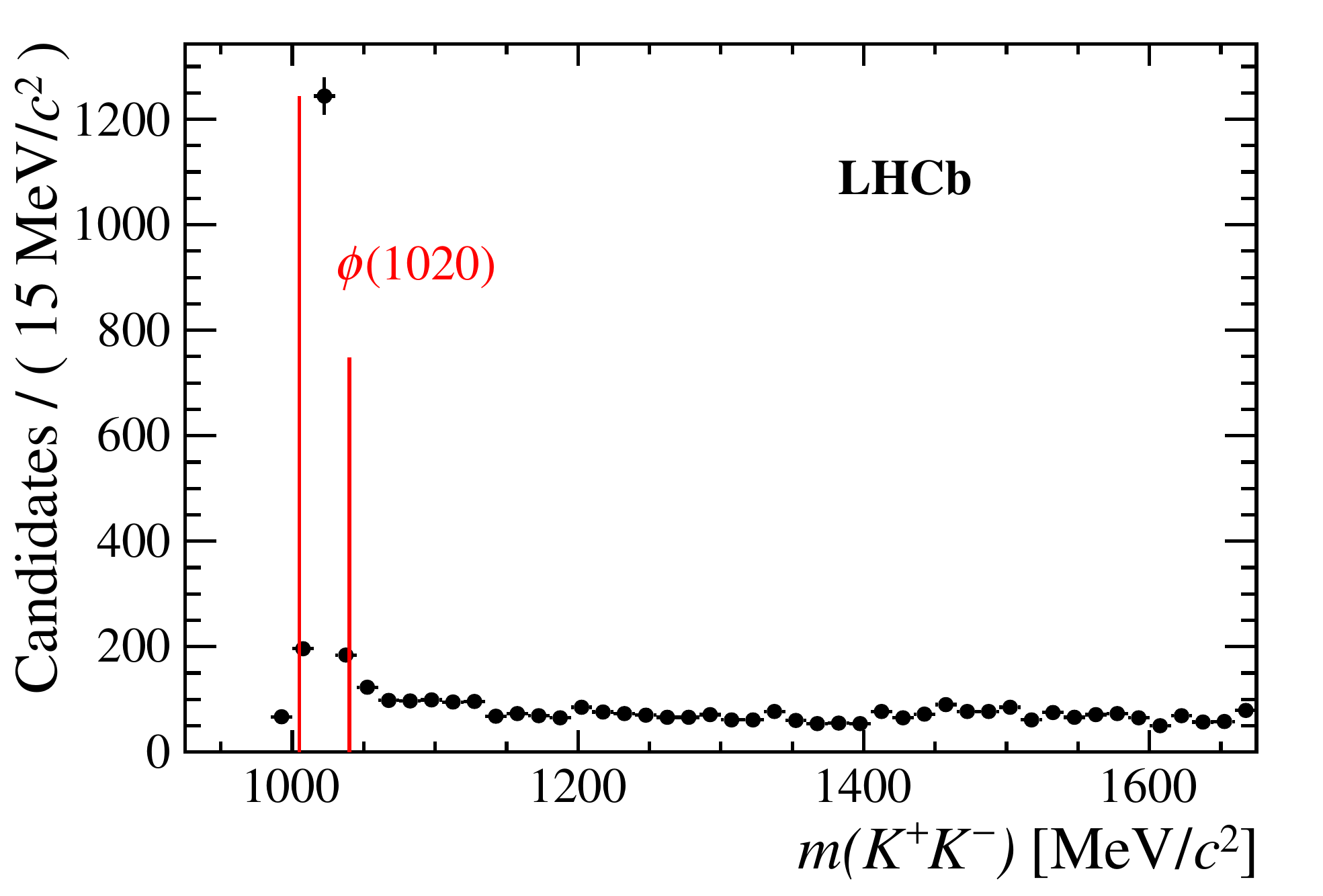} & \includegraphics[width = .49\textwidth]{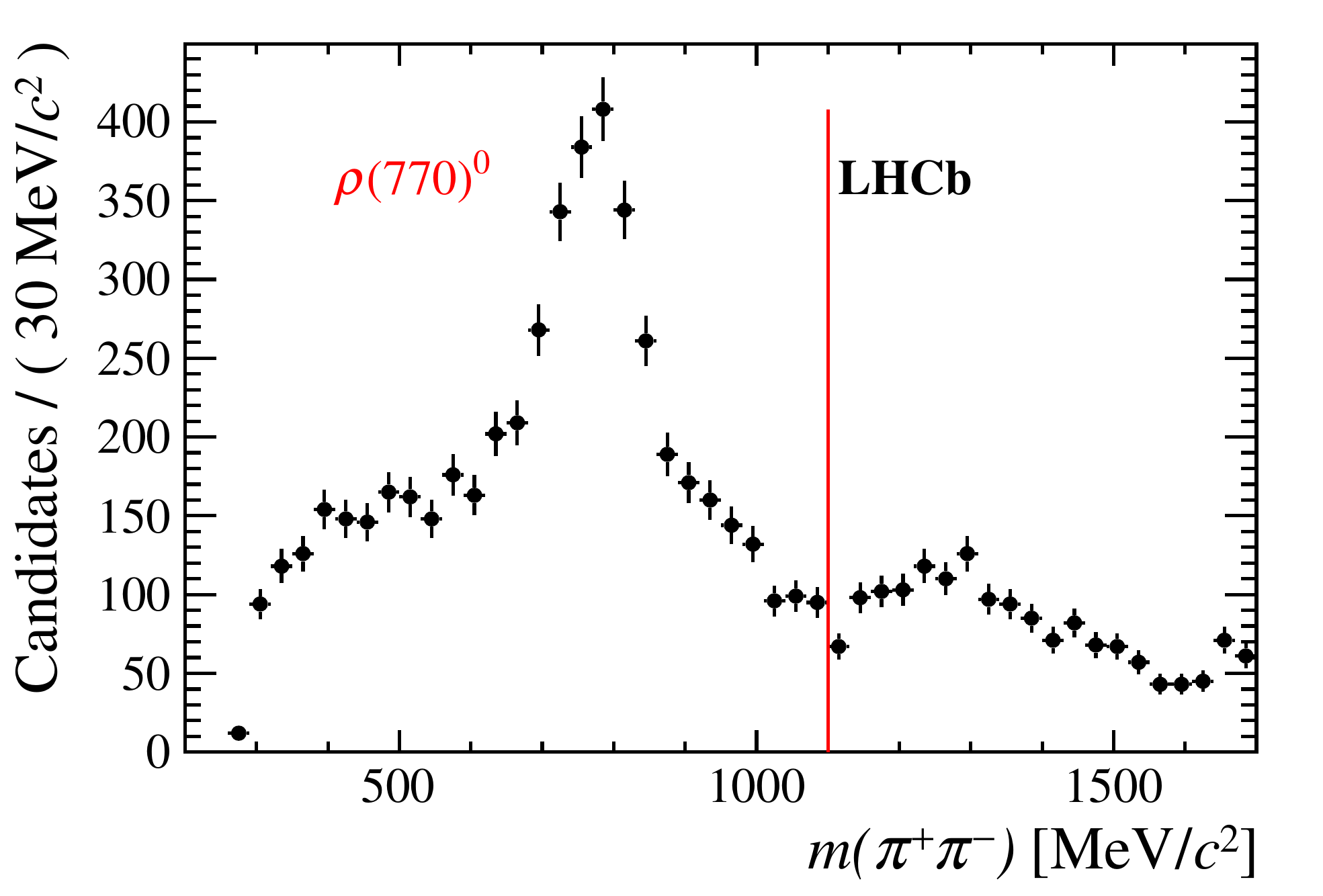} \\
\small (e) & \small (f) \\
\end{tabular}
\caption{Distributions of invariant masses of pairs of final-state particles for the candidates selected in the mass window of $\pm 3\sigma$ around the measured \Lb mass. Figures (a), (b) and (c) show the two-body invariant-mass distributions of baryonic $p\pi^-$, $p\pi^+$ pairs from \LbTopKpipi decays and $pK^-$ pairs from \LbTopKKK decays, respectively. Structures around known the masses of the $N(1520)$, $\Delta(1232)^{++}$ and $\Lambda(1520)$ baryons are observed. Figures (d), (e) and (f) show the invariant-mass distributions of $K^-\pi^+$, $K^-K^+$ and $\pi^+\pi^-$ pairs from \LbTopKpipi, \LbTopKKK and \LbToppipipi decays, respectively. Structures corresponding to the $K^*(892)^0$, $\phiz$ and $\rhoz$ resonances are visible. The red lines correspond to the invariant-mass requirements applied to the selection of the main quasi-two- or three-body decays analysed.}
\label{fig:phsp}
\end{figure}

\item \CP asymmetries are measured for regions of the phase space that contain specific quasi-two-body decays, ${\LbTopaone}$, ${\LbToNstarRhoOrFz}$, ${\LbTopKone}$, ${\LbToLstarRhoOrFz}$, ${\LbToNstarKstar}$, ${\LbToLstarPhi}$ or three-body decays,   ${\LbToDeltapipi}$,  ${\LbToDeltaKpi}$,  and ${\LbTopKPhi}$, where in the latter only the high $pK^{-}$ mass region is selected. Invariant-mass requirements for these measurements are reported in Table~\ref{tab:phspcuts}. Only the narrower baryons or the well-known baryon and meson resonances have been considered, with the noticeable exception of the  $\aone(1260)$ meson. Although the $\aone(1260)$ meson is a broad resonance, the analogous \B-meson decay $\B^0 \to \aone(1260)^{\pm} \pi^{\mp}$ has been studied at the \B-factories~\cite{Aubert:2009ab,Dalseno:2012hp} and could serve as a benchmark comparison in the interpretation of the results obtained for the \LbTopaone decay. 

\end{itemize}

\begin{table}[tb]
  \centering
  \caption{Invariant-mass requirements applied for the different phase-space selections for each final state considered.}
  \label{tab:phspcuts}
  \setlength{\tabcolsep}{5pt}
  \renewcommand{\arraystretch}{1.2}
  \begin{tabular}{lr}
    \hline
    Decay mode              &     Invariant-mass requirements (in \mevcc) \\
    \hline
    \LbToppipipi & \\
    \cline{1-1}
    LBM                     & \small{$m(p\pi^{-})<2000$ and  $m(\pi^+\pi^-)<1640$} \\
    \LbTopaone              & \small{$419<m(\pi^{+}\pi^{-}\pi^{+})<1500$} \\
    \LbToNstarRhoOrFz       & \small{$1078<m(p\pi^-)<1800$ and $m(\pi^{+}\pi^-)<1100$} \\
    \LbToDeltapipi          & \small{$1078<m(p\pi^+)<1432$} \\
    \hline
    \LbTopKpipi & \\
    \cline{1-1}
    LBM                     & \small{$m(pK^{-})<2000$ and $m(\pi^+\pi^-)<1640$} \\
    \LbToNstarKstar         & \small{$1078<m(p\pi^-)<1800$ and $750<m(\pi^{+}K^-)<1100$} \\
    \LbToLstarRhoOrFz       & \small{$1460<m(pK^-)<1580$ and $m(\pi^{+}\pi^-)<1100$} \\
    \LbToDeltaKpi           & \small{$1078<m(p\pi^+)<1432$} \\
    \LbTopKone              & \small{$1200<m(K^-\pi^+\pi^-)<1600$} \\
    \hline
    \LbTopKKK & \\
    \cline{1-1}
    LBM                  & \small{$m(pK^{-})<2000$ and $m(K^+K^-)<1675$} \\
    \LbToLstarPhi        &  \small{$1460<m(pK^-)<1600$ and $1005<m(K^{+}K^-)<1040$} \\
    \LbTopKPhi           &  \small{$m(pK^-)>1600$ and $1005<m(K^{+}K^-)<1040$} \\
    \hline
  \end{tabular}
\end{table}

\section{Corrections for experimental detection asymmetries and related systematic uncertainties}
\label{sec:systematics}

Tracking reconstruction, trigger selection and particle identification requirements can generate charge-dependent selection efficiencies of the decays of interest. Most of these charge-dependent effects are however cancelled out in the \dACP observables, up to the kinematical differences between signal and control channels. The remaining impact is addressed by evaluating corrections to the \dACP observables. These correction factors are determined from calibration samples as discussed in this Section.  Systematic uncertainties are estimated for each correction factor and propagated to the \dACP measurements. A summary of the systematic uncertainties is reported in Table~\ref{tab:allsyst} for all modes. 

\begin{table}[htbp]
    \centering
    \renewcommand{\arraystretch}{1.2}
    \setlength{\tabcolsep}{6.5pt}
    \caption{Systematic uncertainties for each decay mode. The uncertainties related to the kaon and proton detection asymmetry, the difference of triggering efficiency, the PID asymmetries and the production asymmetry are respectively reported as $\sigma_K$, $\sigma_p$, $\sigma_{\rm L0}$, $\sigma_{\rm PID}$ and $\sigma_{A_P}$.}
    \label{tab:allsyst}
    \begin{tabular}{lccccc|c}

\hline
Decay mode       & \multicolumn{5}{c}{Absolute uncertainties (\%)} & Total (\%) \\
                    & $\sigma_K$ & $\sigma_p$ &  $\sigma_{\rm L0}$ & $\sigma_{\rm PID}$ & $\sigma_{A_P}$  &  \\
\hline                                                            
\LbToppipipi        & ---        &  0.20 &  0.06         &  0.42         &  0.28 &   0.54 \\
\LbTopKpipi         &  0.17 &  0.20 &  0.06         &  0.41         &  0.24 &   0.55 \\
\LbTopKKpi          & ---        &  0.21 &  0.06         &  0.40         &  0.55 &   0.72 \\
\LbTopKKK           &  0.15 &  0.20 &  0.07         &  0.41         &  0.33 &   0.59 \\
\XibzTopKpipi       &  0.17 &  0.20 &  0.05         &  0.42         &  0.24 &   0.55 \\
\XibzTopKpiK        &  0.15 &  0.20 &  0.05         &  0.41         &  0.55 &   0.73 \\
\hline                                                                                    
\LbToppipipi (LBM)  &  ---       &  0.16 &  0.06         &  0.36         &  0.28 &  0.49 \\
\LbTopKpipi  (LBM)  &  0.17 &  0.17 &  0.05         &  0.34         &  0.24 &  0.48 \\
\LbTopKKK    (LBM)  &  0.16 &  0.17 &  0.05         &  0.37         &  0.33 &  0.55 \\
\hline                                                                                   
\LbTopaone          & ---        &  0.20 &  0.09         &  0.48         &  0.28 &  0.60 \\
\LbToNstarRhoOrFz   & ---        &  0.12 &  0.05         &  0.23         &  0.28 &  0.39 \\
\LbToDeltapipi      & ---        &  0.18 &  0.05         &  0.47         &  0.28 &  0.59 \\
\hline                                                                                  
\LbTopKone          &  0.16 &  0.14 &  0.11         &  0.58         &  0.24 &  0.74 \\
\LbToLstarRhoOrFz   &  0.12 &  0.12 &  0.04         &  0.36         &  0.24 &  0.49 \\
\LbToNstarKstar     &  0.16 &  0.14 &  0.04         &  0.32         &  0.24 &  0.45 \\
\LbToDeltaKpi       &  0.22 &  0.19 &  0.05         &  0.48         &  0.24 &  0.61 \\
\hline                                                                                  
\LbToLstarPhi       &  0.11 &  0.10 &  0.05         &  0.30         &  0.33 &  0.34 \\
\LbTopKPhi          &  0.15 &  0.14 &  0.06         &  0.58         &  0.33 &  0.64 \\
\hline

\end{tabular}
\end{table}

\begin{itemize}


\item Tracking detection efficiency: differences between the interactions of oppositely charged pions, kaons or protons in the material of the spectrometer induce detection charge asymmetries. The difference in $\pi^{\pm}$ and $K^{\pm}$ tracking efficiency has been quantified with  calibration samples, as a function of the transverse momentum of the tracks~\cite{LHCb-PAPER-2012-009,LHCb-PAPER-2014-013}. The simulated signal and control channels kinematics is used to weight the simulation track efficiency in order to match the hadron detection efficiencies as measured in those calibration samples. The \dACP values (as measured by the difference of signal yields) are then corrected for these efficiencies and the uncertainty on the detection efficiency determination itself is propagated as a systematic uncertainty to the final \dACP measurements, taking into account the correlation between signal and control channel induced by the use of the same calibration samples. The systematic uncertainty arises from the size of the simulated samples used in the weighting, the statistical and systematic uncertainties on the charge asymmetry of the data calibration samples and the knowledge of the kinematical distributions generated in the simulated samples. The latter is determined by taking two different kinematic configurations of the final state (saturated by quasi two-body modes on one hand and phase-space decay on the other hand) and using the difference as the systematic estimate. The difference between the \Pp and $\overline\Pp$ particles is not measured to date. Simulation is used to obtain the reconstruction efficiencies as a function of the momentum of the proton or antiproton track. An additional systematic uncertainty related to the knowledge of the material budget in the simulation is added, as reported in Ref.~\cite{LHCb-PAPER-2018-025}. The proton detection correction follows the same procedure as $\pi^{\pm}$ and $K^{\pm}$ detection asymmetry correction.    


\item The same methodology is used  to correct for the difference of triggering efficiency between oppositely charged hadrons of the signal candidate, at the hardware stage of the trigger system. The trigger asymmetry effects are quantified as a function of the transverse momentum of the tracks of interest, by studying the triggering efficiency of $K^-$ and $\pi^+$ from the decay $D^0 \to K^-\pi^+$~\cite{LHCb-DP-2012-004} and protons from \LbToLcpiLcTopKpi decays.


\item The production asymmetry can depend on the kinematical properties of the reconstructed \Xb candidates, though the actual dependence has not been observed yet~\cite{LHCb-PAPER-2016-062}.  Differences between signal and control channel \Xb candidates kinematics would reflect in an incomplete cancellation of the production asymmetry in the \dACP observable. This effect has been estimated by considering the \Lb production asymmetry measured in Ref.~\cite{LHCb-PAPER-2016-062} as a function of its \pt and pseudorapidity.  


\item The PID requirements set on the tracks of the final state can induce asymmetries. Efficiencies for the final-state particles are determined from \Lc decays selected in data, and are parameterised by their momentum and electric charge. 
The correction factors to apply to the value of \dACP are here again determined by performing a weighting of the simulated signal and control channel events to match the efficiencies measured in the data. The uncertainties coming from the finite size of the calibration samples are propagated as a systematic uncertainty for the final \dACP measurements. 

\end{itemize}

The first three corrections on the value of \dACP are found to be at the few per mille level, commensurate with their uncertainties. The lattermost source is dominating the systematic uncertainty budget, and can reach the percent level. The correction factors are however consistent with zero.    
The design of the fit model and the simultaneous fit strategy allow the direct measurement of the combinatorial background and the \B-meson decay asymmetries. 
No significant asymmetries are observed and the results are presented in Section~\ref{sec:conclusions}.  
Systematic uncertainties can be induced by the fit model and the fit complexity and it is evaluated by means of pseudoexperiments reproducing the nominal fit results. No significant biases are obtained. The normalised residuals of the signal yields are computed and the uncertainties on their pull mean value are propagated as a systematic uncertainty to each relevant \dACP measurement. The largest uncertainty is determined to be at the level of few $10^{-4}$, hence negligible in comparison to the aforementioned systematic uncertainty estimate.

\section{Fit results}
\label{sec:fitresults}

The results of the simultaneous fits to the five experimental spectra split by year of data taking, magnet polarity and trigger conditions are discussed in this section. The fit results are reported for each final state in the following subsections, and the summary of the measured yields is reported in Table~\ref{tab:yields}.

\begin{table}[tb]
  \renewcommand{\arraystretch}{1.2}
  \setlength{\tabcolsep}{14pt}
  
  \caption{Signal yields for each decay mode, summed over all trigger configurations and years of data taking.}
  \label{tab:yields}
  \begin{center}
    \begin{tabular}{l c c}
      \hline
   Decay mode             &  \multicolumn{2}{c}{Signal yields} \\
                          &  \Xb             &   \Xbbar       \\
   \hline
   \LbToppipipi           & \phantom{0}2335 $\pm$ \phantom{0}56  & \phantom{0}2264 $\pm$ \phantom{0}55 \\
   \LbTopKpipi            & \phantom{0}6807 $\pm$ \phantom{0}92  & \phantom{0}6232 $\pm$ \phantom{0}89 \\
   \LbTopKKpi             & \phantom{00}555 $\pm$ \phantom{0}38  & \phantom{00}630 $\pm$ \phantom{0}38 \\
   \LbTopKKK              & \phantom{0}2312 $\pm$ \phantom{0}54  & \phantom{0}2248 $\pm$ \phantom{0}54 \\
   \XibzTopKpipi          & \phantom{00}180 $\pm$ \phantom{0}28  & \phantom{00}252 $\pm$ \phantom{0}29 \\
   \XibzTopKpiK           & \phantom{00}265 $\pm$ \phantom{0}25  & \phantom{00}305 $\pm$ \phantom{0}26 \\
   \LbToLcpiLcToppipi     & \phantom{0}1607 $\pm$ \phantom{0}40  & \phantom{0}1586 $\pm$ \phantom{0}40 \\
   \LbToLcpiLcTopKpi      & 24687           $\pm$           159  & 24052 $\pm$ 157 \\
   \XibzToXicpiXicTopKpi  & \phantom{00}259 $\pm$ \phantom{0}18  & \phantom{00}260 $\pm$ \phantom{0}18 \\
   \hline
   \LbToppipipi (LBM)     & \phantom{00}498 $\pm$ \phantom{0}25 & \phantom{00}455 $\pm$ \phantom{0}24 \\
   \LbTopKpipi  (LBM)     & \phantom{0}3217 $\pm$ \phantom{0}61 & \phantom{0}2929 $\pm$ \phantom{0}58 \\
   \LbTopKKK    (LBM)     & \phantom{0}1240 $\pm$ \phantom{0}38 & \phantom{0}1146 $\pm$ \phantom{0}36 \\
   \hline                                                                               
   \LbTopaone             & \phantom{00}422 $\pm$ \phantom{0}23  & \phantom{00}425 $\pm$ \phantom{0}23 \\
   \LbToDeltapipi         & \phantom{00}783 $\pm$ \phantom{0}30  & \phantom{00}771 $\pm$ \phantom{0}29 \\
   \LbToNstarRhoOrFz      & \phantom{00}241 $\pm$ \phantom{0}16  & \phantom{00}230 $\pm$ \phantom{0}16 \\
   \LbTopKone             & \phantom{00}548 $\pm$ \phantom{0}26  & \phantom{00}488 $\pm$ \phantom{0}25 \\
   \LbToDeltaKpi          & \phantom{00}998 $\pm$ \phantom{0}37  & \phantom{00}895 $\pm$ \phantom{0}34 \\
   \LbToLstarRhoOrFz      & \phantom{00}167 $\pm$ \phantom{0}14  & \phantom{00}160 $\pm$ \phantom{0}14 \\
   \LbToNstarKstar        & \phantom{00}977 $\pm$ \phantom{0}33  & \phantom{00}856 $\pm$ \phantom{0}31 \\
   \LbToLstarPhi          & \phantom{00}192 $\pm$ \phantom{0}15  & \phantom{00}172 $\pm$ \phantom{0}14 \\
   \LbTopKPhi             & \phantom{00}548 $\pm$ \phantom{0}25  & \phantom{00}542 $\pm$ \phantom{0}25 \\
   \hline
    \end{tabular}
  \end{center}
\end{table}


\begin{itemize}

\item {\bf \ppipipib final state}: Figures~\ref{fitresults_pipipi_1} and \ref{fitresults_pipipi_2} show the results of the simultaneous fits to the invariant-mass spectra of the \ppipipi spectra for the inclusive, LBM and quasi two-body measurements. 
The high-mass region of the \ppipipi spectrum is only populated by either \B-meson decays or combinatorial background.  The good agreement between the data and the fit model, especially in this region, validates the chosen modelling of these components. The same comment is in order for the fit in the different phase-space regions. 
The combinatorial component becomes negligible in the quasi two-body case.  

\begin{figure}[ht]
  \centering
  \includegraphics[width=.495\columnwidth]{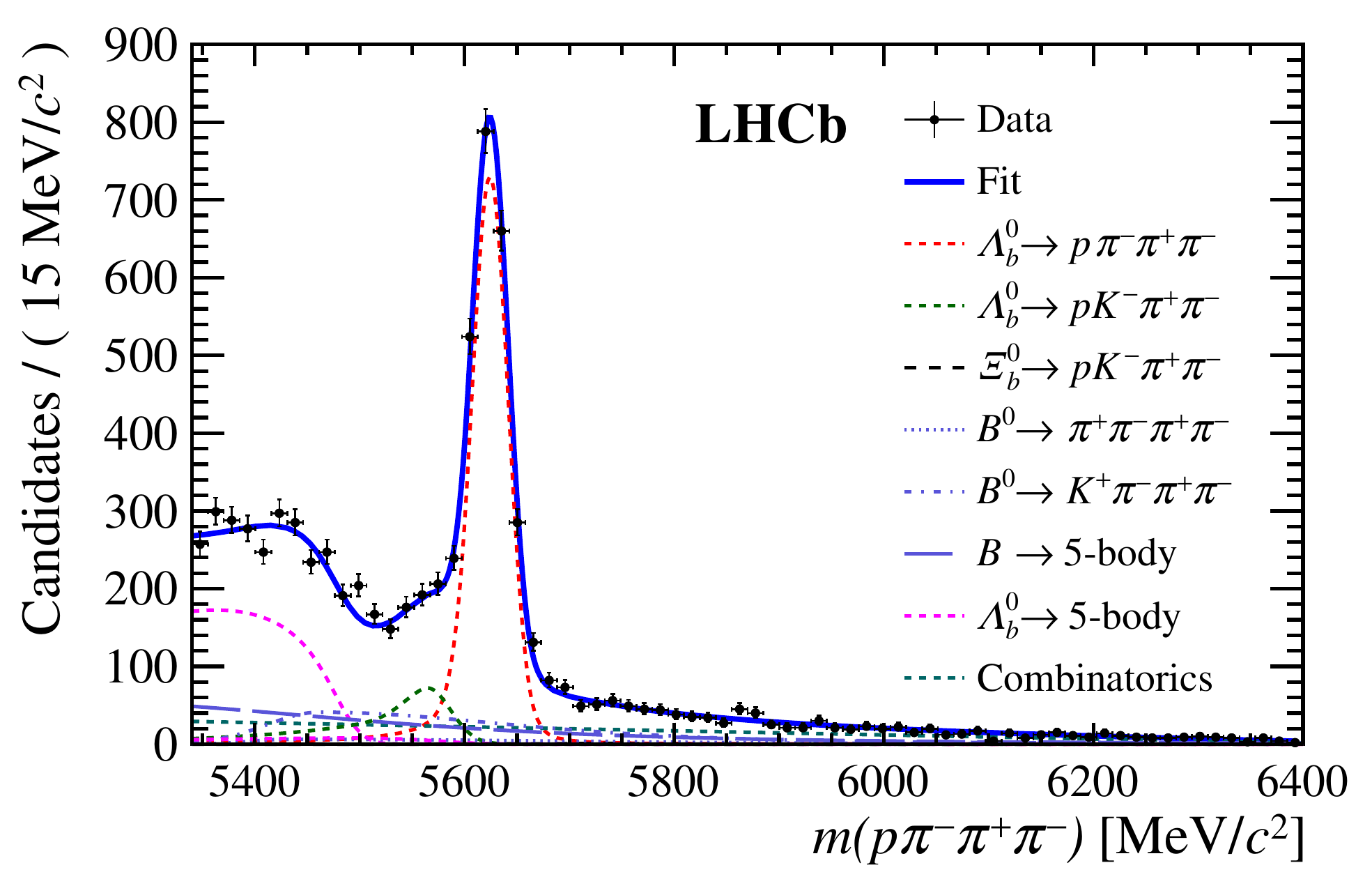}
  \includegraphics[width=.495\columnwidth]{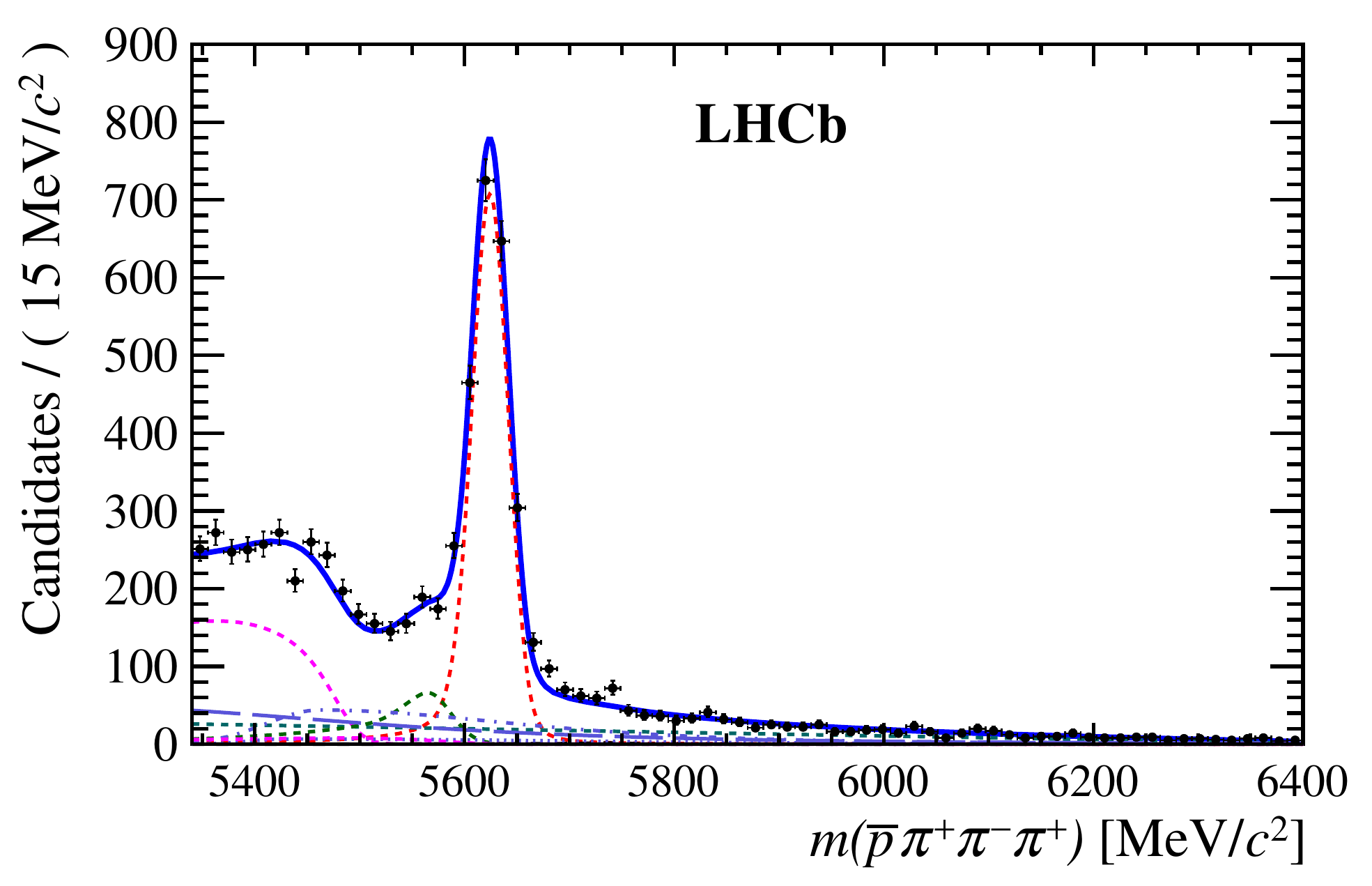}
  \includegraphics[width=.495\columnwidth]{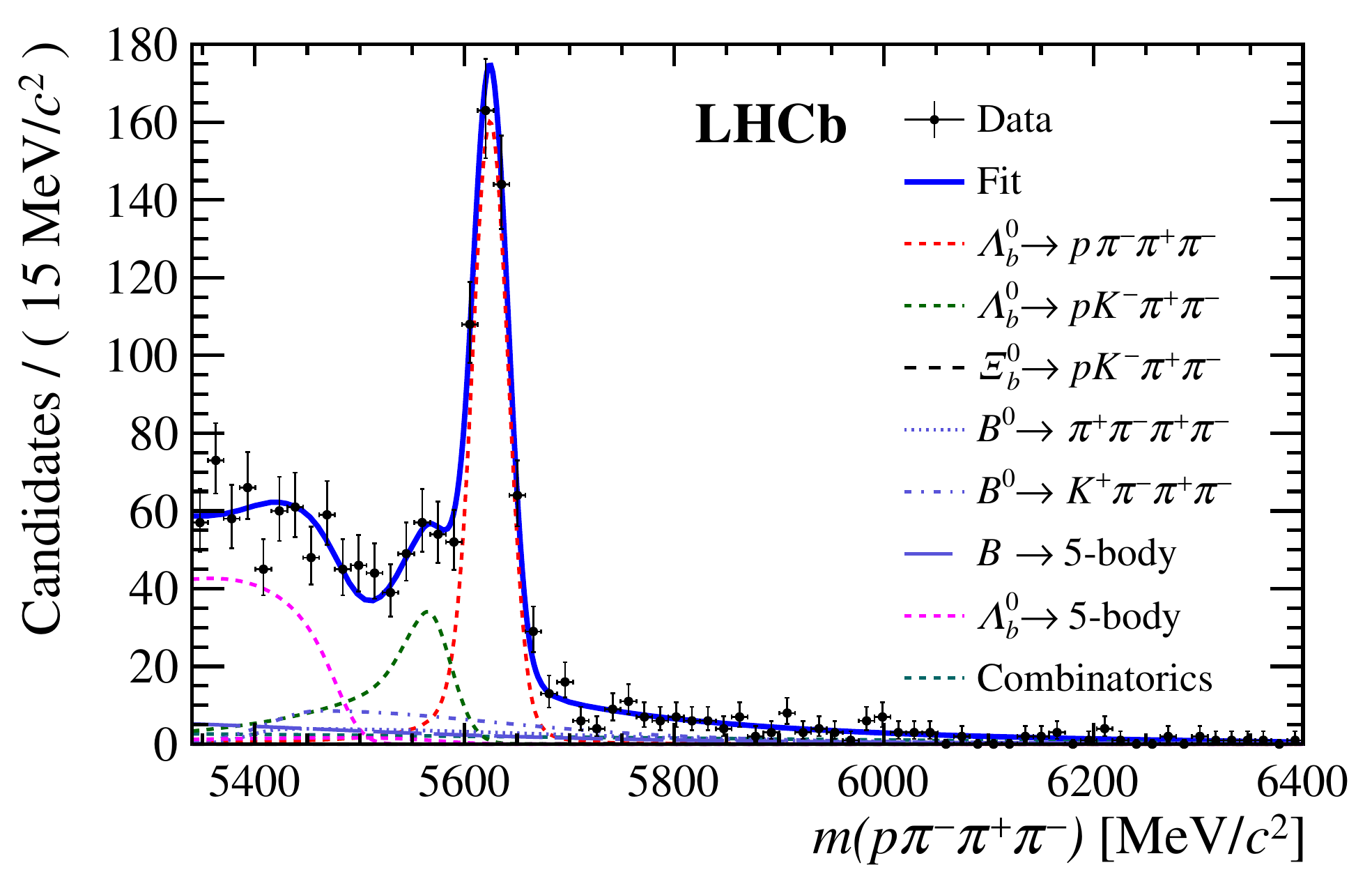}
  \includegraphics[width=.495\columnwidth]{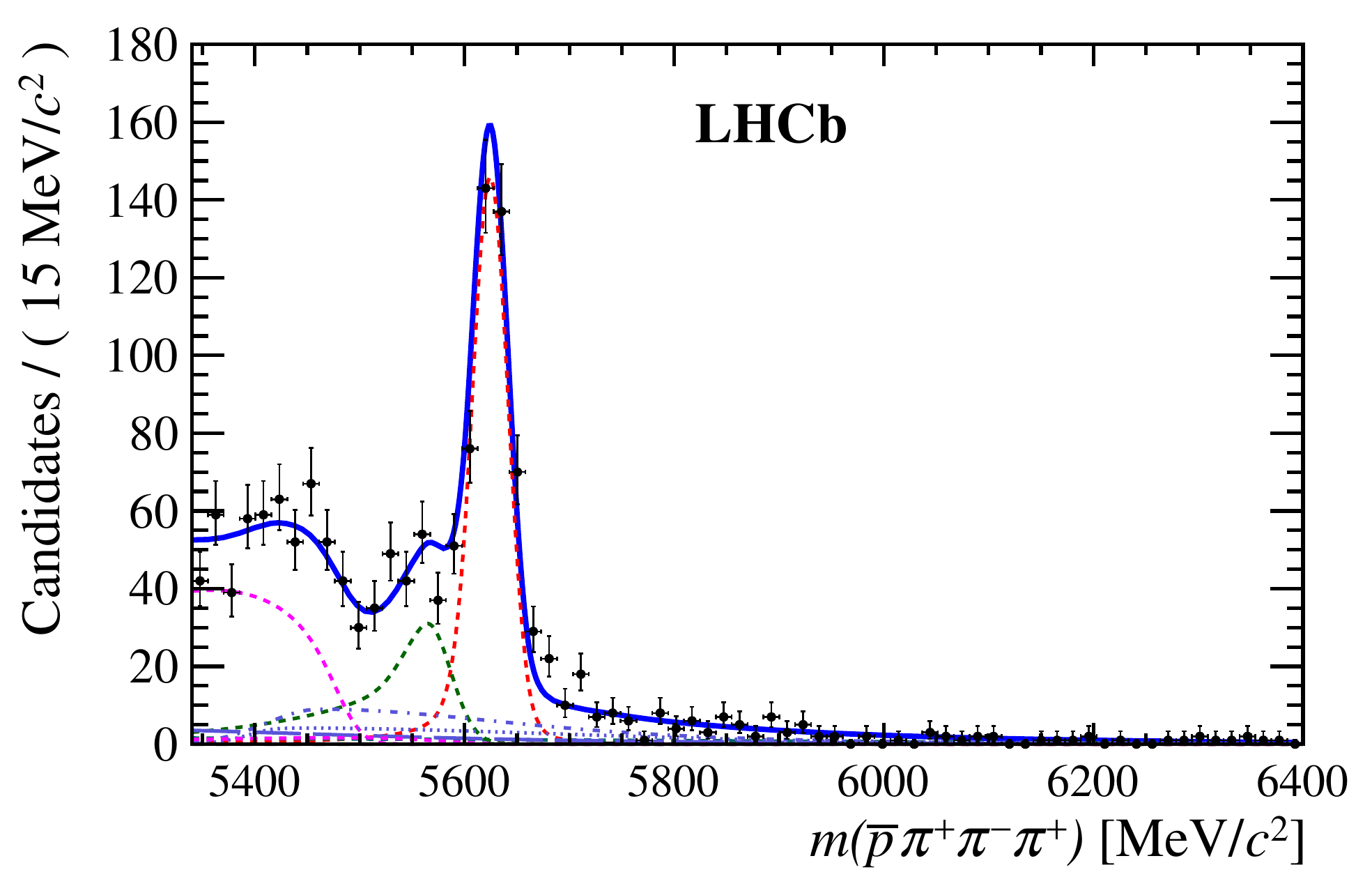}
  \includegraphics[width=.495\columnwidth]{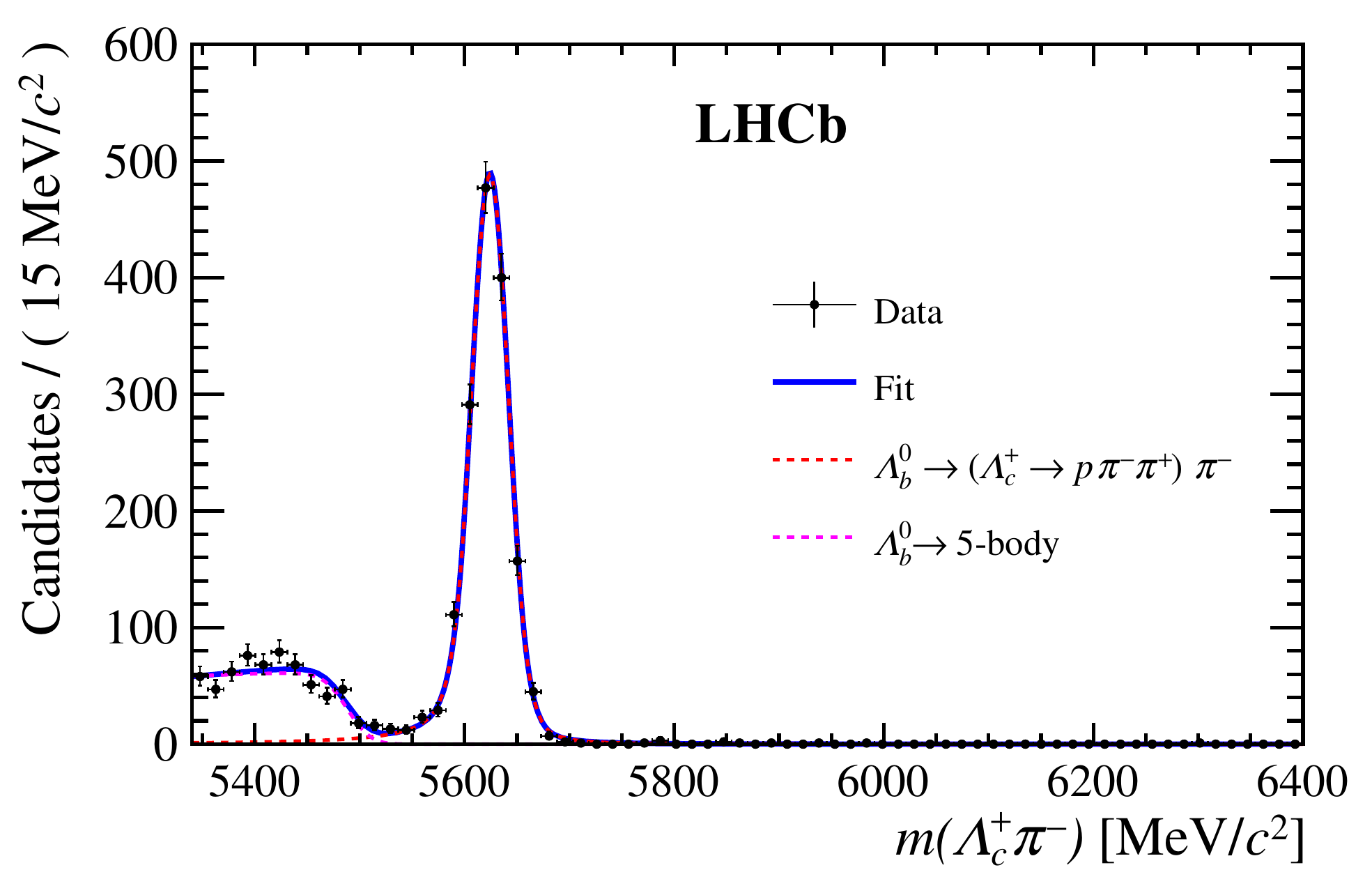}
  \includegraphics[width=.495\columnwidth]{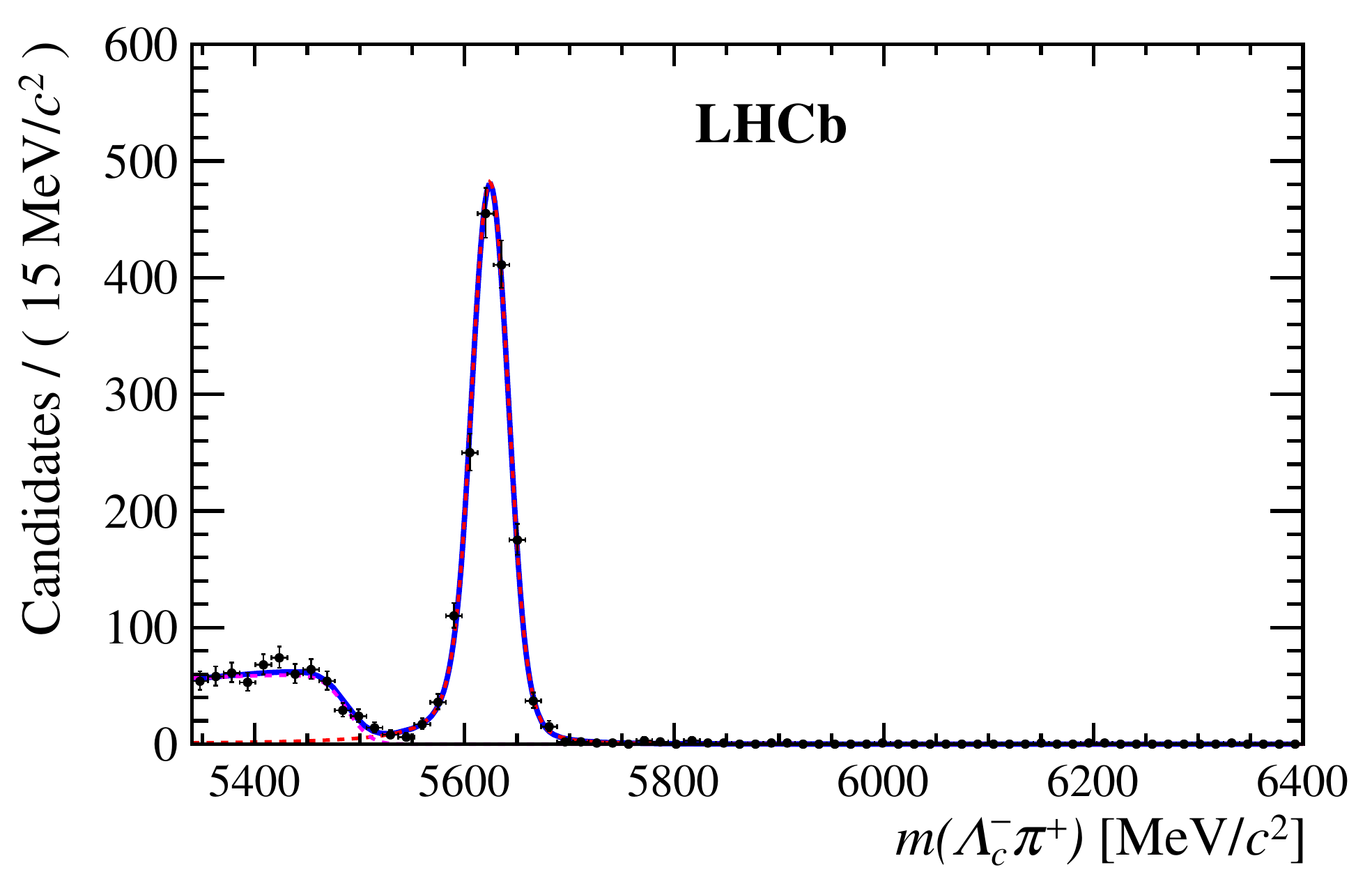}
  \caption{Invariant \ppipipi mass distributions with the results of the fit superimposed: (first row) full phase space, (second row) LBM and (third row) \LbToLcpiLcToppipi control channel. The two columns correspond to the charge-conjugate final states: (left) baryon, (right) antibaryon. The different components employed in the fit model are indicated in the legends. The $\Lb\to$ five-body legend includes two components: the partially reconstructed \LbToppietap and \LbToppipipipiz decays where a $\gamma$ or $\pi^0$ is not reconstructed. The latter has a lower-mass endpoint.}
  \label{fitresults_pipipi_1}
\end{figure}

\begin{figure}[ht]
  \centering
  \includegraphics[width=.495\columnwidth]{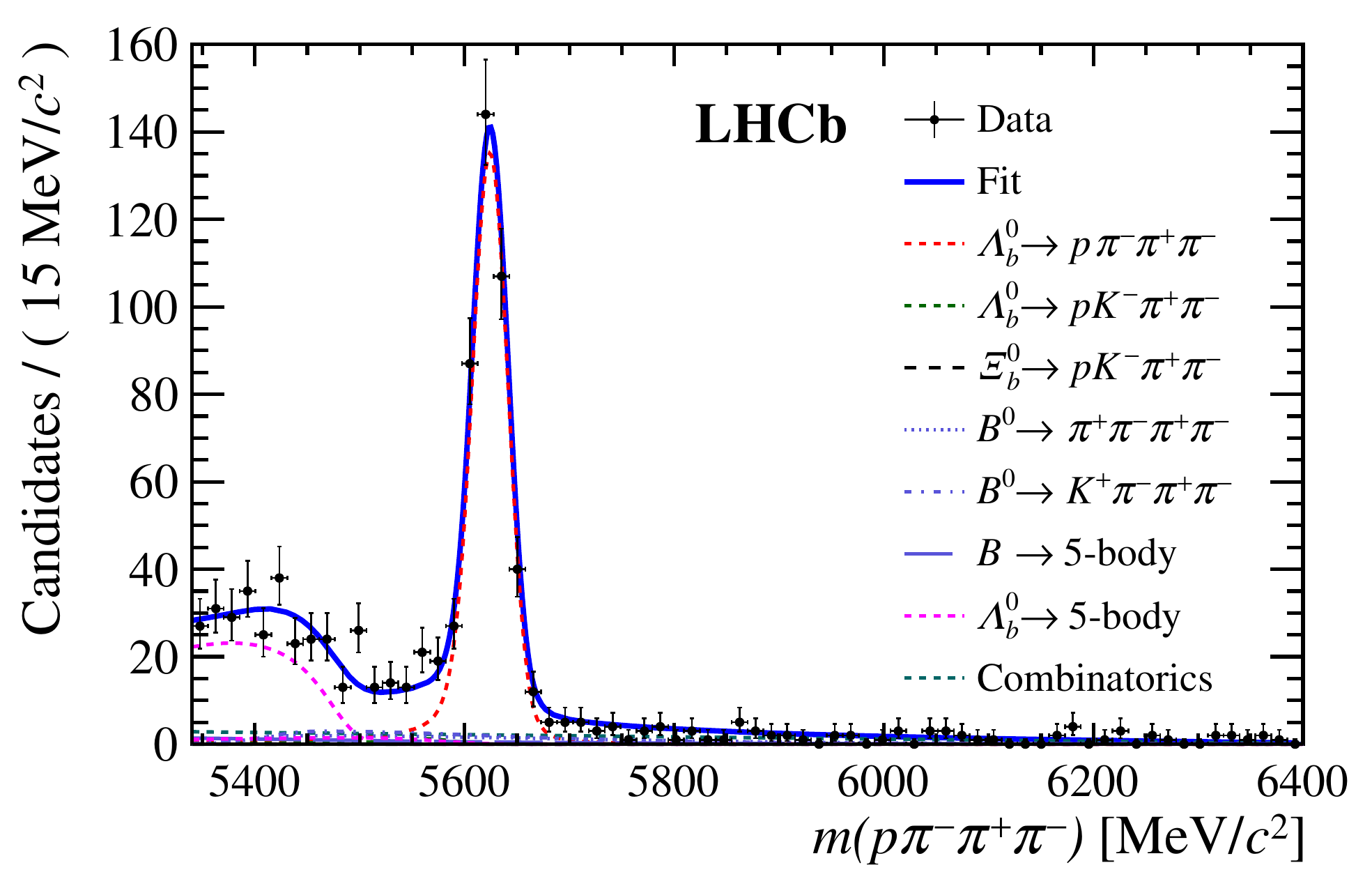}
  \includegraphics[width=.495\columnwidth]{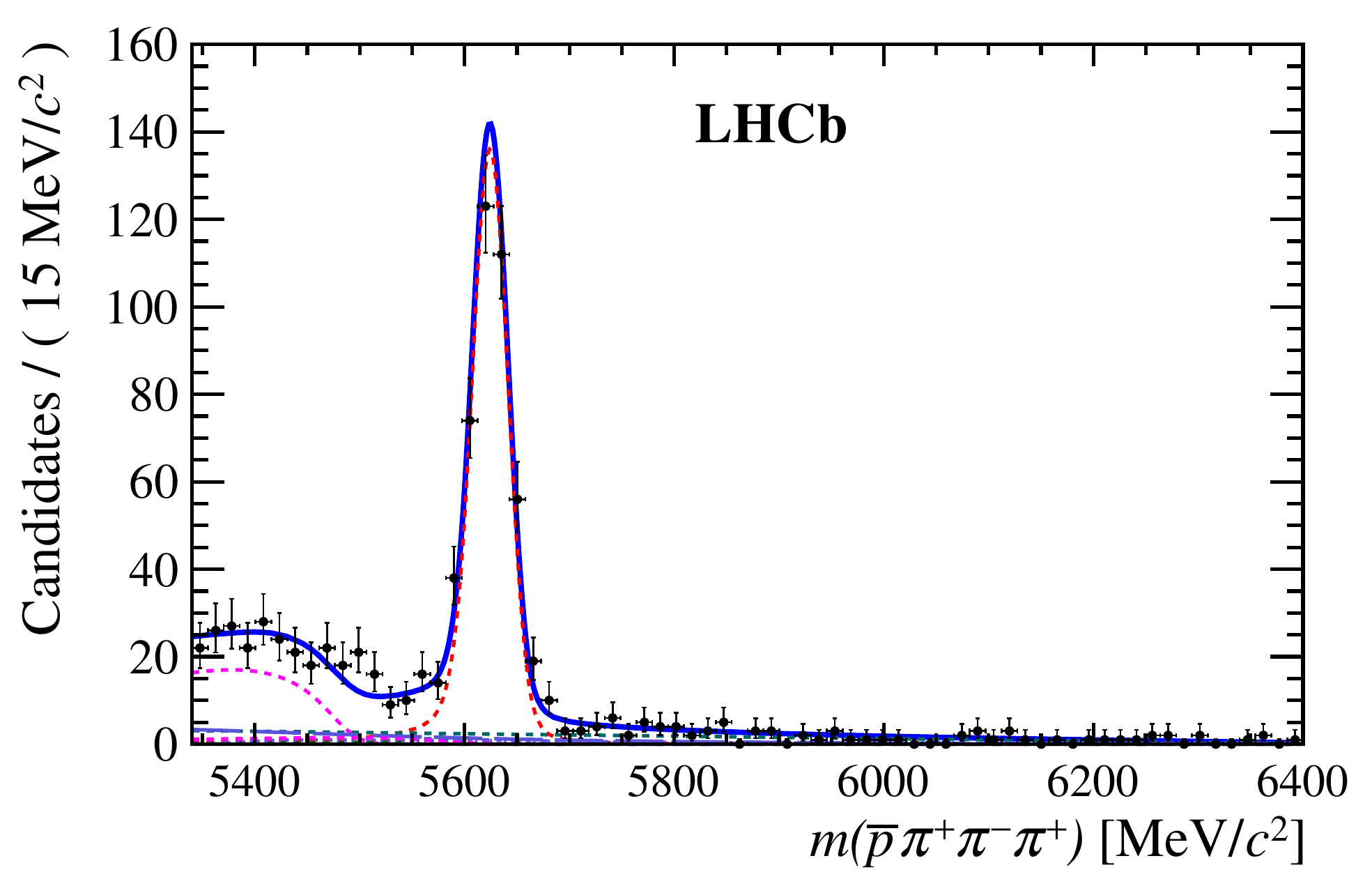}
  \includegraphics[width=.495\columnwidth]{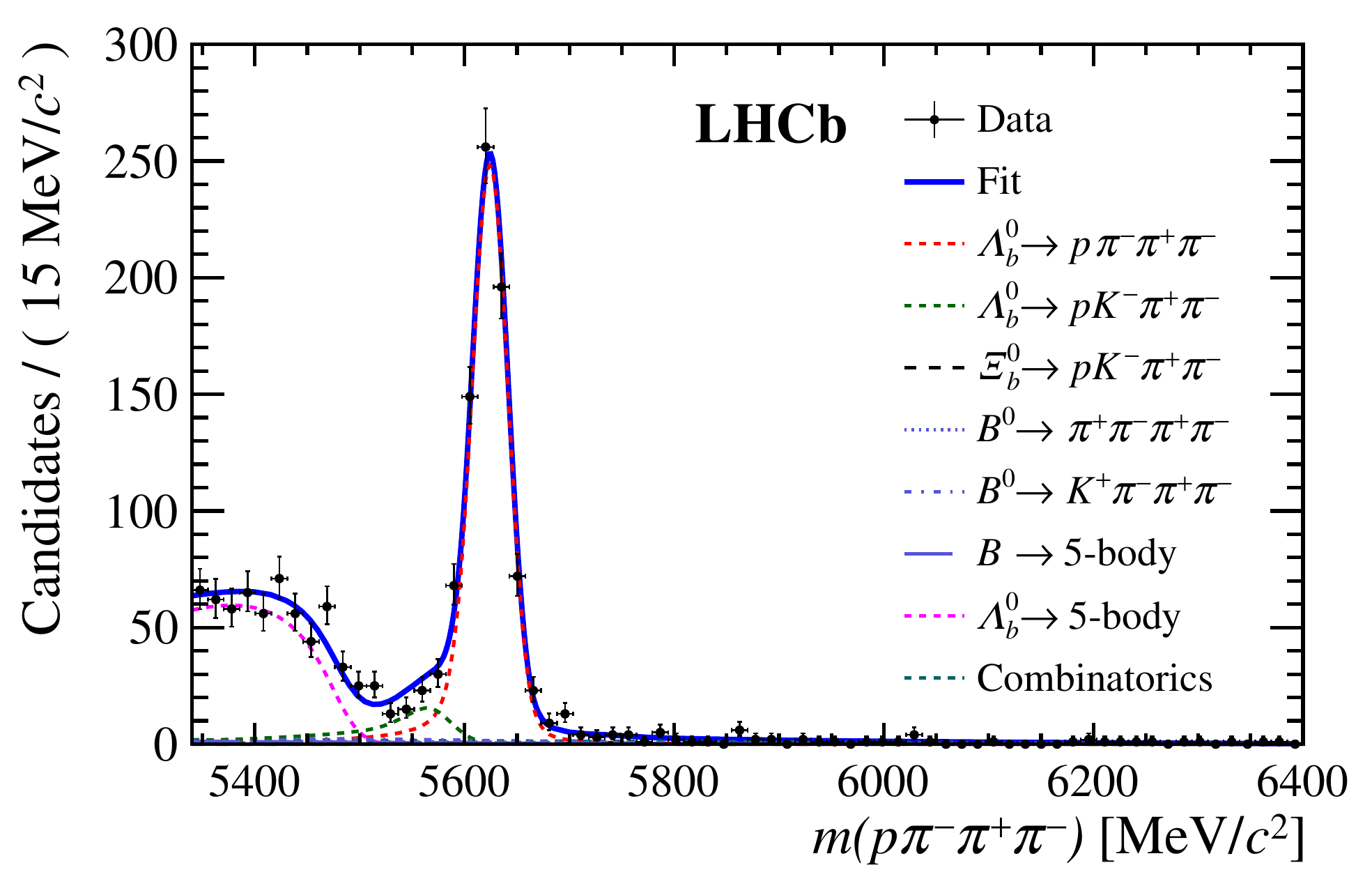}
  \includegraphics[width=.495\columnwidth]{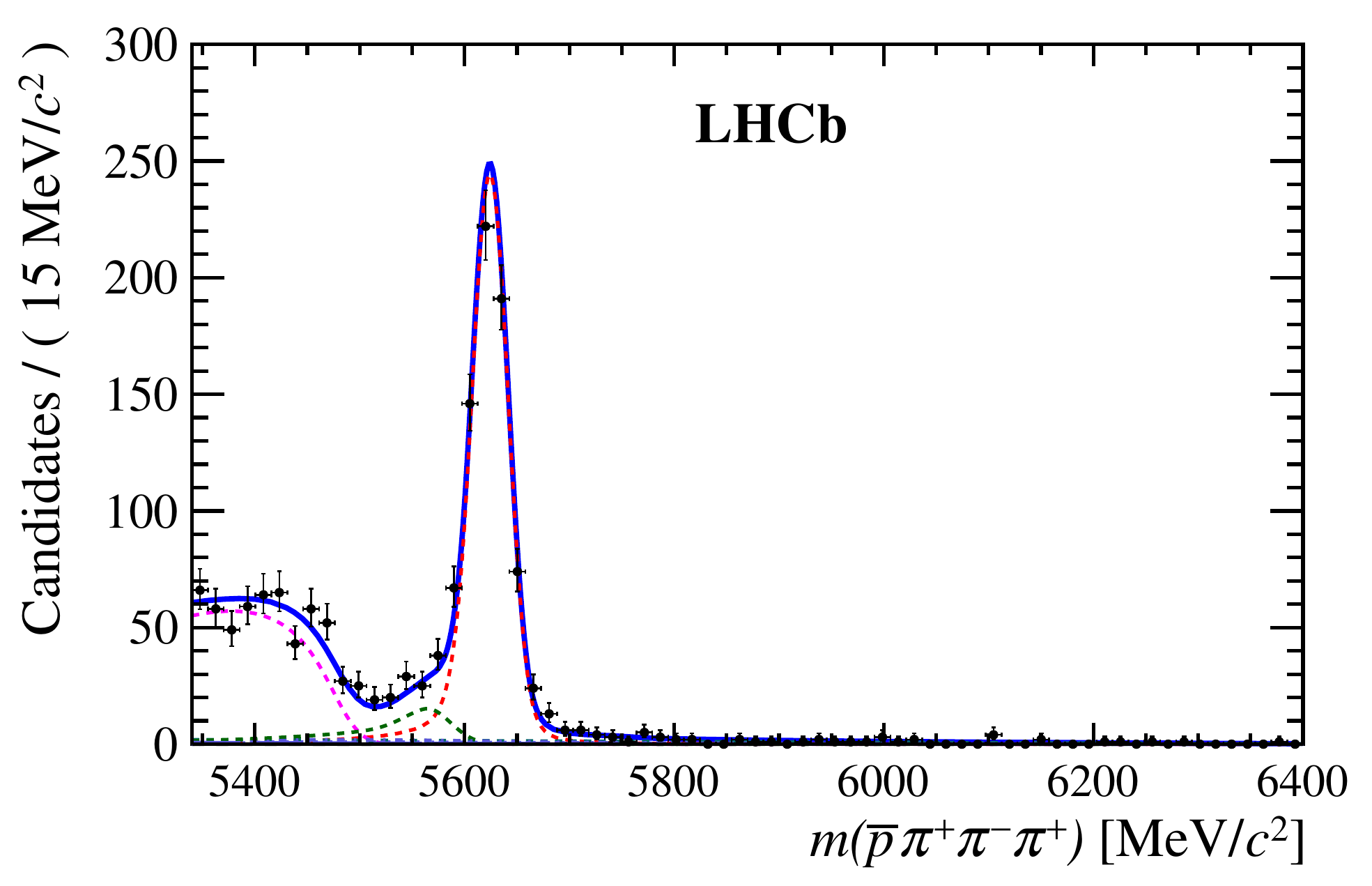}
  \includegraphics[width=.495\columnwidth]{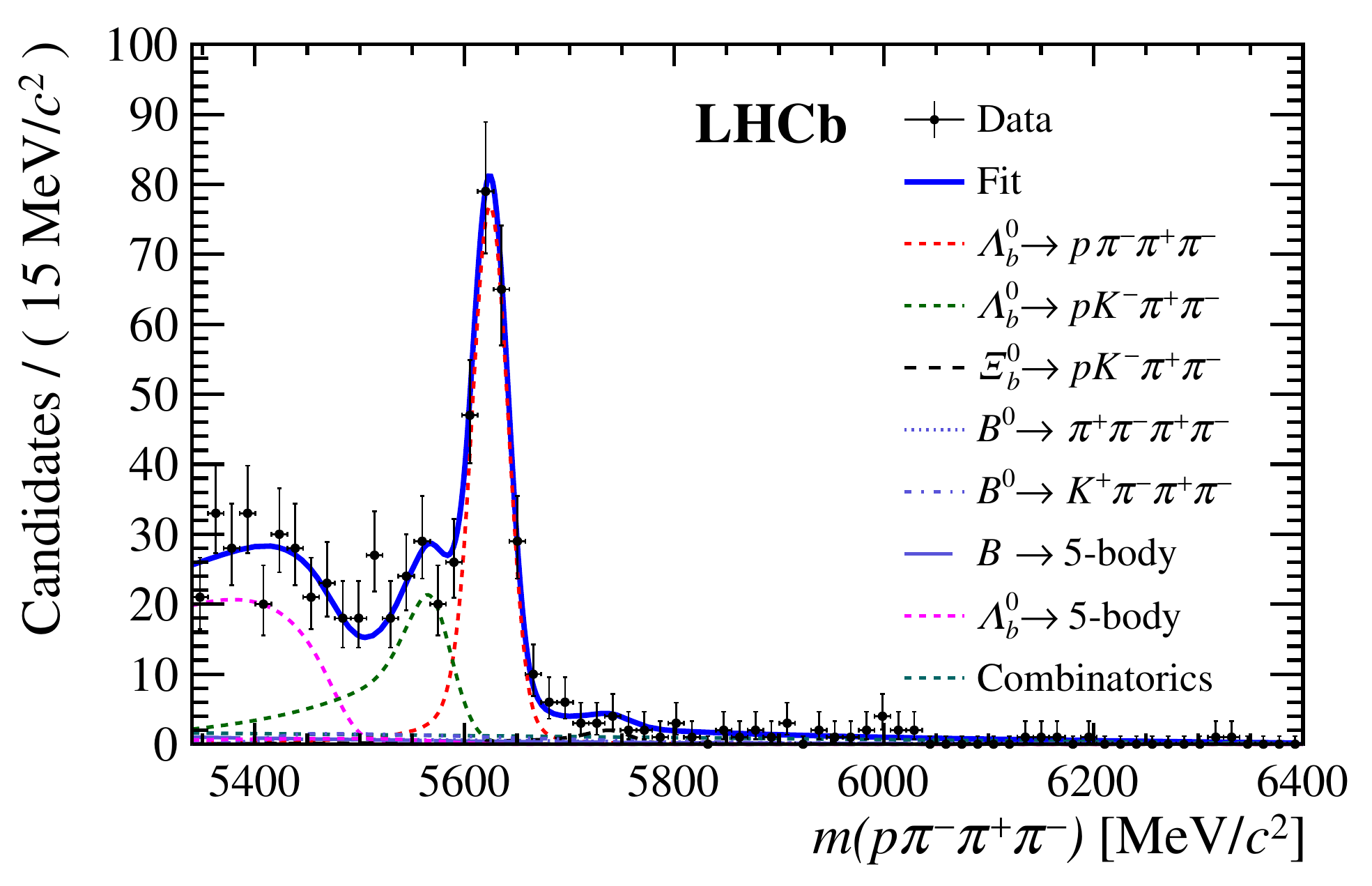}
  \includegraphics[width=.495\columnwidth]{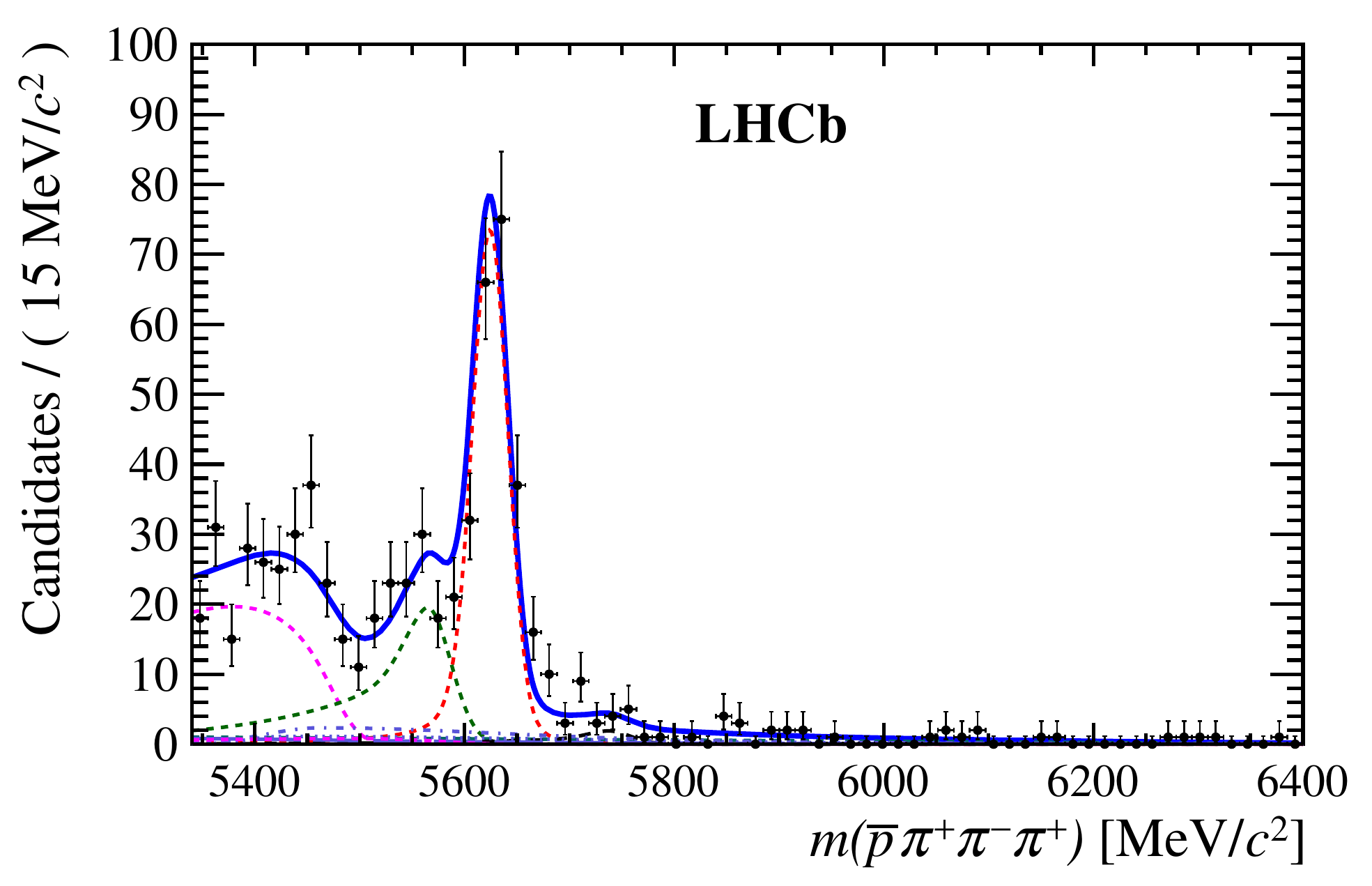}
  \caption{Invariant \ppipipi mass distributions with the results of the fit superimposed: region of the phase space containing (first row)  \LbTopaone, (second row) \LbToDeltapipi and (third row) \LbToNstarRhoOrFz quasi two-body decays. The two columns correspond to the charge-conjugate final states: (left) baryon, (right) antibaryon. The different components employed in the fit model are indicated in the legends. The $\Lb\to$ five-body legend includes two components: the partially reconstructed \LbToppietap and \LbToppipipipiz decays where a $\gamma$ or $\pi^0$ is not reconstructed. The latter has a lower-mass endpoint.}
  \label{fitresults_pipipi_2}
\end{figure}

\item {\bf \pKpipib final state}: Figures~\ref{fitresults_pKpipi_1} and \ref{fitresults_pKpipi_2} show the results of the simultaneous fits to the \pKpipi mass spectrum for the inclusive, LBM and quasi two-body measurements. The fit model provides also in this case a satisfactory description of the data, despite the very different background contributions depending on the phase-space selection. Raw asymmetries at the level of several percent are observed.  


\begin{figure}[ht]
  \centering
  \includegraphics[width=.495\columnwidth]{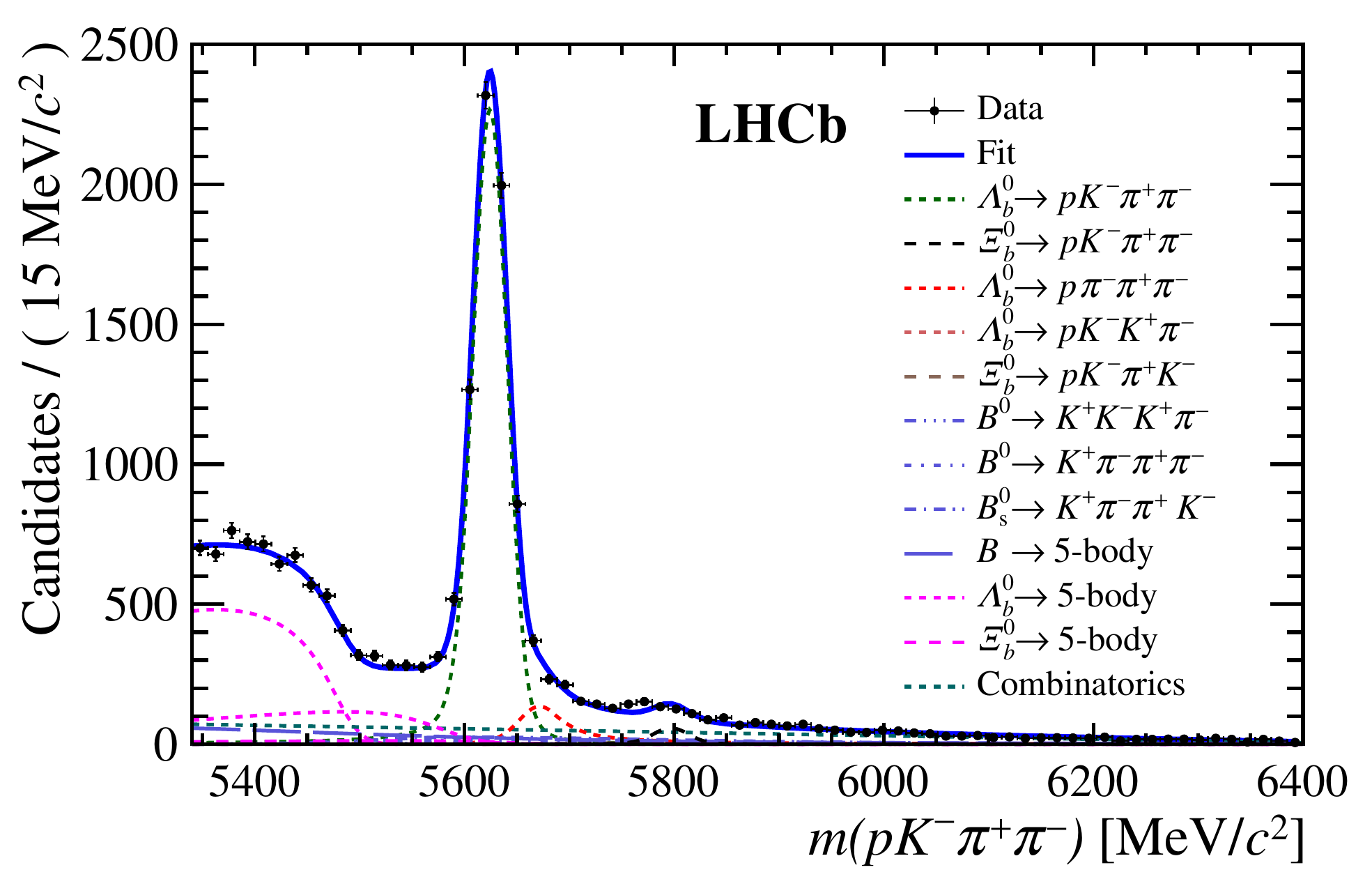}
  \includegraphics[width=.495\columnwidth]{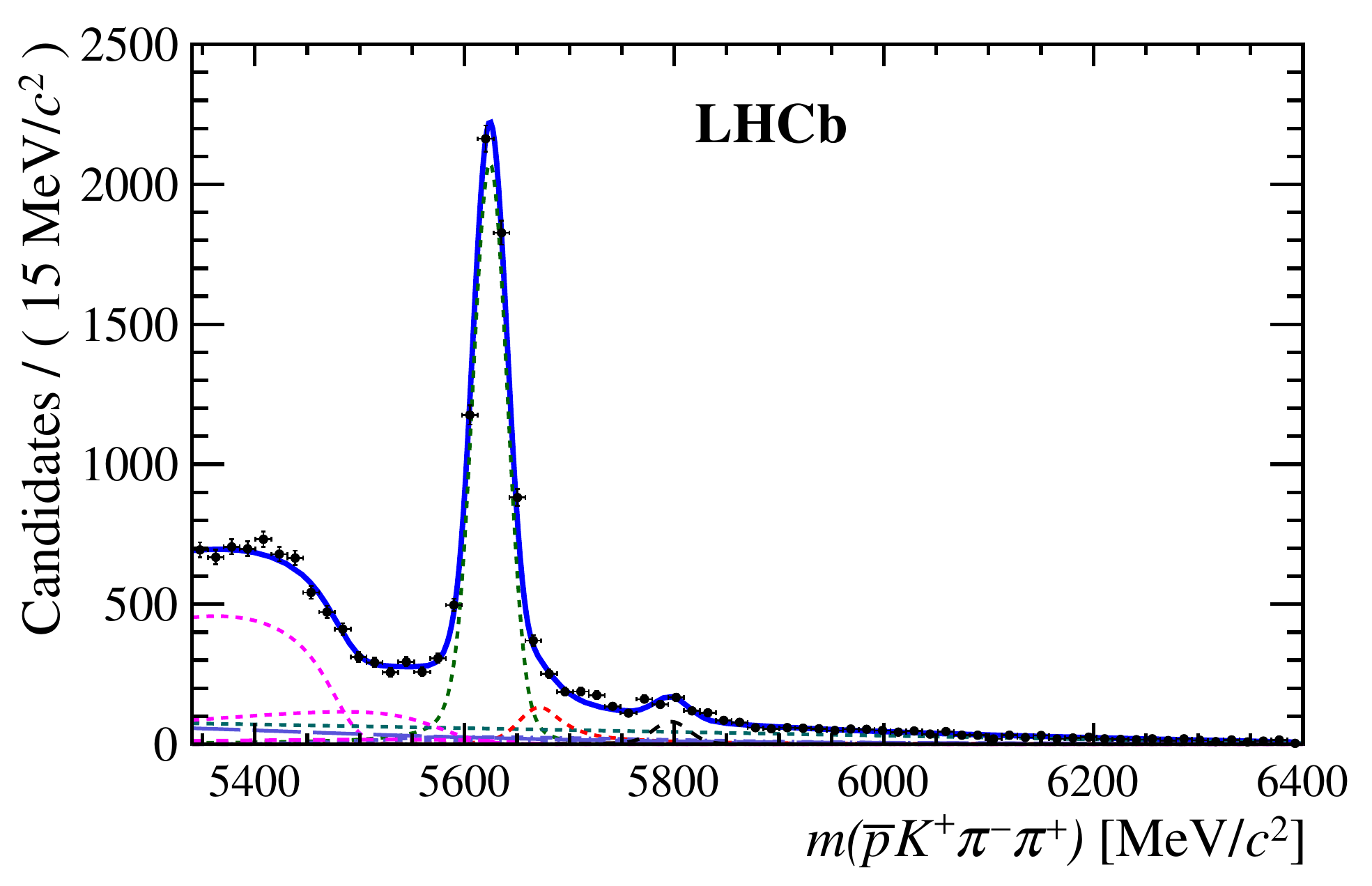}
  \includegraphics[width=.495\columnwidth]{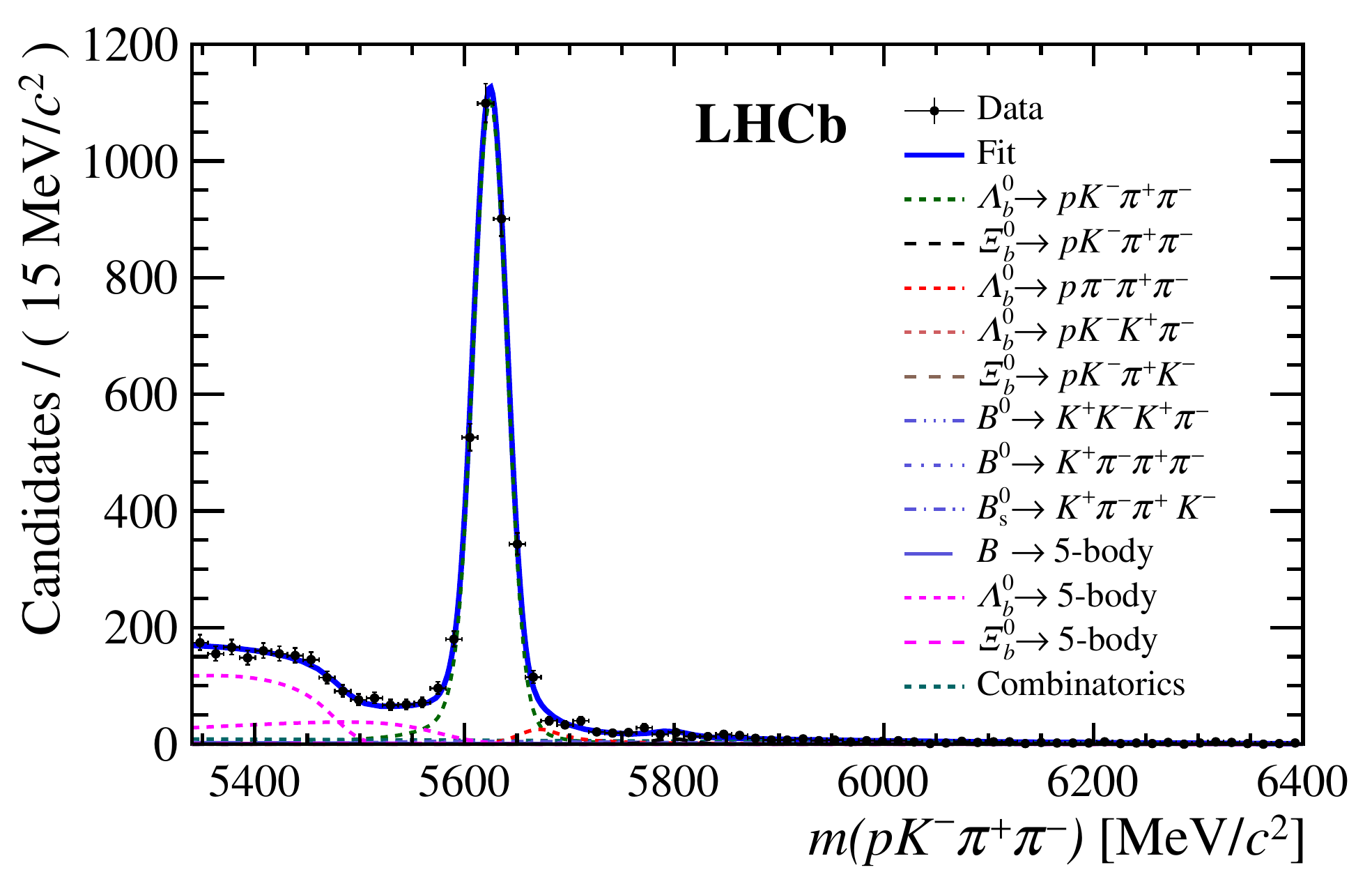}
  \includegraphics[width=.495\columnwidth]{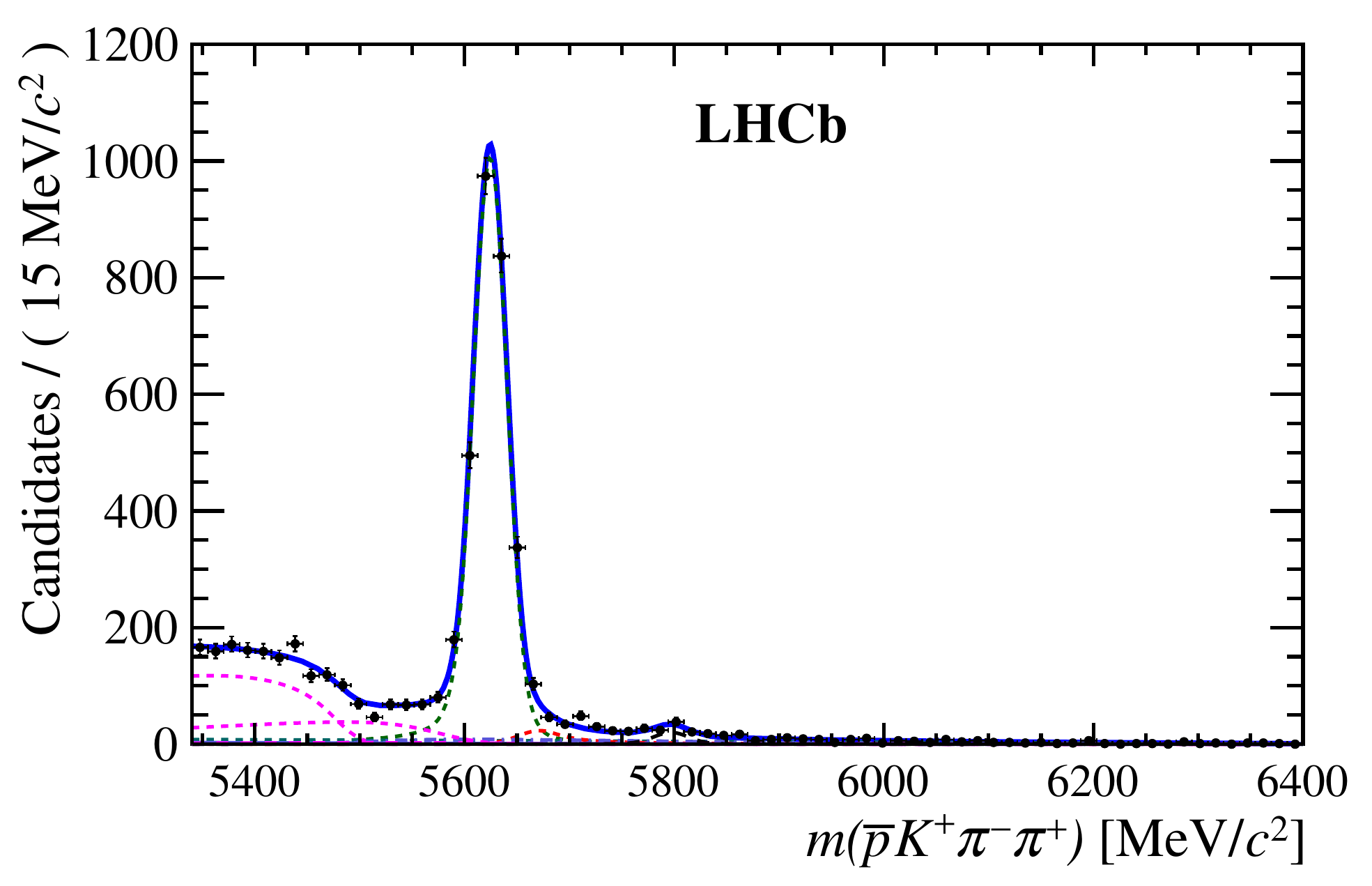}
  \includegraphics[width=.495\columnwidth]{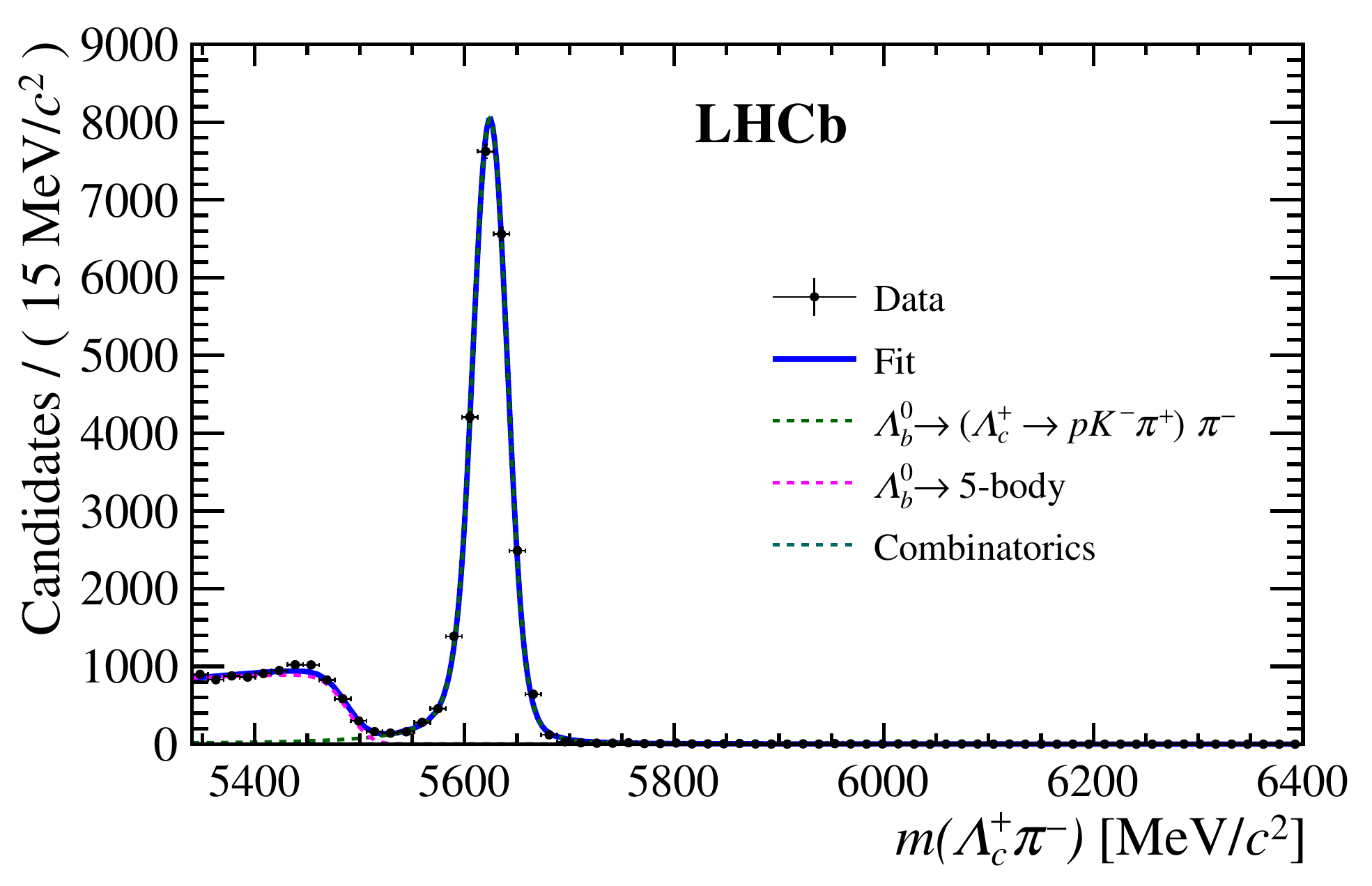}
  \includegraphics[width=.495\columnwidth]{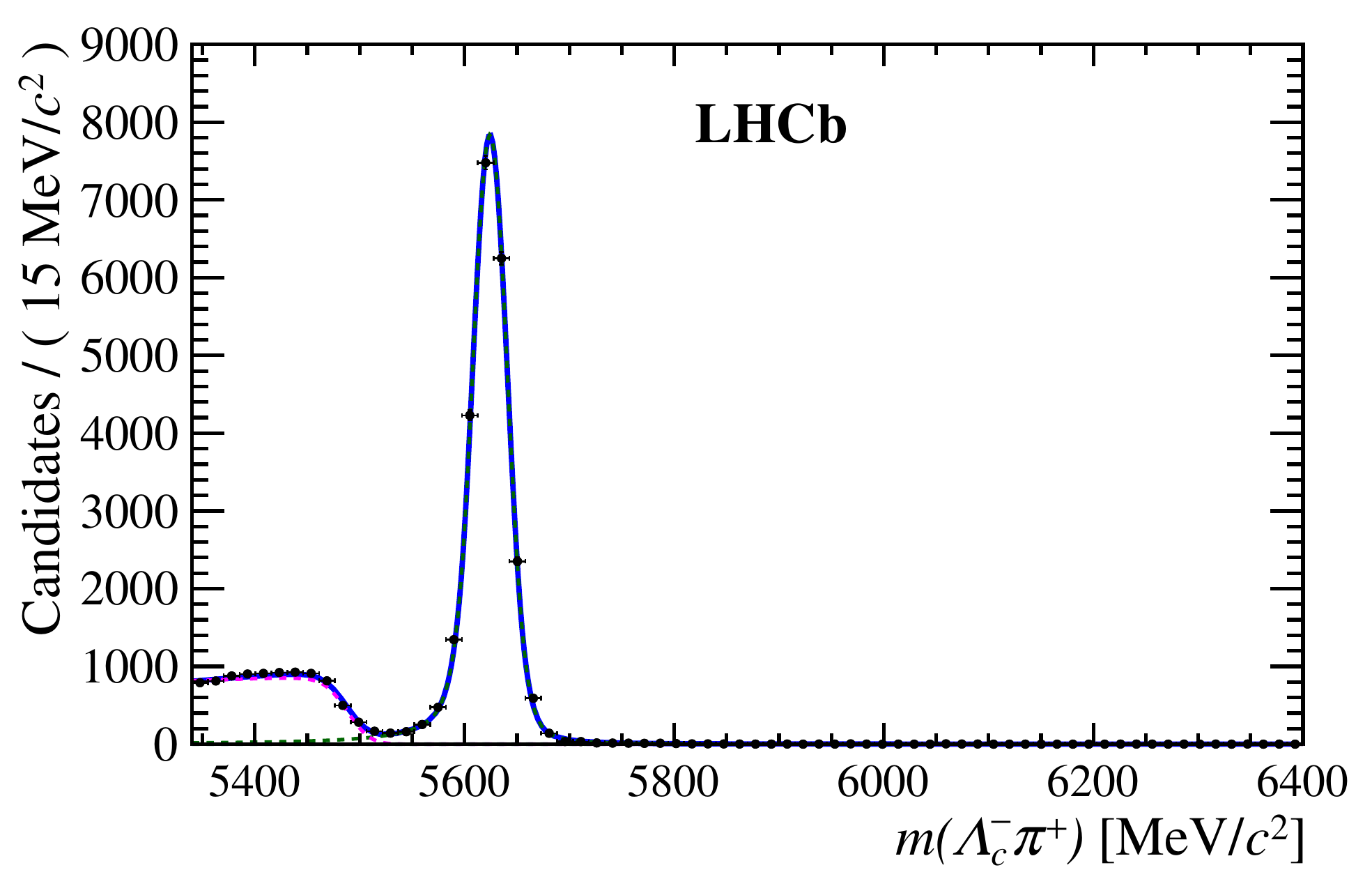}
  \caption{Invariant \pKpipi mass distributions with the results of the fit superimposed: (first row) full phase space, (second row) LBM and (third row) \LbToLcpiLcTopKpi control channel. The two columns correspond to the charge-conjugate final states: (left) baryon, (right) antibaryon. The different components employed in the fit are indicated in the legends. The $\Lb\to$ five-body legend includes two components: the partially reconstructed \LbTopKetap and \LbTopKpipipiz decays where a $\gamma$ or $\pi^0$ is not reconstructed. The latter has a lower-mass endpoint.}
  \label{fitresults_pKpipi_1}
\end{figure}

\begin{figure}[hpbt]
  \centering
  \includegraphics[width=.495\columnwidth]{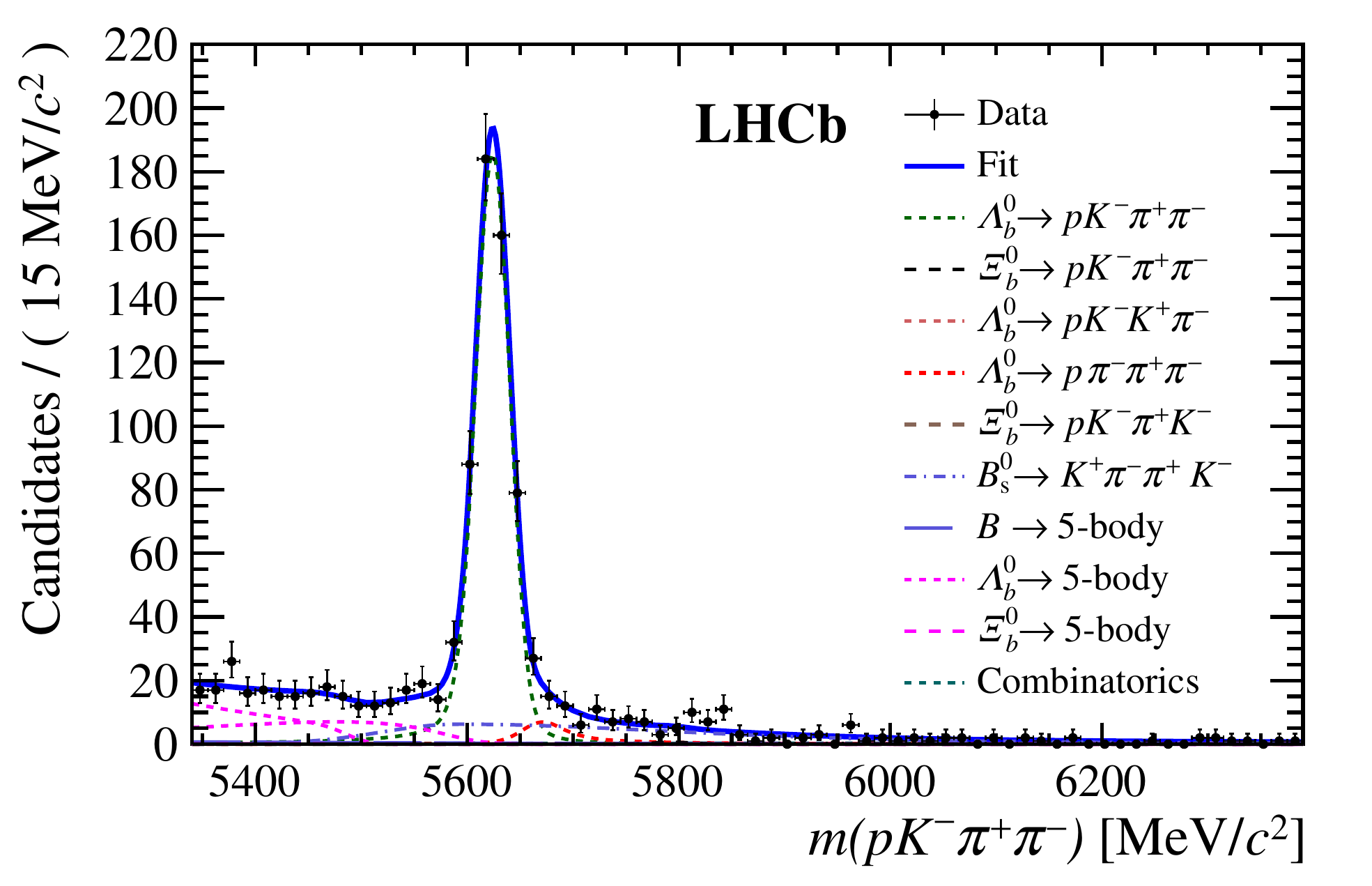}
  \includegraphics[width=.495\columnwidth]{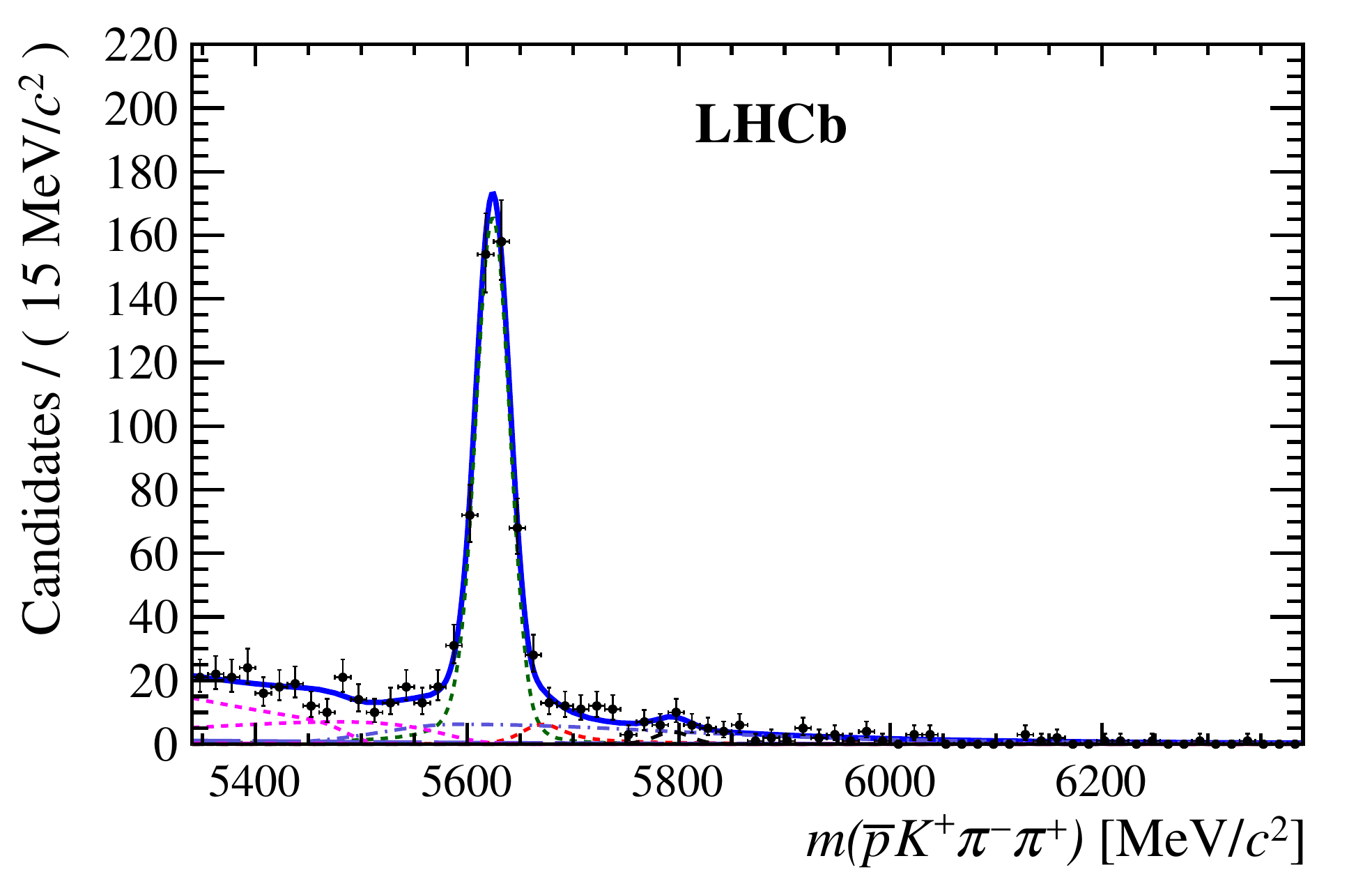}
  \includegraphics[width=.495\columnwidth]{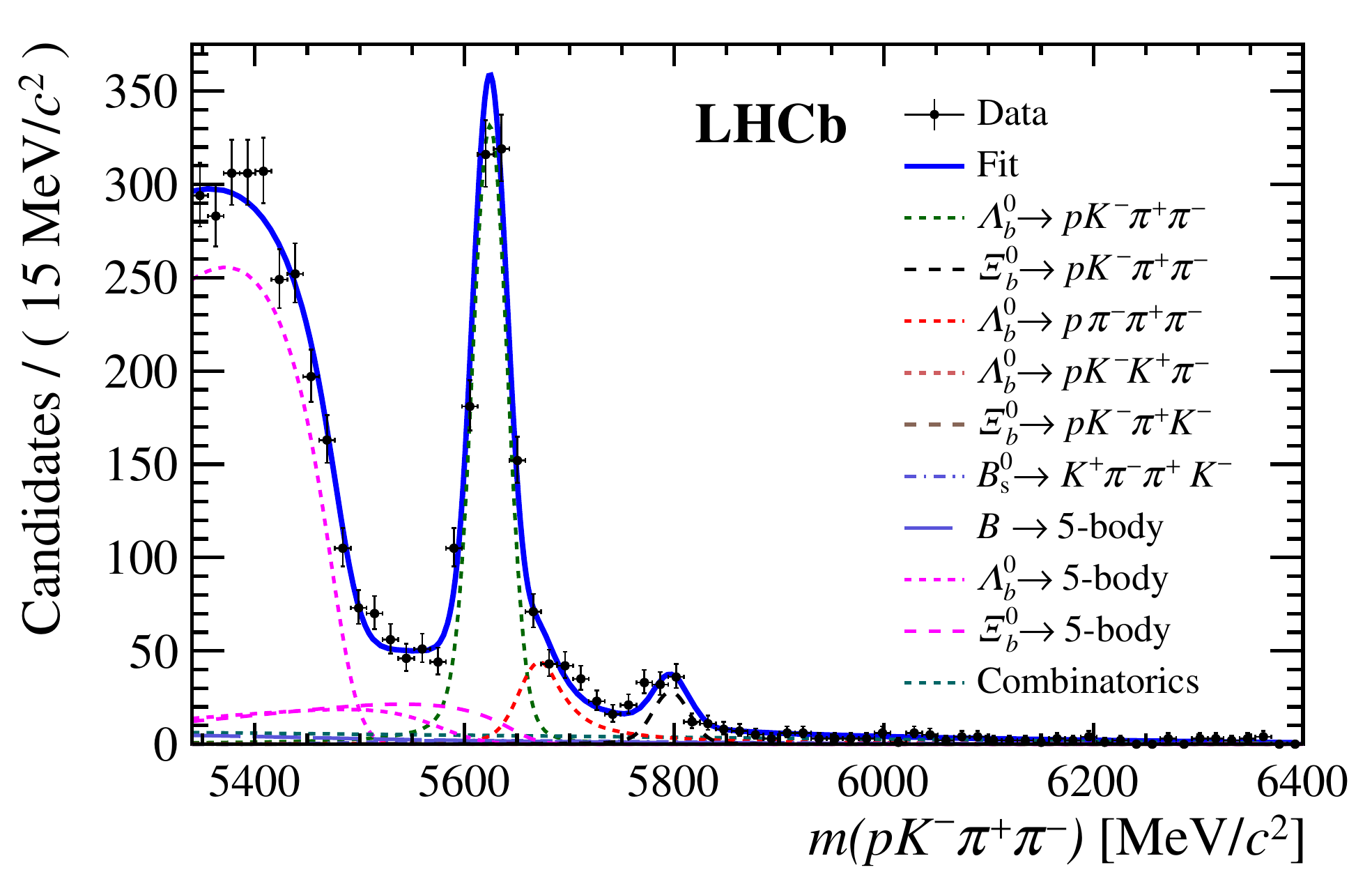}
  \includegraphics[width=.495\columnwidth]{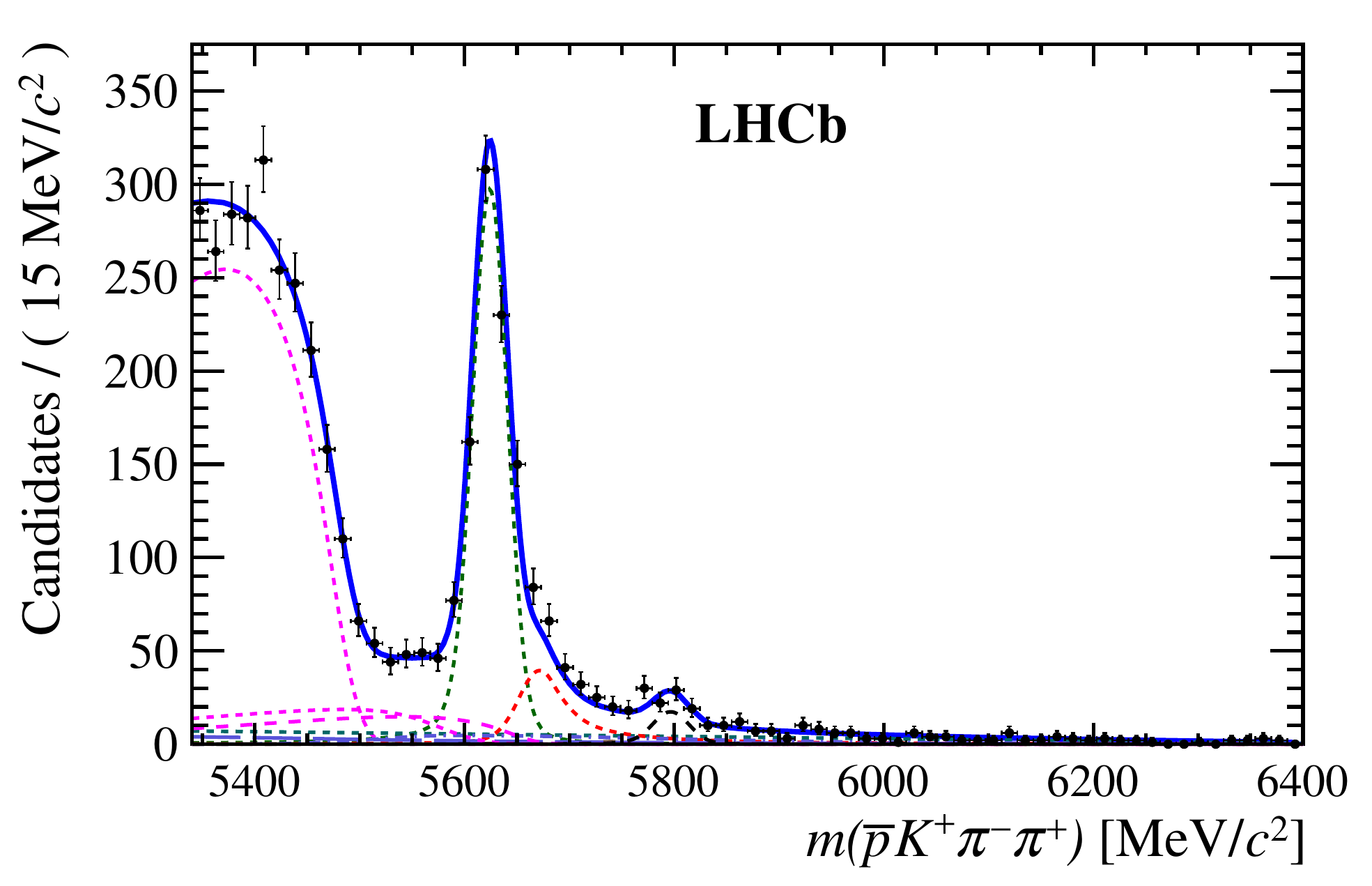}
  \includegraphics[width=.495\columnwidth]{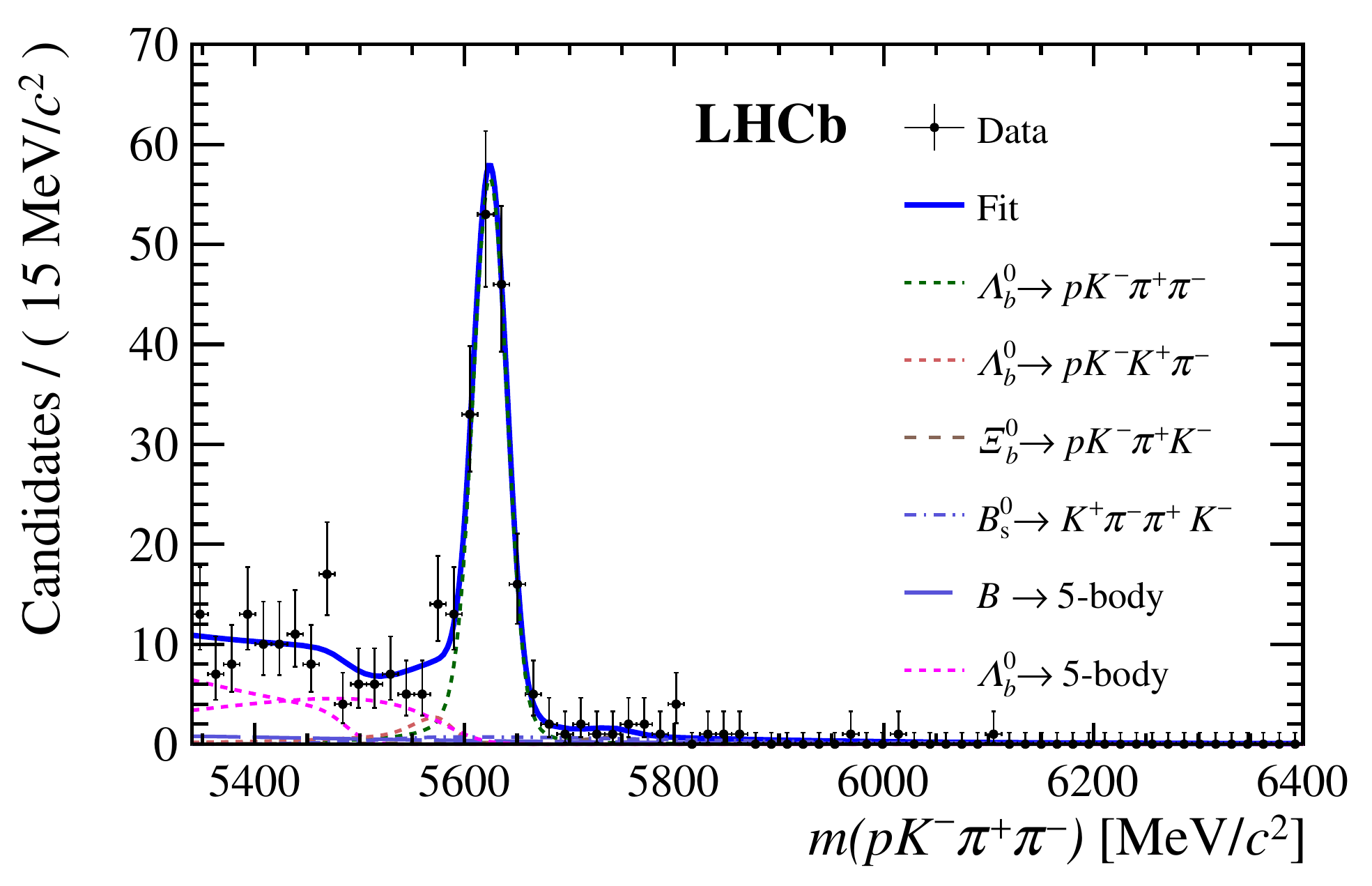}
  \includegraphics[width=.495\columnwidth]{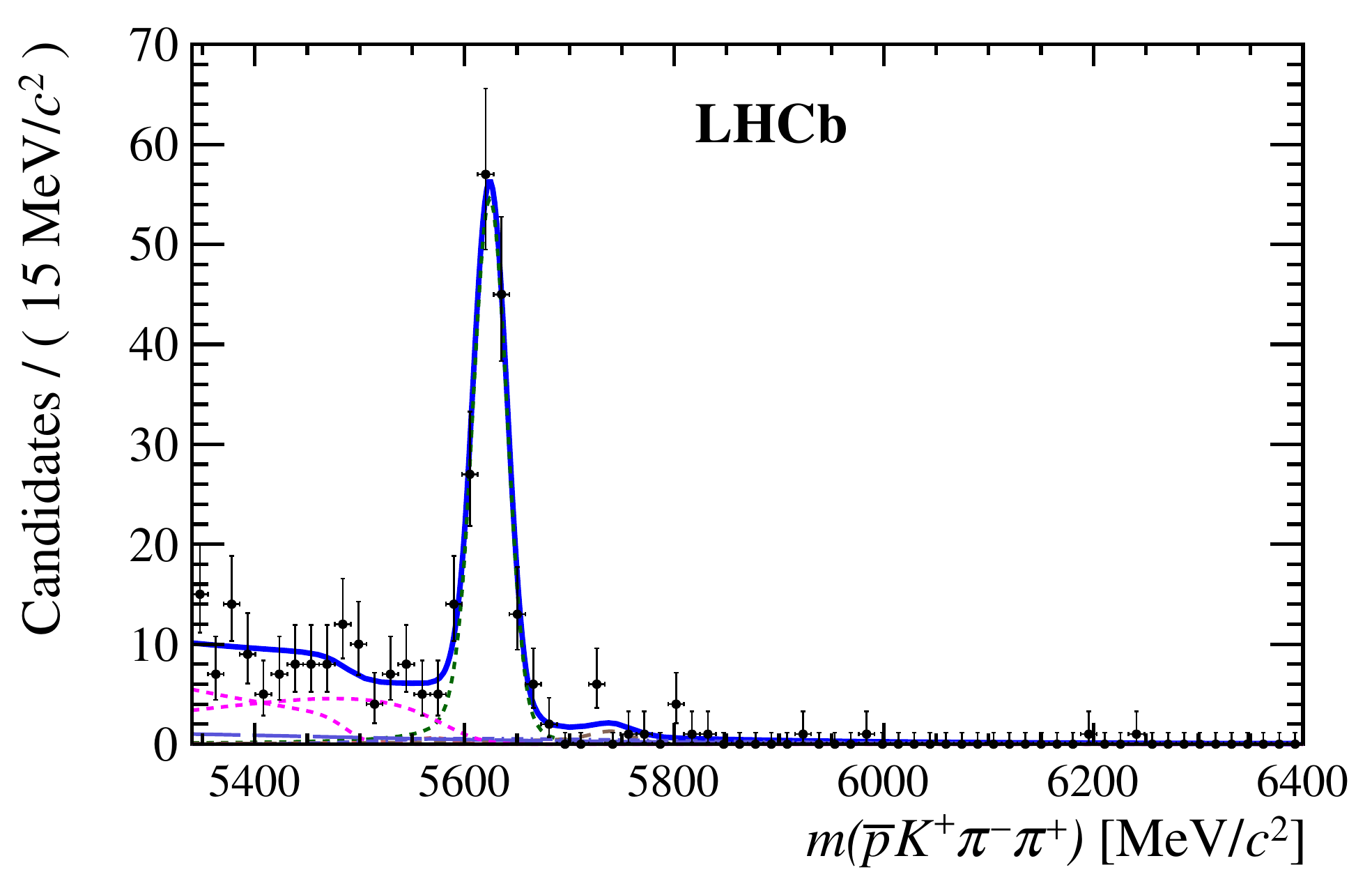}
  \includegraphics[width=.495\columnwidth]{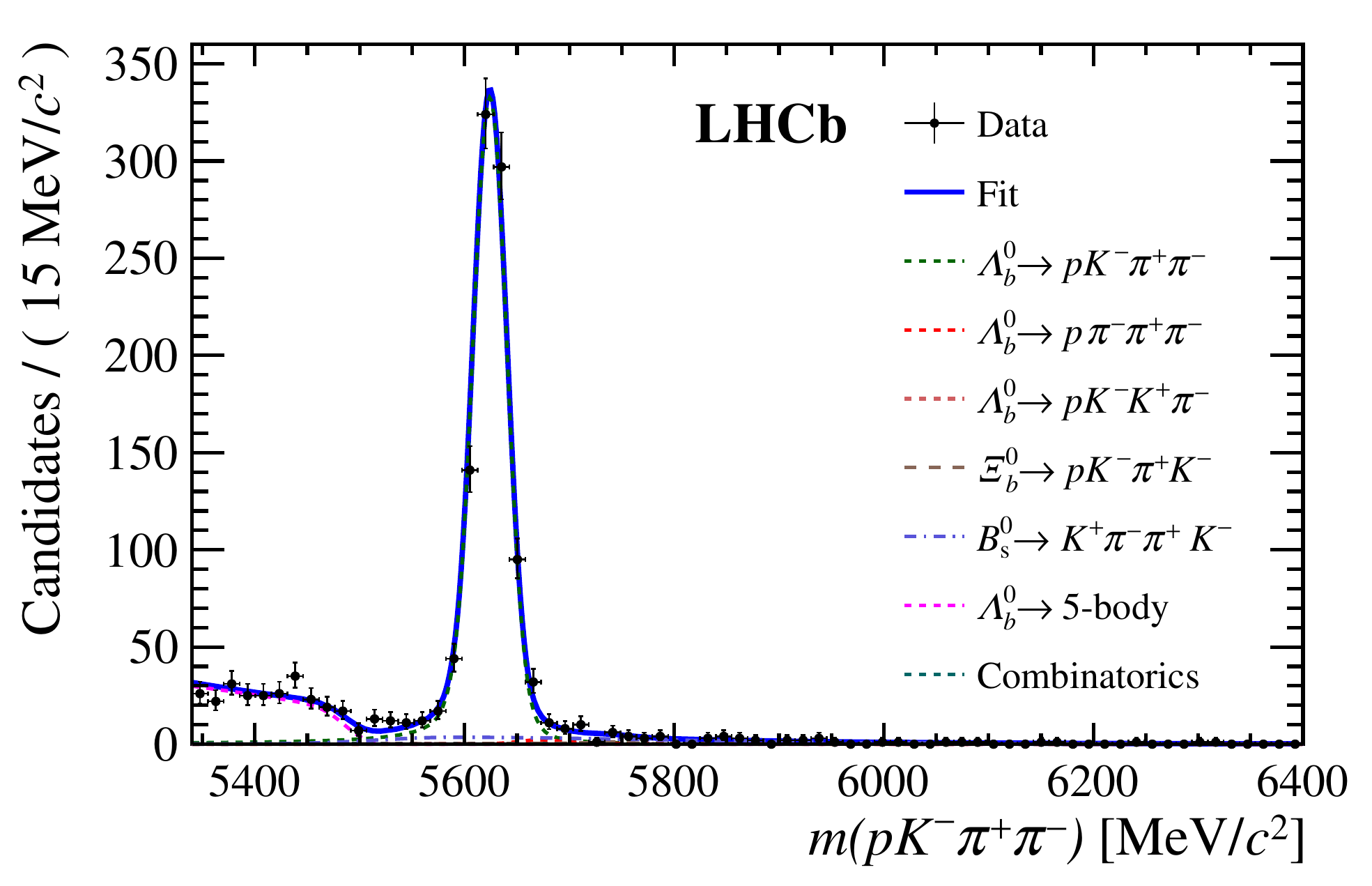}
  \includegraphics[width=.495\columnwidth]{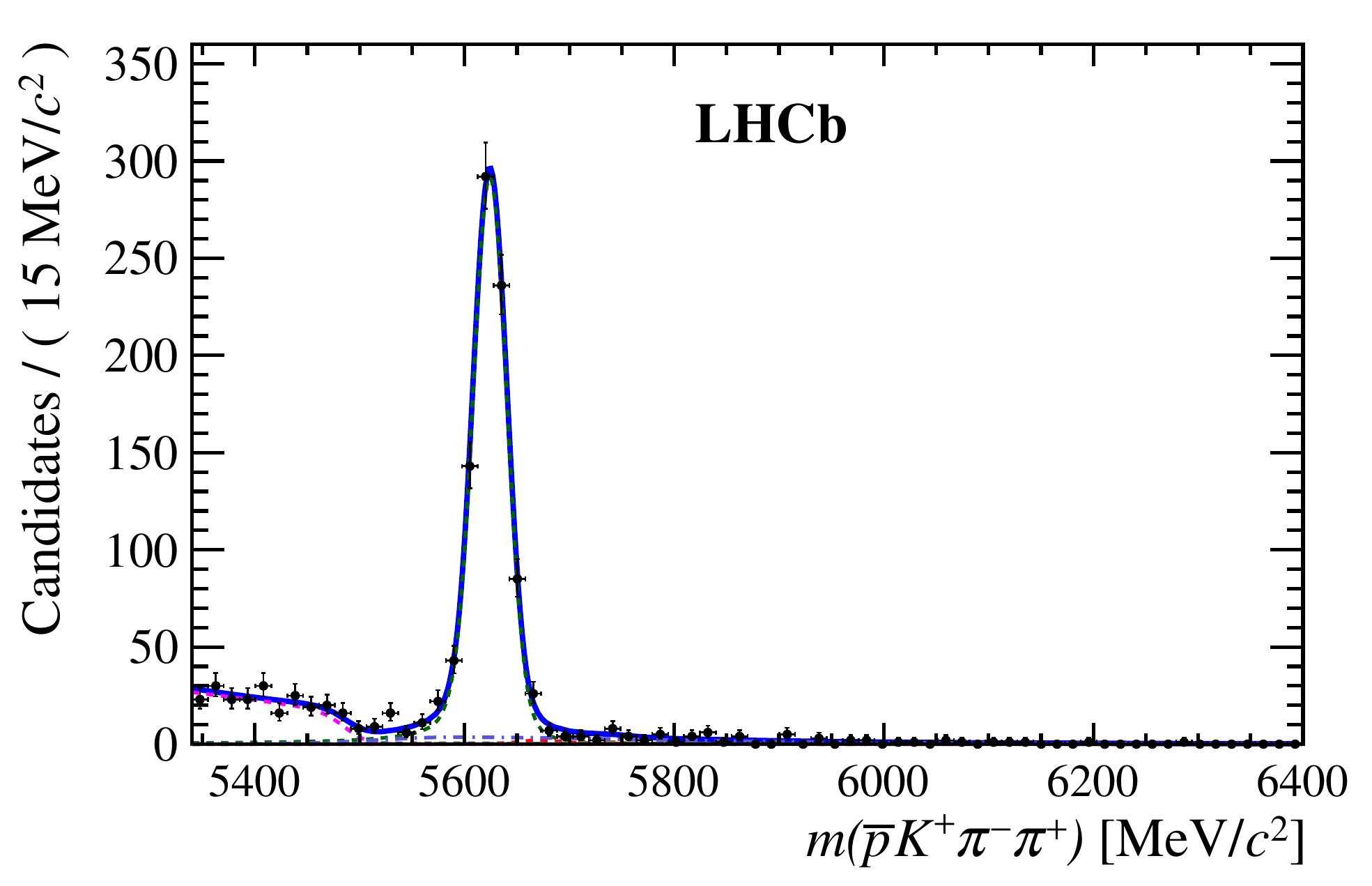}
  \caption{Invariant \pKpipi distributions, with the results of the fit superimposed: region of the phase space containing (first row) \LbTopKone, (second row) \LbToDeltaKpi, (third row) \LbToLstarRhoOrFz and (last row)  \LbToNstarKstar quasi two-body decays. The two columns correspond to the charge-conjugate final states: (left) baryon, (right) antibaryon. The different components employed in the fit are indicated in the legends. 
  }
  \label{fitresults_pKpipi_2}
\end{figure}

\item {\bf \pKKKb final state}: Figure~\ref{fitresults_all_KKK} shows the results of the simultaneous fits to the reconstructed \pKKK  mass spectrum for the inclusive, LBM and quasi two-body measurements. Negligible raw asymmetries are obtained.  

\begin{figure}[ht]
  \centering
  \includegraphics[width=.495\columnwidth]{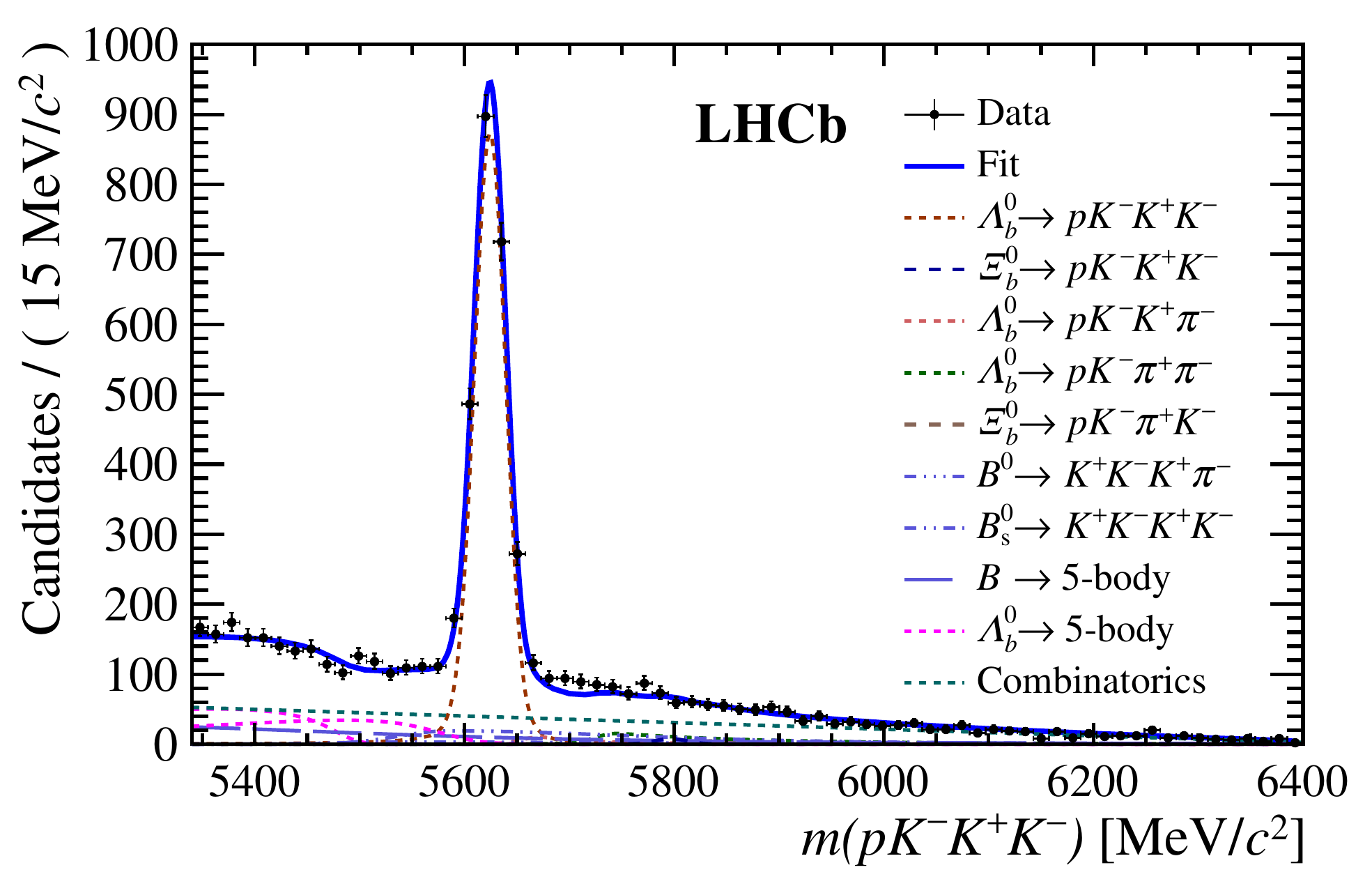}
  \includegraphics[width=.495\columnwidth]{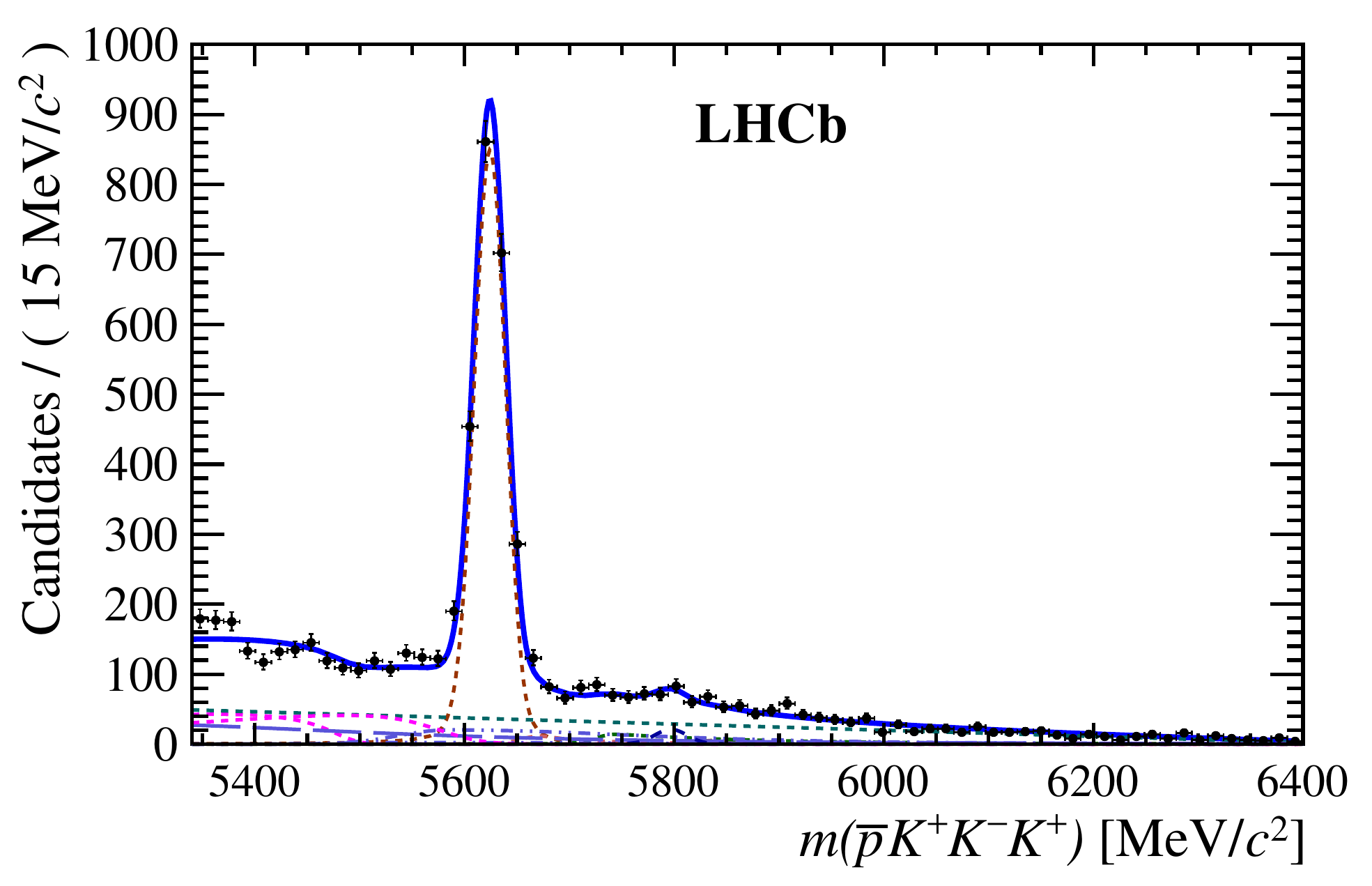}
  \includegraphics[width=.495\columnwidth]{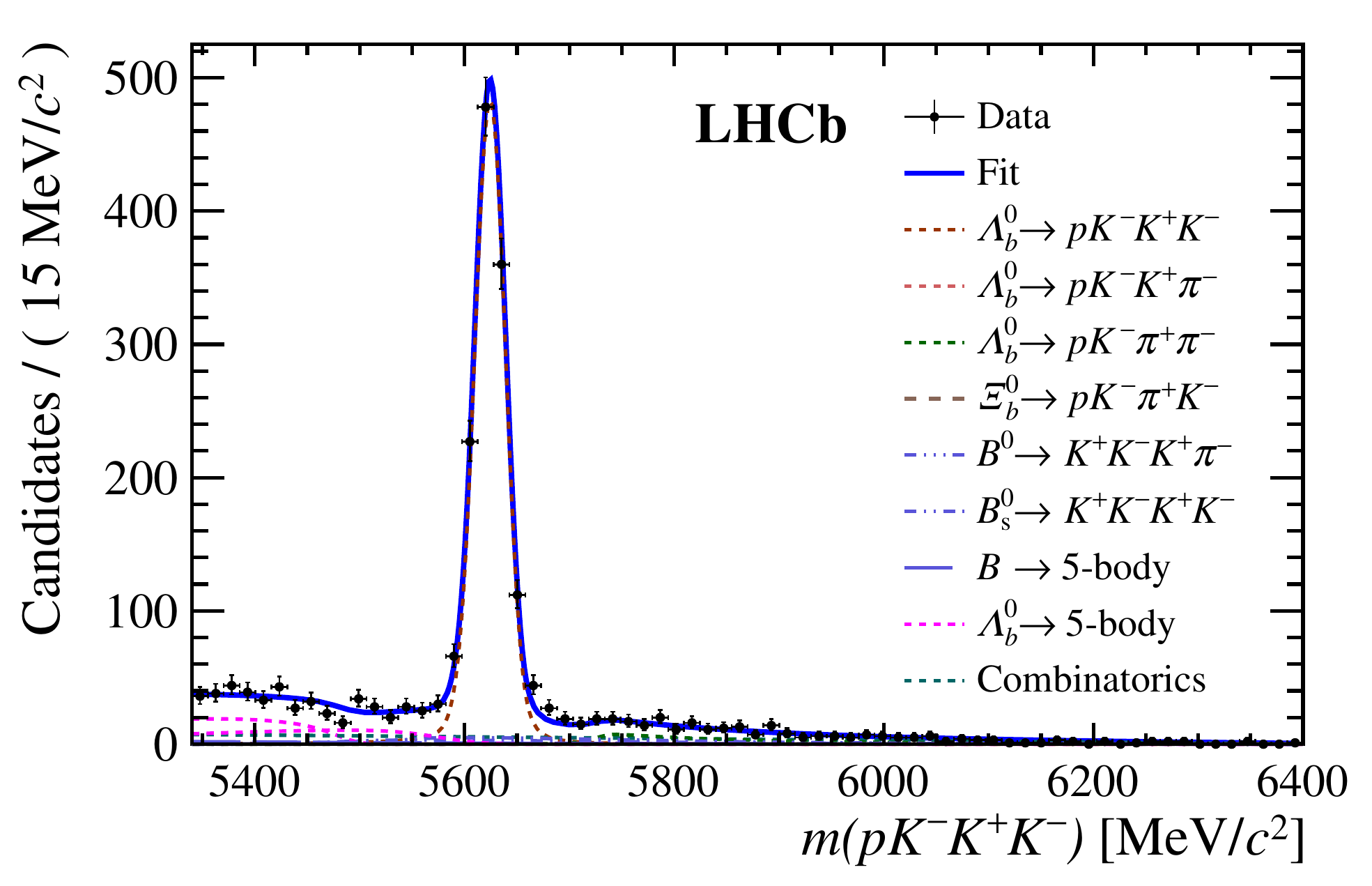}
  \includegraphics[width=.495\columnwidth]{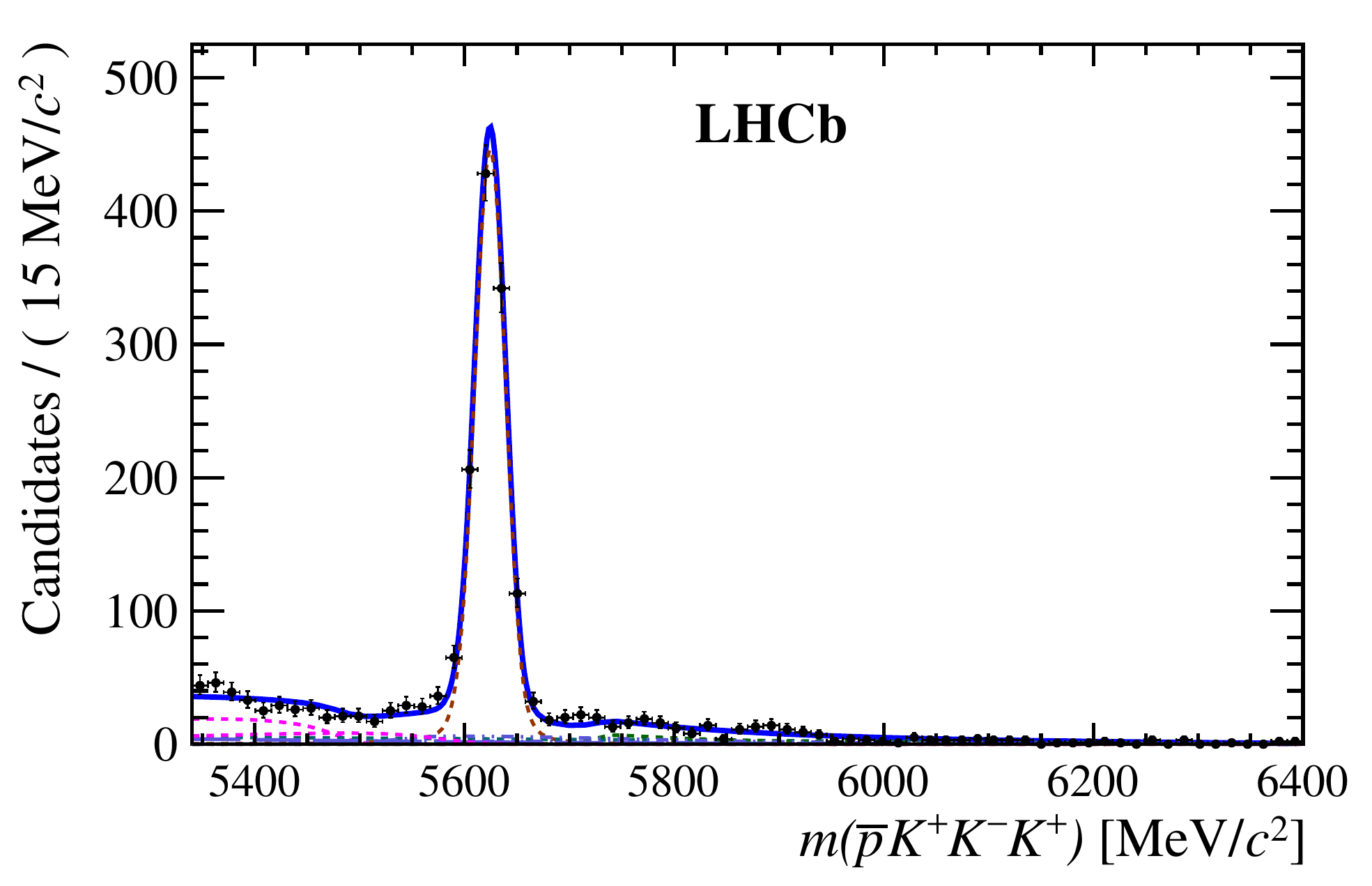}
  \includegraphics[width=.495\columnwidth]{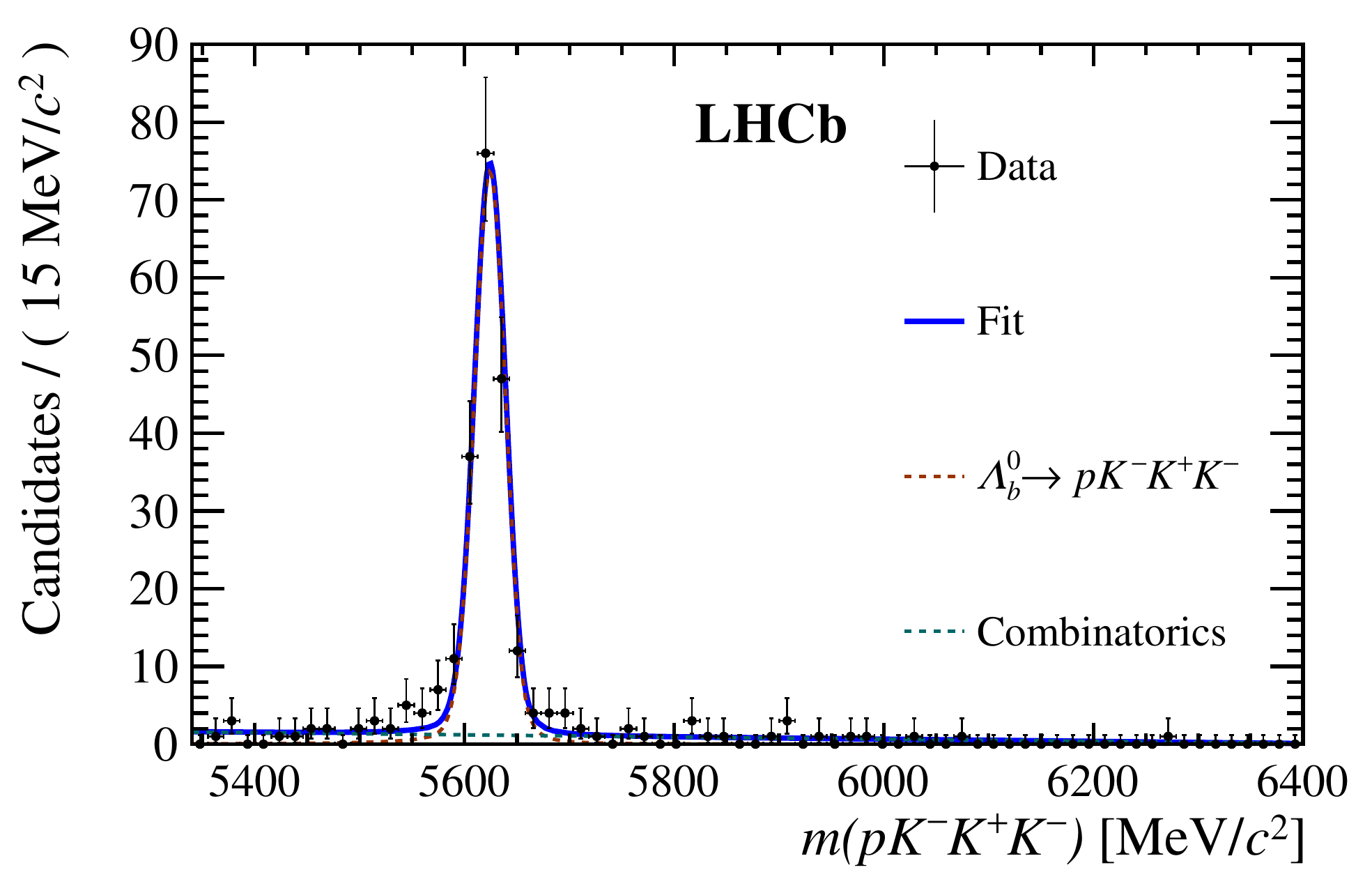}
  \includegraphics[width=.495\columnwidth]{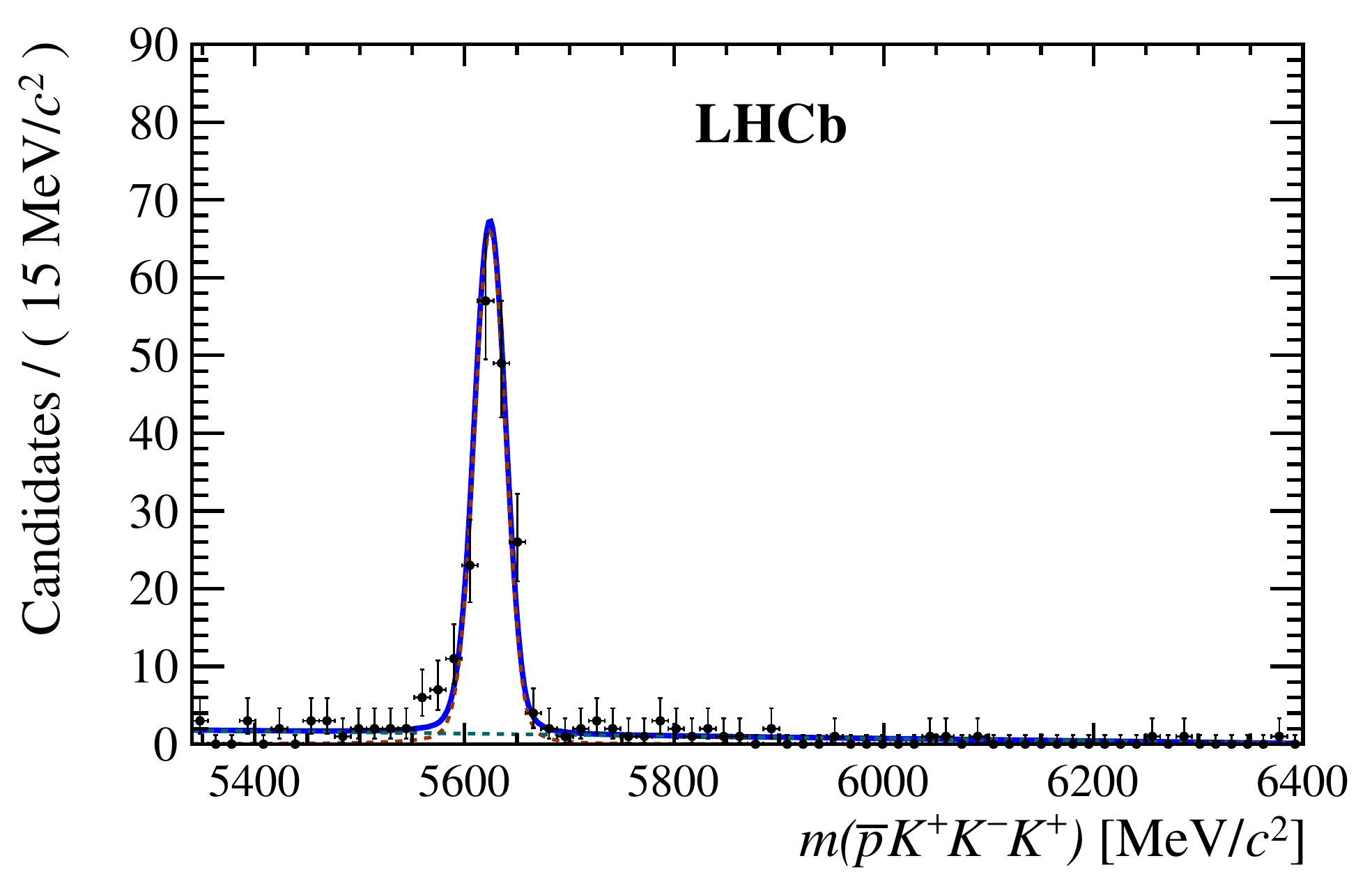}
  \includegraphics[width=.495\columnwidth]{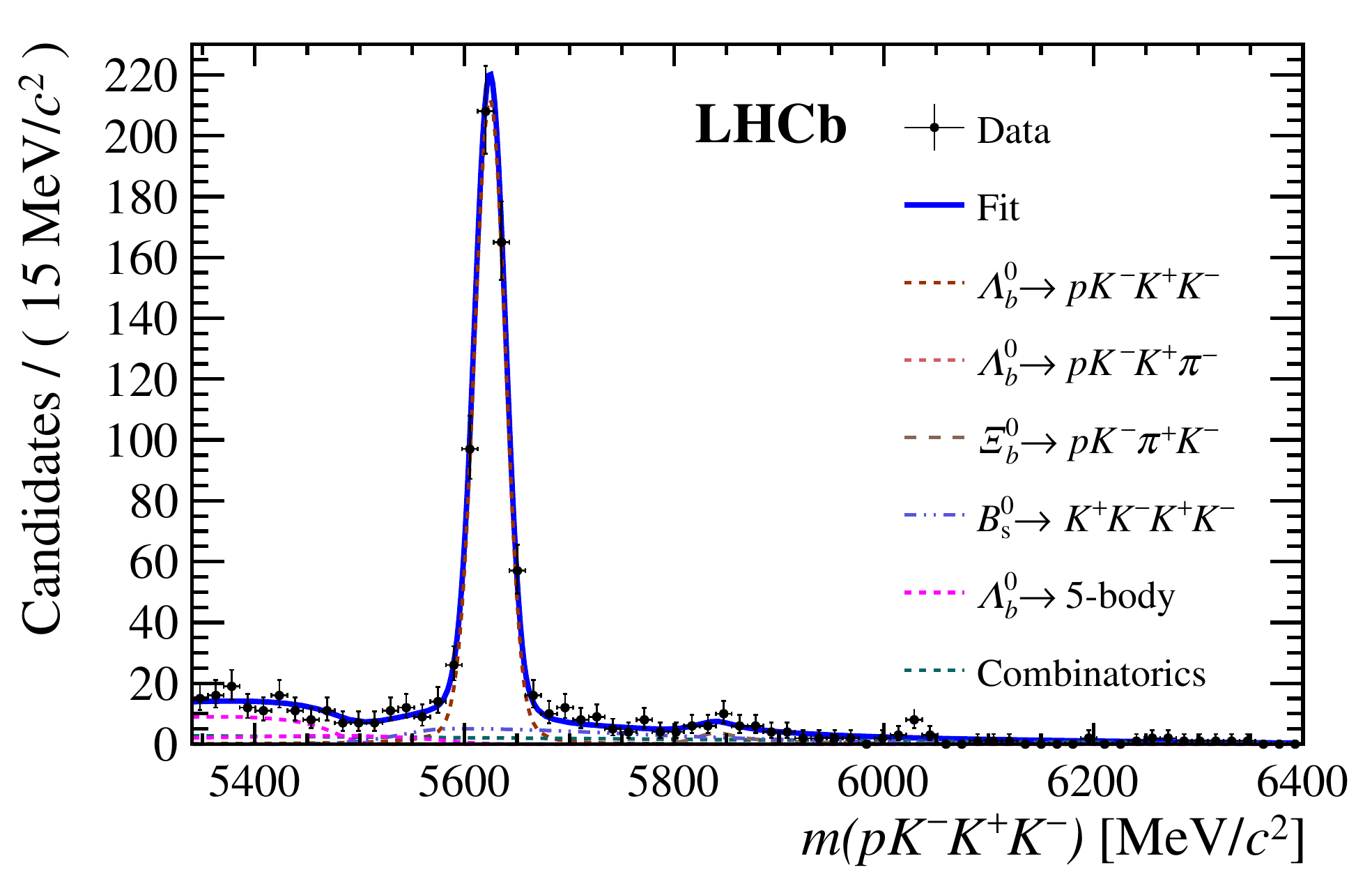}
  \includegraphics[width=.495\columnwidth]{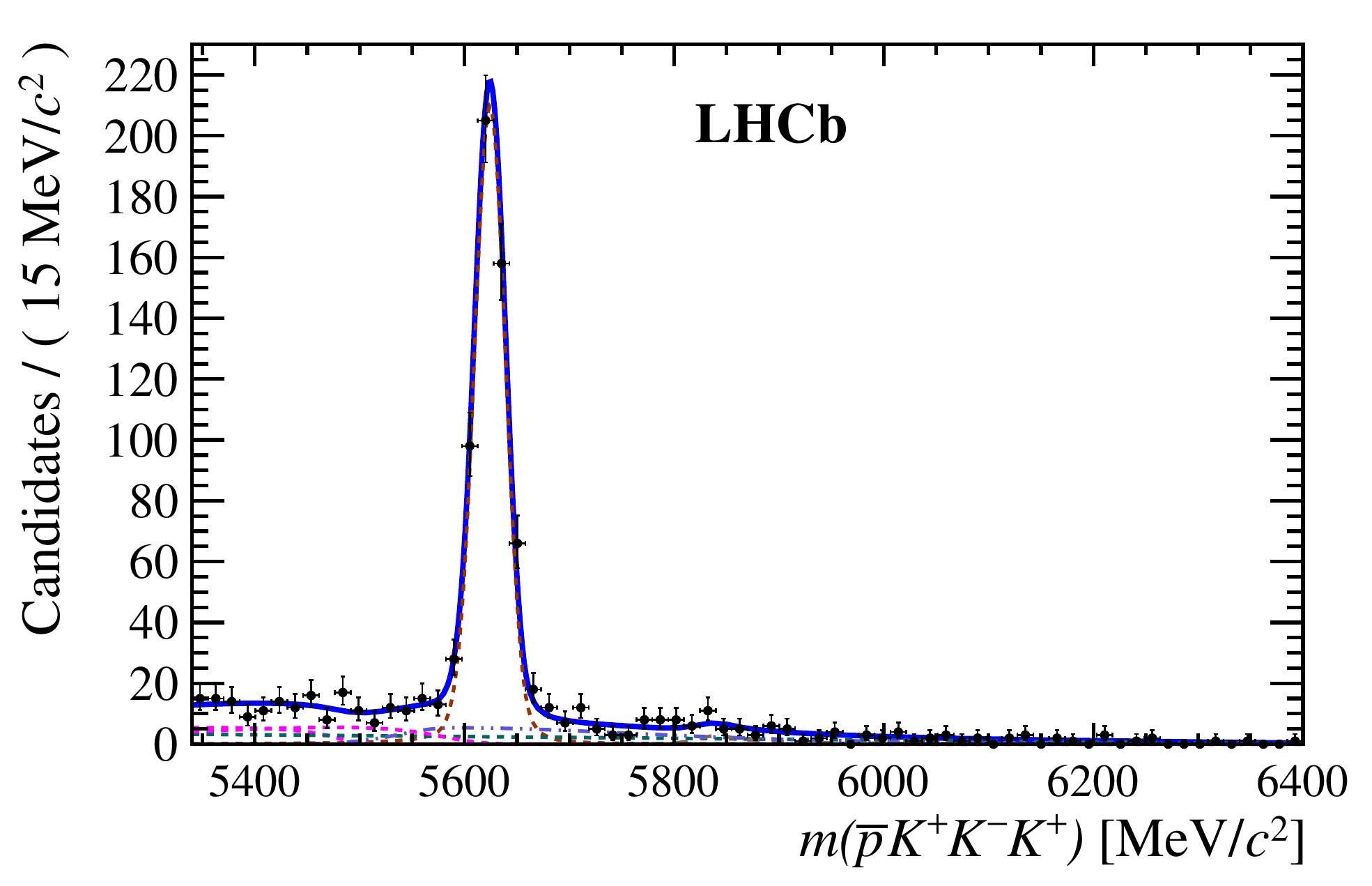}
  \caption{Invariant \pKKK mass distributions, with the results of the fit superimposed: (first row) full phase space and (second row) LBM, (third row) \LbToLstarPhi and (fourth row) \LbTopKPhi. The two columns correspond to the charge-conjugate final states: (left) baryon, (right) antibaryon. The different components employed in the fit are indicated in the legends. The $\Lb\to$ five-body legends includes two decays: partially reconstructed \LbTopKKKg and \LbTopKKKpiz, where the \g and $\pi^0$ are not reconstructed.}
  \label{fitresults_all_KKK}
\end{figure}

\item {\bf \pKKpib and \pKpiKb final states}: The simultaneous fit results for the two remaining final states are shown in Fig.~\ref{fitresults_all_KKpi}. The result of the fit for the control channel \XibzToXicpiXicTopKpi is also displayed and shows a good description of the spectrum. This control channel is used to account for the production asymmetry of the \Xibz modes. 

\begin{figure}[hptb]
  \centering
  \includegraphics[width=.495\columnwidth]{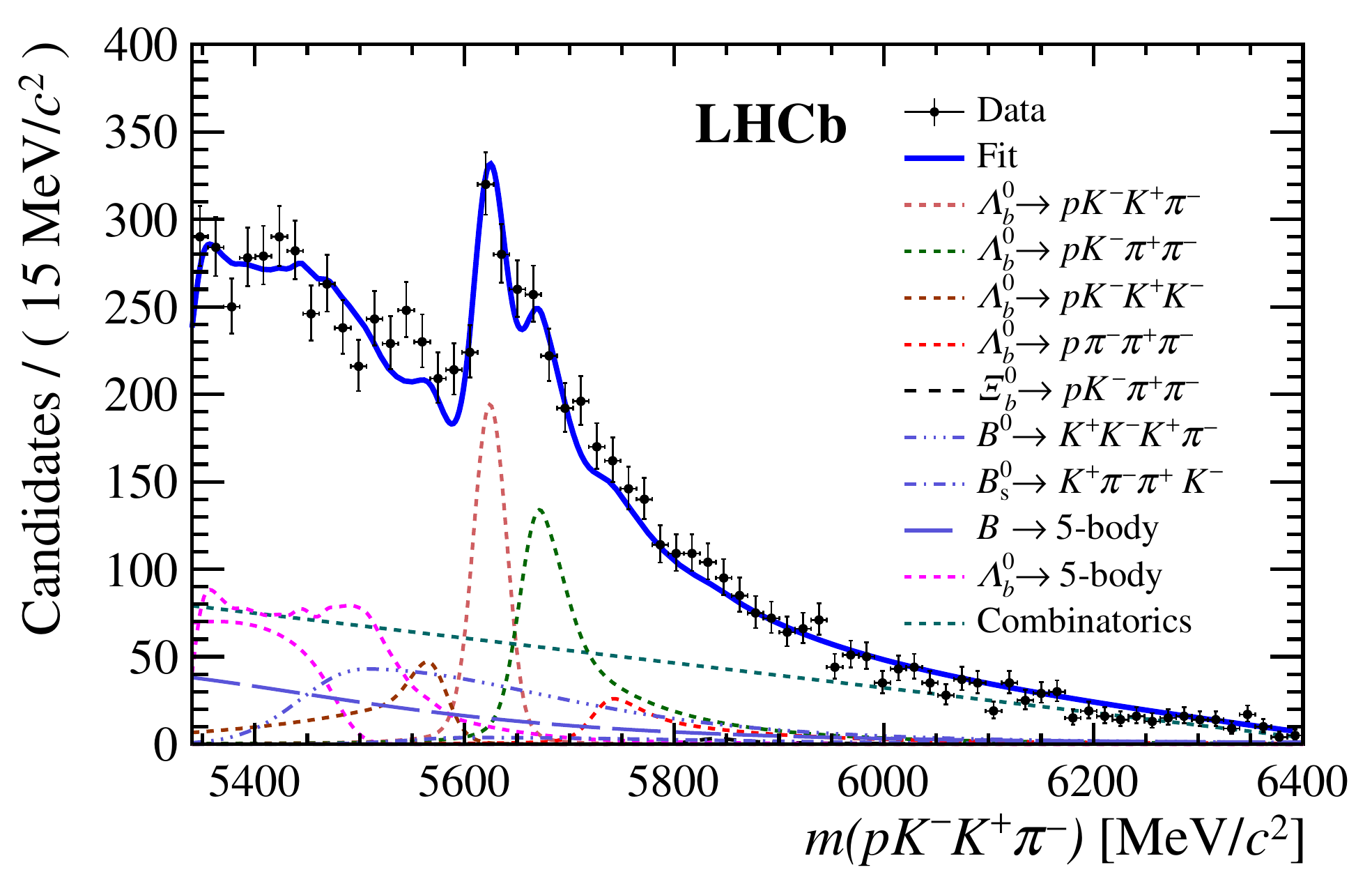}
  \includegraphics[width=.495\columnwidth]{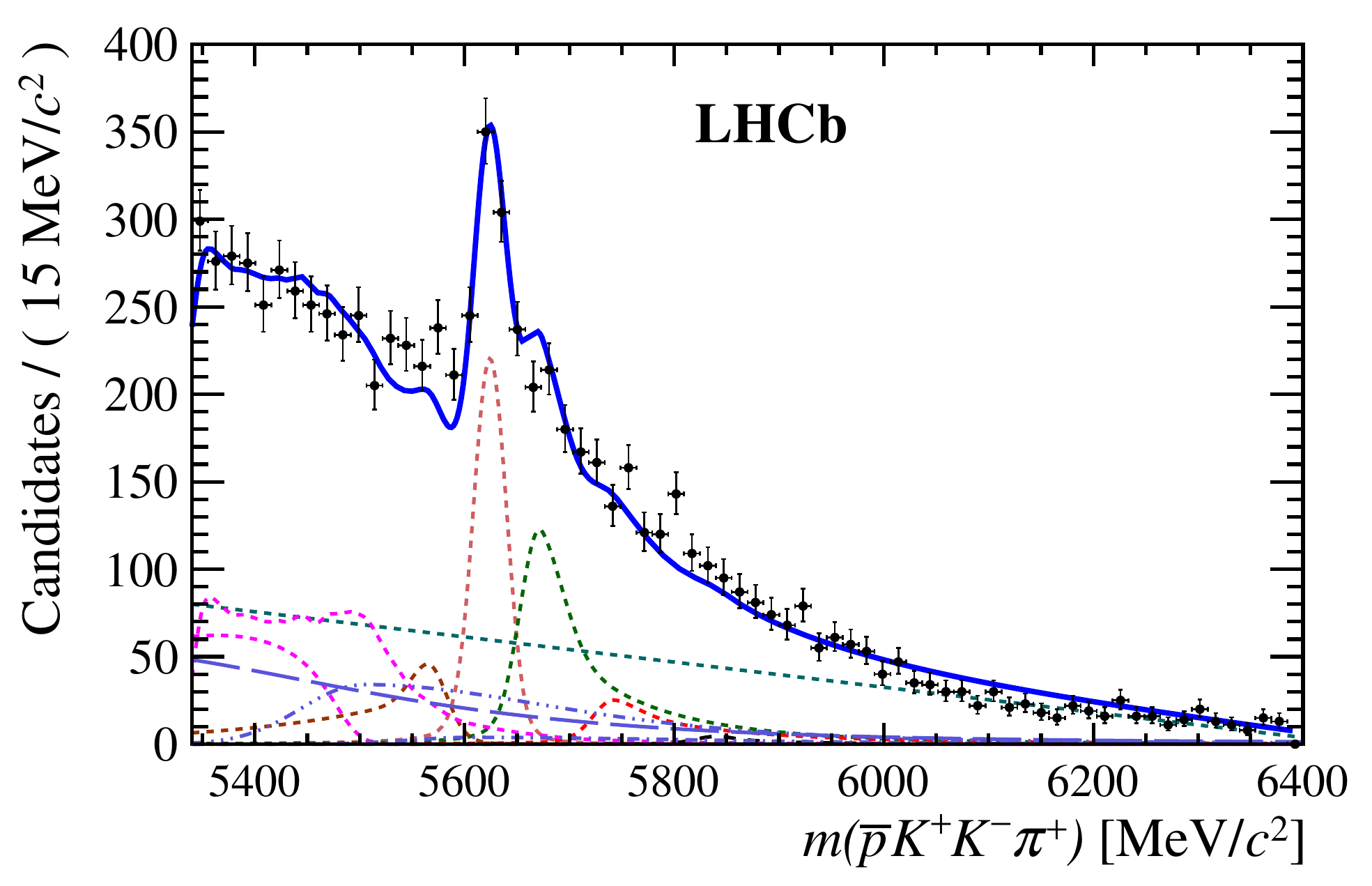}
  \includegraphics[width=.495\columnwidth]{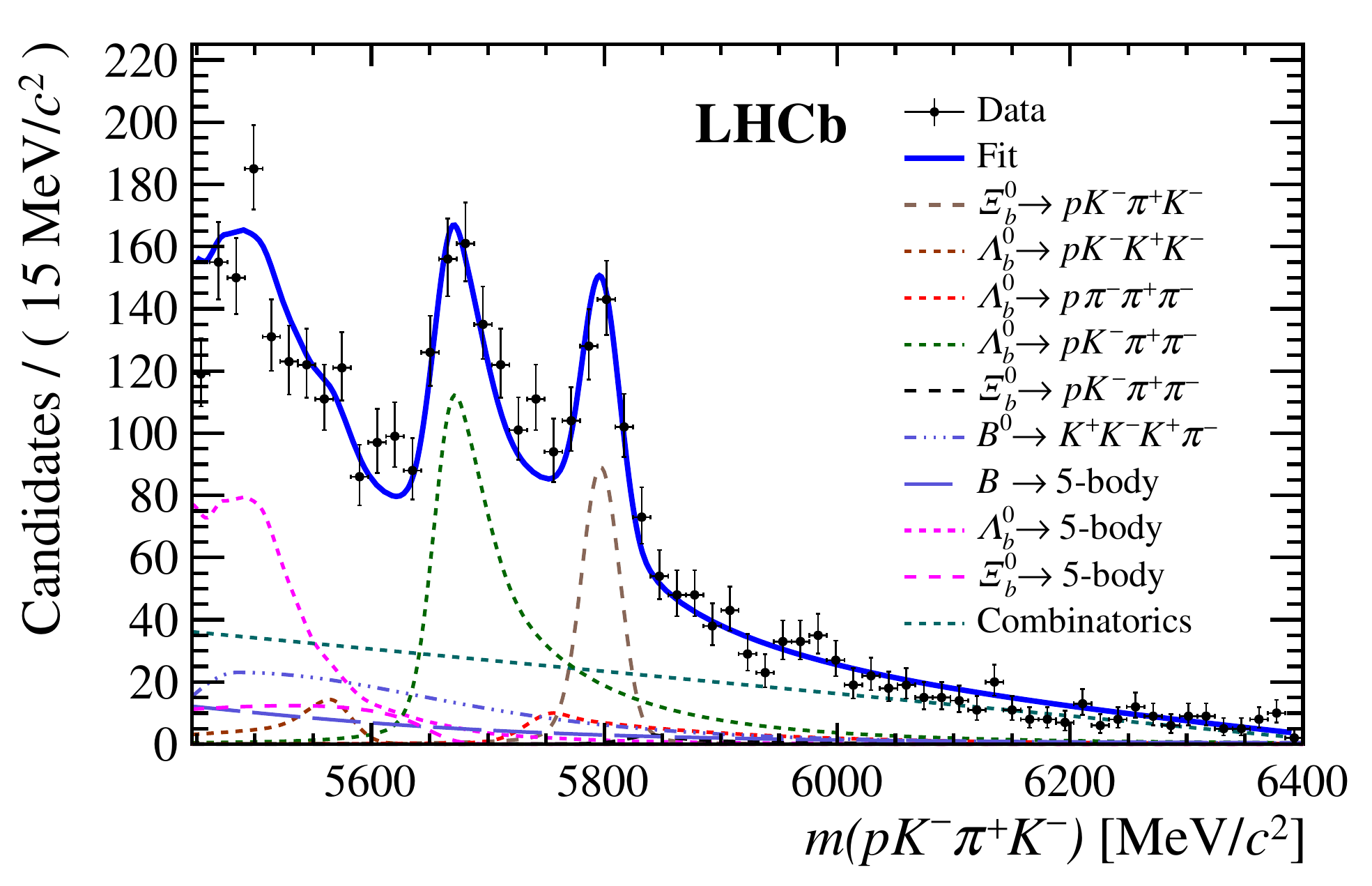}
  \includegraphics[width=.495\columnwidth]{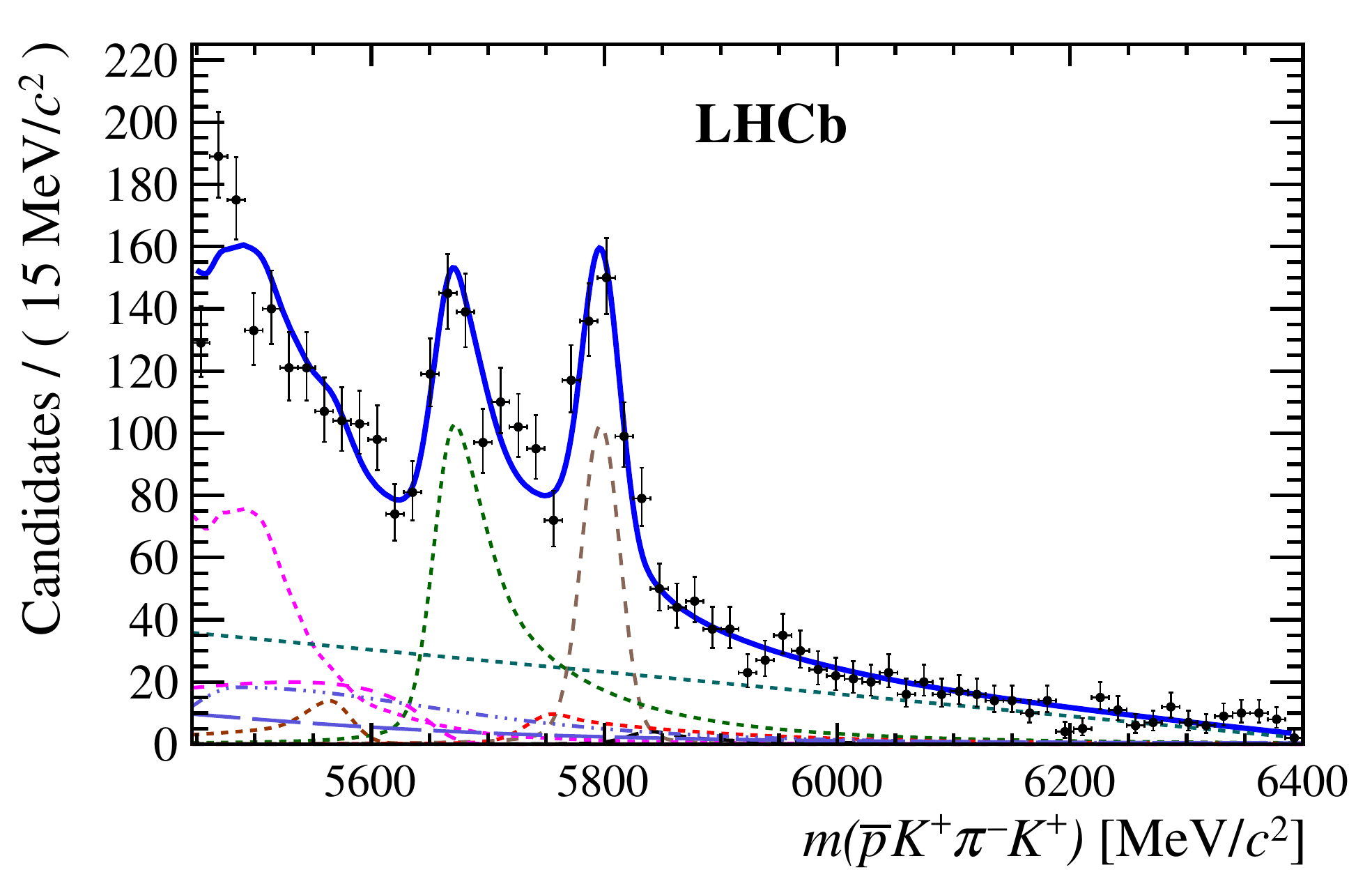}
  \includegraphics[width=.495\columnwidth]{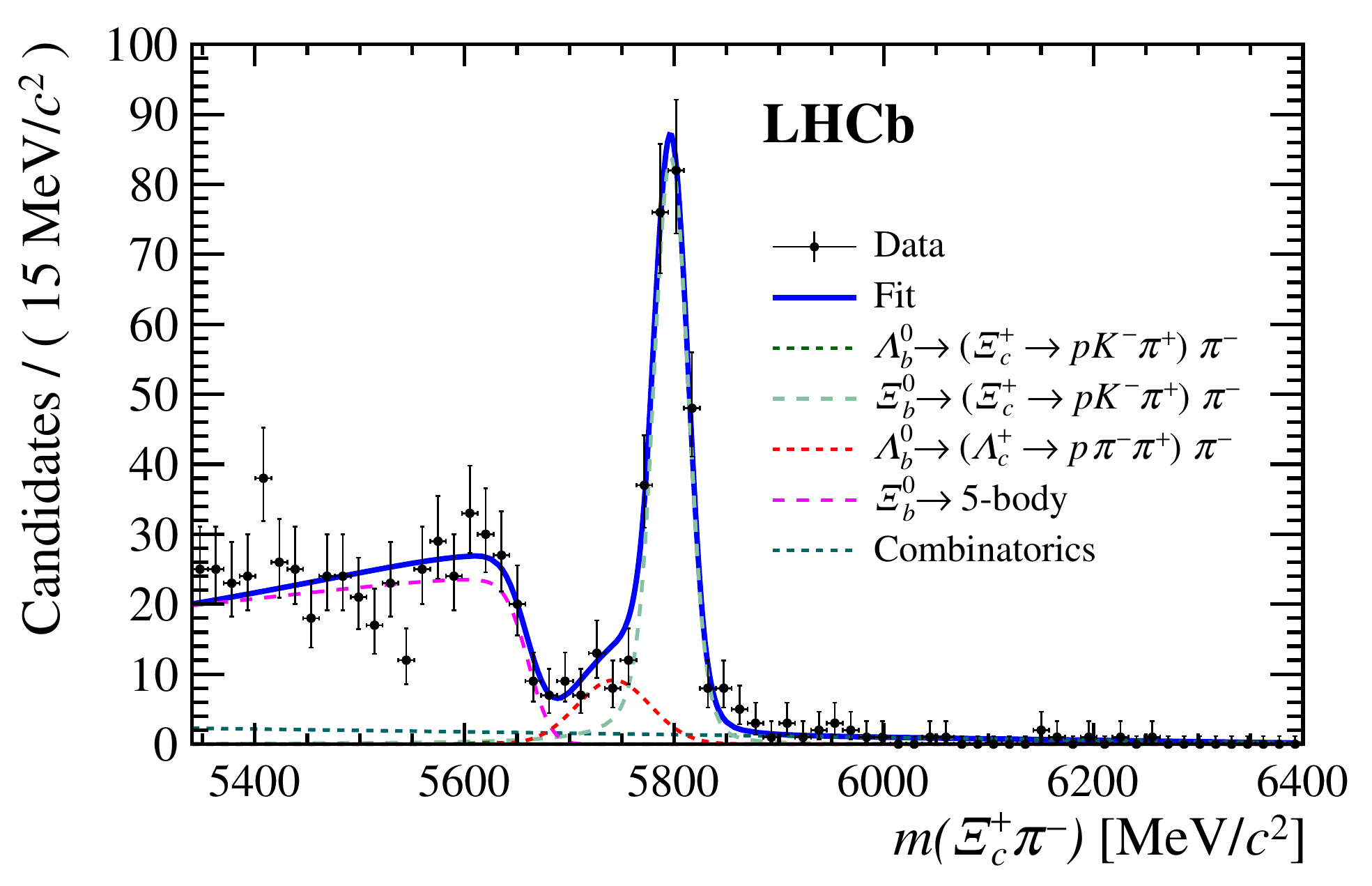}
  \includegraphics[width=.495\columnwidth]{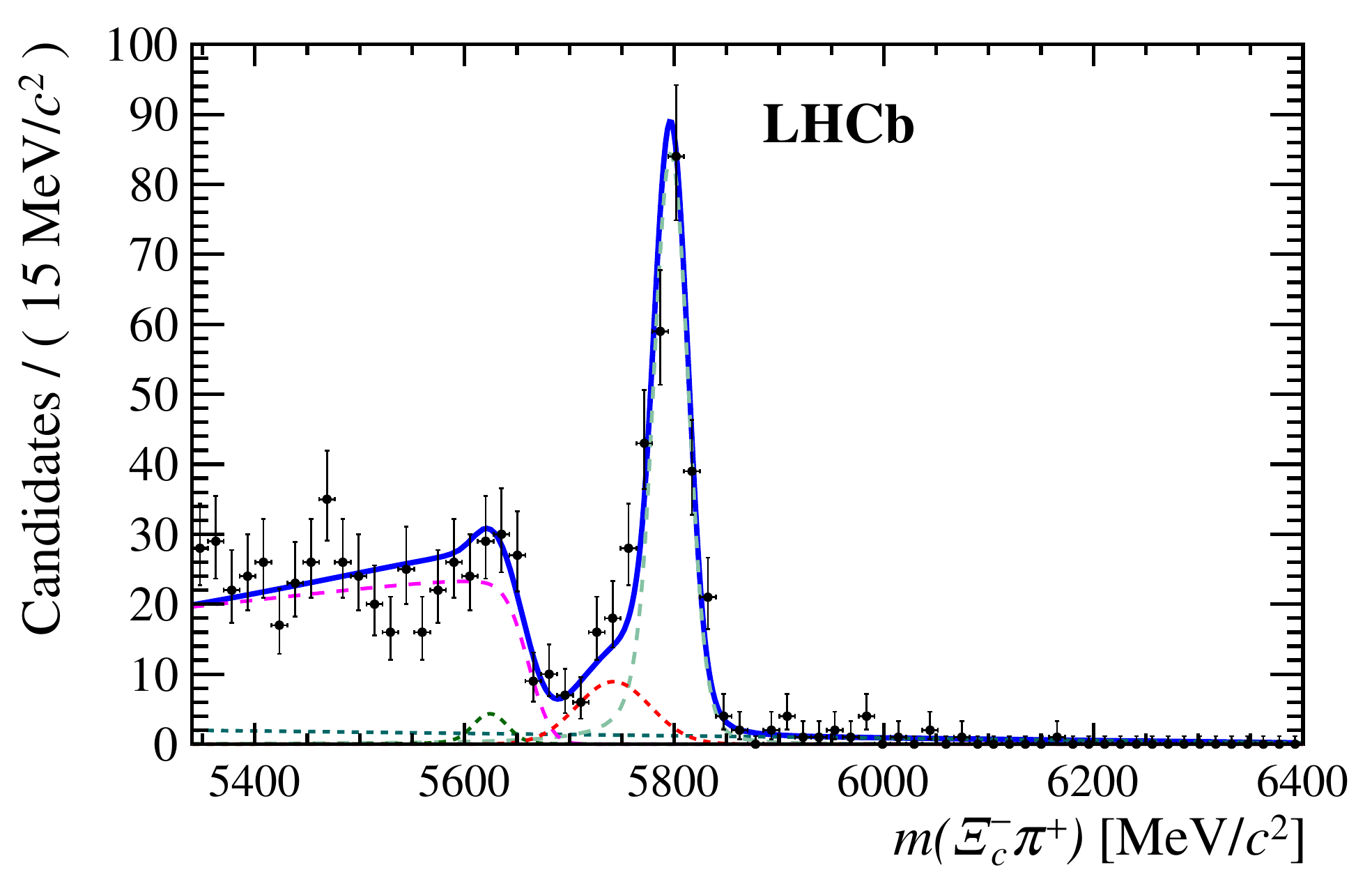}
  \caption{Invariant (first row) \pKKpi and (second row) \pKpiK mass distributions, with the results of the fit superimposed. The two bottom plots are the results of the fit to the \XibzToXicpiXicTopKpi control channel. The two columns correspond to the charge-conjugate final states: (left) baryon, (right) antibaryon. The different components employed in the fit are indicated in the legends. The $\Lb\to$ five-body legend includes two components where a $\pi^0$ is not reconstructed: the partially reconstructed background \LbTopKpipipiz where a pion is misidentified as a kaon and the partially reconstructed background \LbTopKKpipiz.}
  \label{fitresults_all_KKpi}
\end{figure}

\end{itemize}

\FloatBarrier

\section{\boldmath Measurements of \CP asymmetries and concluding remarks}
\label{sec:conclusions}

Five charmless final states of \Lb and \Xibz four-body hadronic decays are examined in this paper. Specific regions of their phase space have been selected to search for local \CP asymmetries in addition to the integrated \CP-asymmetry. A total of eighteen measurements of \CP asymmetries are reported in this paper.

\par

A simple counting experiment allows the measurement of a \CP asymmetry up to the corrections due to instrumental and \bquark-baryon production asymmetries. These corrections are mitigated by establishing the differences (denoted \dACP) between the raw \ACP values of the signals and those of the decay modes with intermediate charmed baryons comprising the same final-state particles. The asymmetries \dACP are further corrected for residual experimental charge asymmetries due to kinematic differences between signal and control modes. The integrated \dACP asymmetry differences are measured to be
\begin{align*}
 \dACP  (\LbToppipipi)    &= (+1.1 \pm 2.5  \pm 0.6)\,\%, \\
 \dACP  (\LbTopKpipi)     &= (+3.2 \pm 1.1  \pm 0.6)\,\%, \\
 \dACP  (\LbTopKKpi)      &= (-6.9 \pm 4.9  \pm 0.8 )\,\%, \\
 \dACP  (\LbTopKKK)       &= (+0.2 \pm 1.8  \pm 0.6 )\,\%, \\
 \dACP  (\XibzTopKpipi)   &= (-17\phantom{.} \pm \phantom{.}11  \pm \phantom{.6}1 )\,\%, \\
 \dACP  (\XibzTopKpiK)    &= (-6.8 \pm 8.0  \pm 0.8)\,\%.
\intertext{The measurements for the two-body low invariant-mass regions are}
 \dACP  (\LbToppipipi)    &= (+3.7 \pm 4.1  \pm 0.5)\,\%, \\
 \dACP  (\LbTopKpipi)     &= (+3.5 \pm 1.5  \pm 0.5)\,\%, \\
 \dACP  (\LbTopKKK)       &= (+2.7 \pm 2.3  \pm 0.6)\,\%. 
\intertext{Finally, the measurements for the quasi two-body decays are}
 \dACP  (\LbTopaone)             &= (-1.5 \pm 4.2  \pm 0.6)\,\% \, , \\
 \dACP  (\LbToNstarRhoOrFz)      &= (+2.0 \pm 4.9  \pm 0.4)\,\%, \\
 \dACP  (\LbToDeltapipi)         &= (+0.1 \pm 3.2  \pm 0.6)\,\%, \\
 \dACP  (\LbTopKone)             &= (+4.7 \pm 3.5 \pm 0.8)\,\%,  \\
 \dACP  (\LbToLstarRhoOrFz)      &= (+0.6 \pm 6.0 \pm 0.5)\,\%, \\
 \dACP  (\LbToNstarKstar)        &= (+5.5 \pm 2.5  \pm 0.5)\,\%,  \\
 \dACP  (\LbToDeltaKpi)          &= (+4.4 \pm 2.6  \pm 0.6)\,\%, \\
 \dACP  (\LbToLstarPhi)          &= (+4.3 \pm 5.6 \pm 0.4)\,\%,   \\
 \dACP  (\LbTopKPhi)             &= (-0.7 \pm 3.3 \pm 0.7)\,\%.
\end{align*}
In all cases the first uncertainties are statistical and the second systematic. No significant \CP violation is observed. The \dACP measurements for the independent samples of the two magnet polarities, the two categories of trigger requirements and the two distinct data-taking samples are found to be consistent. In addition, the measured asymmetries for the combinatorial background in all spectra are consistent with zero. The background contributions coming from \B-meson decays (that could potentially exhibit nonzero \CP violation) are also consistent with null asymmetries. 
\par

In a previous analysis, the \lhcb collaboration reported evidence for \CP violation in a specific region of the phase space of the decay \LbToppipipi, by measuring triple-product asymmetries~\cite{LHCb-PAPER-2016-030}. By contrast, in the present analysis, no indication of a significant \CP-violating asymmetry is obtained with the same data sample, providing complementary insights about the origin of this potential \CP-symmetry breaking effect.  
The quest for the first observation of \CP violation in baryon decays continues. \lhcb Run 2 data provides about five times larger yields  allowing for a more sensitive search of smaller \CP-violating effects.


\section*{Acknowledgements}
%
%
\noindent We express our gratitude to our colleagues in the CERN
accelerator departments for the excellent performance of the LHC. We
thank the technical and administrative staff at the LHCb
institutes.
We acknowledge support from CERN and from the national agencies:
CAPES, CNPq, FAPERJ and FINEP (Brazil); 
MOST and NSFC (China); 
CNRS/IN2P3 (France); 
BMBF, DFG and MPG (Germany); 
INFN (Italy); 
NWO (Netherlands); 
MNiSW and NCN (Poland); 
MEN/IFA (Romania); 
MSHE (Russia); 
MinECo (Spain); 
SNSF and SER (Switzerland); 
NASU (Ukraine); 
STFC (United Kingdom); 
NSF (USA).
We acknowledge the computing resources that are provided by CERN, IN2P3
(France), KIT and DESY (Germany), INFN (Italy), SURF (Netherlands),
PIC (Spain), GridPP (United Kingdom), RRCKI and Yandex
LLC (Russia), CSCS (Switzerland), IFIN-HH (Romania), CBPF (Brazil),
PL-GRID (Poland) and OSC (USA).
We are indebted to the communities behind the multiple open-source
software packages on which we depend.
Individual groups or members have received support from
AvH Foundation (Germany);
EPLANET, Marie Sk\l{}odowska-Curie Actions and ERC (European Union);
ANR, Labex P2IO and OCEVU, and R\'{e}gion Auvergne-Rh\^{o}ne-Alpes (France);
Key Research Program of Frontier Sciences of CAS, CAS PIFI, and the Thousand Talents Program (China);
RFBR, RSF and Yandex LLC (Russia);
GVA, XuntaGal and GENCAT (Spain);
the Royal Society
and the Leverhulme Trust (United Kingdom);
Laboratory Directed Research and Development program of LANL (USA).



\addcontentsline{toc}{section}{References}
\setboolean{inbibliography}{true}
\bibliographystyle{LHCb}
\bibliography{main,LHCb-PAPER,LHCb-CONF,LHCb-DP,LHCb-TDR}
 
\newpage
\newpage
\centerline{\large\bf LHCb collaboration}
\begin{flushleft}
\small
R.~Aaij$^{40}$,
B.~Adeva$^{39}$,
M.~Adinolfi$^{48}$,
Z.~Ajaltouni$^{5}$,
S.~Akar$^{59}$,
J.~Albrecht$^{10}$,
F.~Alessio$^{40}$,
M.~Alexander$^{53}$,
A.~Alfonso~Albero$^{38}$,
S.~Ali$^{43}$
% ...

{\footnotesize \it
% ...
$ ^{5}$Clermont Universit{\'e}, Universit{\'e} Blaise Pascal, CNRS/IN2P3, LPC, Clermont-Ferrand, France\\
% ...
$ ^{10}$Fakult{\"a}t Physik, Technische Universit{\"a}t Dortmund, Dortmund, Germany\\
% ...
$ ^{38}$ICCUB, Universitat de Barcelona, Barcelona, Spain\\
$ ^{39}$Universidad de Santiago de Compostela, Santiago de Compostela, Spain\\
$ ^{40}$European Organization for Nuclear Research (CERN), Geneva, Switzerland\\
% ...
$ ^{43}$Nikhef National Institute for Subatomic Physics, Amsterdam, The Netherlands\\
% ...
$ ^{48}$H.H. Wills Physics Laboratory, University of Bristol, Bristol, United Kingdom\\
% ...
$ ^{53}$School of Physics and Astronomy, University of Glasgow, Glasgow, United Kingdom\\
% ...
$ ^{59}$University of Cincinnati, Cincinnati, OH, United States\\
% ...
}
\end{flushleft}



\end{document}